\documentclass{pas}

\usepackage{multirow}

\usepackage{graphicx}	% Including figures
\usepackage{tabularx}

\usepackage{amsmath}
\let\gtrsim\undefined
\let\lesssim\undefined
\usepackage{amssymb}	% Extra maths symbols
\usepackage[nopatch]{microtype}
\usepackage{booktabs}
\usepackage{natbib}

\DeclareUnicodeCharacter{2009}{ }
\DeclareUnicodeCharacter{1E3A}{L} 
\DeclareUnicodeCharacter{223C}{$\sim$}

\usepackage{braket}
\usepackage[table]{xcolor}% http://ctan.org/pkg/xcolor
\usepackage{float, makecell, array}
\usepackage{xspace}
\usepackage{pdflscape}
\usepackage{longtable, makecell}
\usepackage{subfigure}
\usepackage{multirow, multicol}
\usepackage{subcaption}

\newcommand{\beq}{\begin{equation}}
\newcommand{\eeq}{\end{equation}}
\newcommand{\bea}{\begin{eqnarray}}
\newcommand{\eea}{\end{eqnarray}}
\newcommand{\multilinecomment}[1]{}

\newcommand{\kmsMpc}{km\,s$^{-1}$Mpc$^{-1}$}
\newcommand{\msol}{\mathrm{M_{\odot}}}
\newcommand{\dayu}{\mathrm{d}}
\newcommand{\angstrom}{\mbox{\normalfont\AA}}

\newcommand{\red}{\textcolor{red}}

\newcommand{\decam}{DECam\xspace}
\newcommand{\ozdes}{OzDES\xspace}
\newcommand{\sdss}{SDSS\xspace}

\newcommand{\hbeta}{H$\beta$\xspace}
\newcommand{\mgii}{Mg\textsc{II}\xspace}
\newcommand{\civ}{C\textsc{IV}\xspace}

\newcommand{\javelin}{\texttt{JAVELIN}\xspace}
\newcommand{\emcee}{\texttt{emcee}\xspace}
\newcommand{\PyCCF}{\texttt{PyCCF}\xspace}
\newcommand{\numpyro}{\texttt{NumPyro}\xspace}
\newcommand{\pyroa}{\texttt{PyROA}\xspace}

\newcommand{\cream}{\texttt{CREAM}\xspace}
\newcommand{\python}{\texttt{python}\xspace}

\newcommand\tablespacing{6pt}
\newcommand{\tableheader}[1]{Source ID  &  Redshift &  \makecell{Monochromatic \\ Luminosity \\ $\log_{10}(\lambda L_{#1 \angstrom}\mathrm{erg/s})$} & \makecell{Rest-Frame Lag \\ (Days)}  & \makecell{ Velocity \\ Dispersion \\  $\sigma_V $ \\ (km/s)  }&  \makecell{Mass \\($10^6 \msol$) }&  \makecell{Dimensionless \\ Accretion Rate \\ $\dot{M}$}  \\[12pt]}
\LTcapwidth=\textwidth

\begin{document}

\lefttitle{Publications of the Astronomical Society of Australia}
\righttitle{Hugh McDougall}

\jnlPage{1}{32}
\jnlDoiYr{2026}
\doival{10.1017/pasa.2026.10219, DES-2025-860 FERMILAB-PUB-25-0824-PPD}

\articletitt{Research Paper}

\title{OzDES Reverberation Mapping of Active Galactic Nuclei: Final Data Release, Black-Hole Mass Results, \& Scaling Relations}

\author{
\gn{Hugh} \sn{McDougall}$^{1}$, \gn{Tamara M.} \sn{Davis}$^{1}$, \gn{Zhefu} \sn{Yu}$^{2}$, \gn{Paul} \sn{Martini}$^{3, 4}$, \gn{Christopher} \sn{Lidman}$^{5, 6}$, \gn{Umang} \sn{Malik}$^{6}$, \gn{Andrew} \sn{Penton}$^{1}$, \gn{Geraint F.} \sn{Lewis}$^{7}$, \gn{Brad E.} \sn{Tucker}$^{6}$, \gn{Benjamin~J.~S. } \sn{Pope}$^{8, 1}$, \gn{Sahar} \sn{Allam}$^{9}$, \gn{Felipe} \sn{Andrade-Oliveira}$^{10}$, \gn{Jacobo} \sn{Asorey}$^{11}$, \gn{David} \sn{Bacon}$^{12}$, \gn{Sebastian} \sn{Bocquet}$^{13}$, \gn{David} \sn{Brooks}$^{14}$, \gn{Aurelio} \sn{Carnero Rosell}$^{15, 16, 17}$, \gn{Daniela} \sn{Carollo}$^{18}$, \gn{Anthony} \sn{Carr}$^{19}$, \gn{Jorge} \sn{Carretero}$^{20}$, \gn{Ting-Yun} \sn{Cheng}$^{21}$, \gn{Luiz} \sn{da Costa}$^{16}$, \gn{Maria Elidaiana} \sn{da Silva Pereira}$^{22}$, \gn{Juan} \sn{De Vicente}$^{23}$, \gn{H. Thomas} \sn{Diehl}$^{9}$, \gn{Peter} \sn{Doel}$^{14}$, \gn{Spencer} \sn{Everett}$^{24}$, \gn{Juan} \sn{Garcia-Bellido}$^{25}$, \gn{Karl} \sn{Glazebrook}$^{26}$, \gn{Daniel} \sn{Gruen}$^{13}$, \gn{Gaston} \sn{Gutierrez}$^{9}$, \gn{Kenneth} \sn{Herner}$^{9}$, \gn{Samuel} \sn{Hinton}$^{1}$, \gn{Devon L.} \sn{Hollowood}$^{27}$, \gn{David} \sn{James}$^{28}$, \gn{Alex} \sn{Kim}$^{29}$, \gn{Kyler} \sn{Kuehn}$^{30}$, \gn{Sujeong} \sn{Lee}$^{31}$, \gn{Marisa} \sn{March}$^{32}$, \gn{Jennifer} \sn{Marshall}$^{33}$, \gn{Juan} \sn{Mena-Fern\'andez}$^{34}$, \gn{Felipe} \sn{Menanteau}$^{35, 36}$, \gn{Ramon} \sn{Miquel}$^{37, 20}$, \gn{Justin} \sn{Myles}$^{38}$, \gn{Robert} \sn{Nichol}$^{39}$, \gn{Ricardo} \sn{Ogando}$^{40}$, \gn{Anna} \sn{Porredon}$^{23, 41}$, \gn{Eusebio} \sn{Sanchez}$^{23}$, \gn{David} \sn{Sanchez Cid}$^{23, 10}$, \gn{Rob} \sn{Sharp}$^{6}$, \gn{Mathew} \sn{Smith}$^{42}$, \gn{Eric} \sn{Suchyta}$^{43}$, \gn{Molly} \sn{Swanson}$^{35}$, \gn{Chun-Hao} \sn{To}$^{44}$, \gn{Douglas} \sn{Tucker}$^{9}$, \gn{Alistair} \sn{Walker}$^{45}$, \gn{Noah} \sn{Weaverdyck}$^{46, 29}$
}

\affil{Affiliations are listed after the references.}

\corresp{Hugh McDougall, Email: hughmcdougallemail@gmail.com}

\citeauth{McDougall et al., OzDES Reverberation Mapping of Active Galactic Nuclei: Final Data Release, Black-Hole Mass Results, \& Scaling Relations. {\it Publications of the Astronomical Society of Australia} {\bf Online}, 1--32. https://doi.org/10.1017/pasa.2026.10219}

\history{(Received 1 Dec 2025; revised 30 April 2026; accepted 8 May 2026)}

\begin{abstract}
Over the last decade, the Australian Dark Energy (\ozdes) collaboration has used Reverberation Mapping to measure the masses of high redshift supermassive black holes. Here we present the final review and analysis of this \ozdes reverberation mapping campaign. These observations use $6-7$ years of photometric and spectroscopic observations of $735$ Active Galactic Nuclei (AGN) in the redshift range $z\in [0.13, 3.85]$ and bolometric luminosity range $\log_{10}(L_{\mathrm{bol}})\in [44.3, 47.5] \; \mathrm{erg/s}$. Both photometry and spectra are observed in visible wavelengths, allowing for the physical scale of the AGN broad line region to be estimated from reverberations of the \hbeta, \mgii and \civ emission lines. We successfully use reverberation mapping to constrain the masses of $62$ super-massive black holes, and combine with existing data to fit a power law to the lag-luminosity relation for the \hbeta and \mgii lines with a scatter of $\sim0.25$ dex, the tightest yet identified, fit specifically for consistency with high redshift AGN. We fit a similarly constrained relation for \civ, resolving a tension with the low luminosity literature AGN by accounting for selection effects arising from finite survey length. We also examine the impact of emission line width and luminosity (related to accretion rate) in reducing the scatter of these scaling relationships and find no significant improvement over the lag-only approach for any of the three lines. Using these  relations, we further estimate the masses and accretion rates of $246$ AGN with single epoch methods. We also use these relations to estimate the relative sizes of the \hbeta, \mgii and \civ emitting regions, and find evidence that the \mgii emission may occur further out than \hbeta. In short, we provide a comprehensive benchmark of high redshift AGN reverberation mapping at the close of this most recent generation of surveys, including light curves, time-delays, and a set of significantly improved radius-luminosity relations for use with high-redshift populations.
\end{abstract}


\maketitle

%%%%%%%%%%%%%%%%%%%%%%%%%%%%%%%%%%%%%%%%%%%%%%%%%%

%%%%%%%%%%%%%%%%% BODY OF PAPER %%%%%%%%%%%%%%%%%%
\section{Introduction}
\label{sec: introduction}
Every massive galaxy is thought to host a super-massive black hole (SMBH) at its core, with these objects believed to play an important role in galaxy evolution \citep{Kormendy_2013}. Despite their ubiquity, these objects remain poorly understood, with the exact mechanisms of their formation and growth within their host galaxies remaining as open questions \citep[e.g.][]{Volonteri_2021}. Measuring the mass of these SMBHs over cosmic time is then of great interest, as it offers us insight into the history and evolution of these important objects. 

As dark and compact objects, our ability to observe SMBHs by electromagnetic means relies on observations of their environment, and that ability drops rapidly with distance. This obstacle would frustrate our attempts to observe them into the cosmic past were it not for Active Galactic Nuclei (AGN), extremely luminous SMBHs with bright accretion disks that can easily outshine their entire host galaxy. Under the standard model \citep{Urry_1995}, these AGN consist of a central black hole / accretion disk `engine' and a complex surrounding structure, including the fast orbiting material in the broad line region (BLR). As the luminosity, kinematics and structure of the AGN are all powered by the driving gravitational force of the central black hole, many of the AGN's physical properties offer means of inferring the mass of the black hole.

AGN in the local universe can have their masses constrained by the kinematics of their surrounding stars, either directly tracking orbits \citep[e.g.][]{Schodel_2002} or the bulk stellar velocity dispersion \citep[i.e. the `$M-\sigma$ relation'][]{Ferrarese_2000,Gebhardt_2000}, but such techniques become challenging at larger distances where entire galaxies appear as AGN-dominated point sources.  The previous decade has also seen considerable progress in imaging two of the nearest SMBHs with interferometry at millimetre wavelengths \citep[e.g.][]{EHT_2019}, and in using Very Large Telescope interferometry to potentially access finer angular resolutions \citep{GravityPlus_2022}. At higher redshifts, the limits on angular resolutions can also be surpassed with reverberation mapping (RM), a technique in which temporal coverage of variability can substitute for a lack of spatial information \citep{Blandford_McKee_1982, Peterson_1993}.

\begin{figure*}
    \centering
    \includegraphics[width=0.8\linewidth]{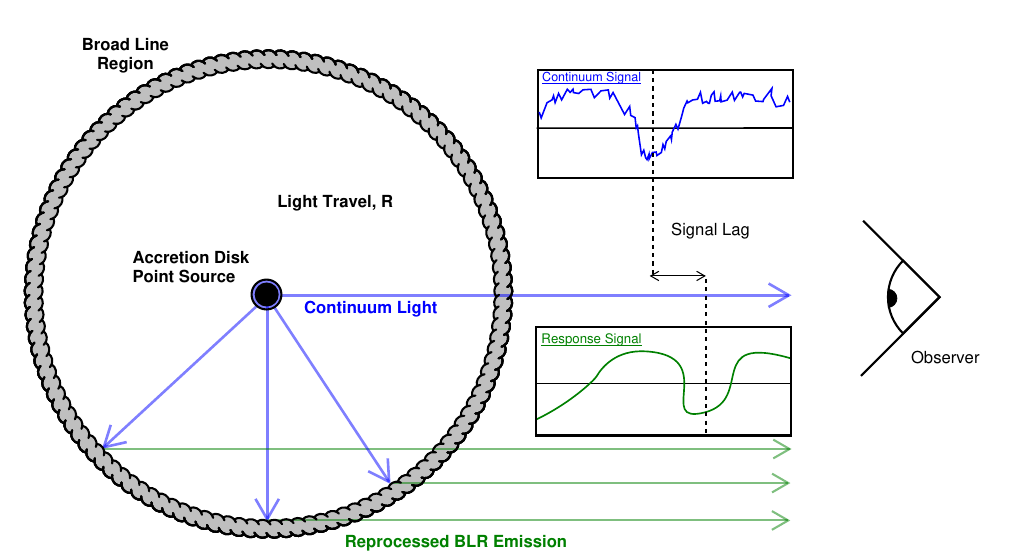}
    \caption{Simplified model of reverberation mapping, showing the different light travel paths for direct and re-processed light. In its simplest `single lag' form, BLR RM relies on the assumption that the accretion disk and BLR are homogeneous, that the accretion disk be reasonably small compared to their angular separation, and the kinematics of the BLR along the line of site be reasonably well characterised by a single representative radius \citep{Shakura_1973, Cackett_2021}. Additional geometric complexity is characterised by the virial factor $\braket{f}$ defined in equation~\ref{eq:RM_mass}.}
    \label{fig: AGN_sketch}
\end{figure*}

Conceptually, reverberation mapping is straightforward: stochastic variations in the brightness of the AGN's central engine drive variations in the brightness of gas in the surrounding broad-line region, but do so with a time delay associated with light travel time between the central engine and the BLR. Where we can distinguish signals from different regions, we can characterise the scale of this time delay and so infer the physical scale of the AGN (see Figure~\ref{fig: AGN_sketch} for a sketch of this process with simplified geometry). Knowledge of this geometric scale, coupled with a measure of the velocity and assumptions about kinematics, provides a means of constraining the mass of the SMBH. As such, reverberation mapping of AGN has become the primary means of constraining SMBH masses at any appreciable redshift, dominating over all other methods beyond $z\approx0.1$ \citep{Cackett_2021}.

For most of the history of reverberation mapping, data have been limited to studies of a hand-full of AGN at a time, predominantly at low redshift \citep[e.g.][]{Peterson_1999, CIV_Peterson_2005, Denney_2006}, limited by the difficulty of continuous photometric and spectroscopic observations over the required time scales. This has changed with the advent of the first generation of `industrial scale surveys', in which reverberation mapping is performed on hundreds of AGN at redshifts probing into the deep cosmological past. Over the last decade, surveys like the Australian Dark Energy Survey \citep[\ozdes;][]{King_2015} and the Sloan Digital Sky Survey \citep[SDSS;][]{SDSS-Shen_2015} have made regular photometric and spectroscopic measurements of well over $1000$ AGN in the redshift range $0<z\lesssim4.5$, dramatically increasing the number of distant AGN observations with optical RM data to a statistically significant sample, and pushing our constraints on SMBH masses well into the high redshift domain.

Reverberation mapping is observationally expensive, requiring many epochs of observation. However, black hole masses can be derived with single epoch spectral observations through an empirically observed power-law relationship between luminosity and reverberation timescale (or lag) for each line, the so-called `$R-L$ Relationship'. The parameters of these relationships are calibrated using RM and they are reasonably well defined at low redshift, but the paucity of measurements of more distant AGN means that there has been difficulty constraining $R-L$ relations for high redshift AGN. One of the aims of \ozdes has been to fill that observational gap. Here we present the complete \ozdes RM sample and also make a comprehensive collation and comparison of contemporary reverberation mapping surveys, and create a catalogue of the SMBH masses derived from reverberation mapping. 

The paper is organised as follows. We discuss the physical principles of reverberation mapping in Section \ref{sec:RM}, including methods and limitations.  In Section~\ref{sec: observations} we introduce the \ozdes and external data sets used in this paper, and then in Section~\ref{sec: methods} we introduce the methodology used by \ozdes, including lag measurement procedures and quality cuts, used to define this data set. We then present the full suite of recovered lags, SMBH masses and other results from the \ozdes RM program in Section~\ref{sec: results-RM}. In Section~\ref{sec: res-RL} we compare the $R-L$ relationship constraints of \ozdes against those of \sdss and a wide range of other surveys for the \hbeta, \mgii and \civ lines, identifying where they agree or are in tension in the $R-L$ plane, and use a combined data set to tightly constrain the $R-L$ slope, offset and scatter for all three lines for samples similar to \ozdes and \sdss. In Section~\ref{sec: res-accretion} we confirm an absence of systematic biases from emission-line width or accretion rate for these $R-L$ fits.  In Section~\ref{sec: res-bolom} we derive a {\em bolometric} $R-L$ relation and use the different lag predictions for \hbeta, \mgii and \civ as a probe of the relative scales of their emission regions. We then apply the updated $R-L$ relationships in Section~\ref{sec: single_epoch} to derive single-epoch SMBH masses for $246$ AGN, and finally discuss the results of this paper, examining possible implications about the evolution of SMBH populations via our single epoch mass estimates, in Section~\ref{sec: discussion}.

\section{Technique and limitations}\label{sec:RM}
\subsection{Reverberation Mapping}
\label{sec: RM_theory}

The core of AGN reverberation mapping is based on arguments of the virialised orbit of the broad line region about the black hole, with the SMBH mass being correlated with the estimated radius of the BLR, $R_\mathrm{BLR}$, and its velocity dispersion $\braket{\sigma_v^2}$ as in  equation~\ref{eq:RM_mass}.  The complexities of the BLR's unresolved geometry are captured in the broad uncertainties of the dimensionless `virial factor', $f$:\footnote{\citet{Mejia_Restrepo_2018} finds that virial factor may be anti-correlated with line width. In this work, we treat $f$ as being independent.}
\begin{equation}    
    M_{\rm BH} = f\frac{ R_{\rm BLR} \braket{\sigma_v^2} }{G}.
    \label{eq:RM_mass}
\end{equation}
Though conceptually simple, each term in the numerator of equation~\ref{eq:RM_mass} is non-trivial to measure for any given source. One hurdle is that the kinematics at play are dependent on the shape and dynamics of the BLR, the nature of which is as yet unresolved \citep{Pancoast_2014a, Pancoast_2014b}. This uncertainty is quantified by calibrating the population average virial factor, $\braket{f}$ from population analysis of nearby AGN, the masses of which are estimated through use of the $M-\sigma$ relation \citep{Woo_2015, Grier_2013b}, or more recently through dynamical modelling via velocity resolved reverberation mapping \citep{Villafana_2023_dynamical, Shen_2024_dynamical}. The $f$ is only loosely constrained for any individual source, as there is an inherent scatter of $\approx0.3-0.4$ dex about $\braket{f}$, and this dominates the uncertainty in BLR RM derived masses \citep{Woo_2015,SDSS-Shen_2023}.

Measurements of $\braket{\sigma_v^2}$ are made from the spectroscopic broadening of the reverberating emission lines, either by their full-width half-maximum (FWHM), or their dispersion (i.e. the second moment of the line profile about its peak), with different virial factors being tuned for each (see Section~\ref{sec: ext_data} for examples of RM papers using either technique).

High redshift AGN are beyond the distances at which we have been able to angularly resolve $R_{\mathrm{BLR}}$, and so BLR RM instead uses the timescale at which light propagates through the AGN geometry to measure this size by proxy (see Figure~\ref{fig: AGN_sketch}). Variations in broad band photometric luminosity are dominated by the light emanating directly from the accretion disk `engine', with these variations being echoed in the driven response in the emission lines of the BLR after some delay, $\Delta t$, typically of order days to months. The BLR spectrum exhibits atomic recombination lines that can be easily distinguished from the broad AGN continuum emission, allowing us to identify this echo through spectroscopic observations even where our photometric  observations are dominated by light from the accretion disk. With sufficiently tight cadence of photometric and spectroscopic observations, the delay (or lag) between driving continuum and driven emission line response, $\Delta t$ can be observed, allowing temporal resolution of observations to substitute for the lack of spatial resolution:
\begin{equation}
    R_{\mathrm{BLR}} = c \Delta t
    \label{eq:RM_radius}.
\end{equation}

To perform reverberation mapping, a source must provide a clearly identifiable reverberating emission line from the BLR. At cosmological scales, the emission from different emission lines redshift in and out of typical spectroscopic wavelength ranges, and so different atomic emission lines are visible at near, moderate, and high redshift ranges. 

For the nearest AGN ($z\lesssim0.6$), optical RM is performed with H$\alpha$ at $6562 \angstrom$ and \hbeta at $4861 \angstrom$  \citep[e.g.][]{HBETA_Bentz_2014}. Beyond measuring the mean lag, some studies have even gone as far as using high signal to noise (SNR) measurements of hydrogen recombination lines to constrain the BLR structure and kinematics in further detail. In this ``velocity resolved reverberation mapping'' \citep[e.g.][]{Denney_2009, Grier_2013a}, the reverberating line is treated as being affected by varying degrees of Doppler broadening from different velocities at different radii within the BLR, resulting in different parts of the line profile reverberating at different times.

At higher redshifts ($0.6\lesssim z\lesssim1.8$), optical RM is performed with the \mgii line at $2798 \angstrom$ \citep[e.g.][]{MgII_Metzroth_2006}. optical RM masses derived from \mgii are complicated by iron contamination from the adjacent FeII emission lines, which can distort estimates of the line width unless accounted for (discussed in detail in Section~\ref{sec: methods}). 

At high redshift ($z\gtrsim 1.8$), optical RM is performed with the \civ line at $1549 \angstrom$. \civ lag measurements have also been made by observing this line in the observer-frame UV range for low redshift and low luminosity sources \citep[e.g.][]{CIV_Rosa_2015, CIV_Peterson_2005, MgII_Metzroth_2006}.

The $R-L$ relationship \citep{Kaspi_2000}, an empirically observed correlation between the luminosity of an AGN and its physical scale as recovered from RM, is parameterised as a power-law expressed as:
\begin{equation}
a
    \label{eq:R-L_relationship-nocalib},
\end{equation}
where $\Delta t$ is the rest-frame lag and $\lambda L_{\lambda}$ is the monochromatic luminosity of the AGN at a line-specific wavelength: typically $5100 \angstrom$ for \hbeta, $3000 \angstrom$ for \mgii, and $1350\angstrom$ for \civ. Though the $R-L$ relationship for \hbeta is well constrained and found to have small scatter  \citep[$\approx 0.1-0.2$ dex;][]{HBETA_Bentz_2013}, recent surveys have suggested this scatter may be higher \citep[e.g.][]{HBETA_Du_2016, HBETA_Du_2018}. The historical $R-L$ relations for \mgii and \civ are found to have higher scatter than \hbeta \citep{MgII_Zajacek_2020, Kaspi_2007}.

The strong dependence of AGN lags on luminosity has given rise to the technique of stacking \citep{Fine_2012_stacking, Fine_2013_stacking, Li_2017_stacking}, in which sources of poor signal to noise but similar luminosity have their lag measurements combined. Through stacking, lag measurements can be made up in aggregate over many sources that would otherwise give poorly constrained results, such as done for the OzDES data in  \citet[][]{OzDES-Malik_2024}.

There have been attempts to explain the diversity of lags through more expressive $R-L$ relationships, reducing the intrinsic scatter by adding predictive variables or dividing the AGN population into sub-groups based on the spectral properties \citep[e.g.][]{Mejia_Restrepo_2018, Mejia_Restrepo_2016}. There is, for example, some evidence that lags are lower for highly accreting sources, above some critical threshold. \citep{HBETA_Du_2016, HBETA_Du_2018}. This is described by the dimensionless accretion rate, $\dot{M}$, given by equation~\ref{eq: accretion}, based on the thin-disk model of \citet{Shakura_1973}.
\begin{equation}
    \dot{M} = 20.1 \left( \frac{L_{5100 \angstrom} / 10^{44} \mathrm{erg/s}}{\cos(i)} \right)^\frac{3}{2} \left( \frac{M}{10^7 \msol}\right)^{-2}.
    \label{eq: accretion}
\end{equation}
Here, $\cos(i)$ is the cosine of the inclination of the AGN, taken to be $0.75$ as an average value for the quasar inclinations in which the BLR and accretion disk are both visible (Seyfert - 1 type AGN), while $L_{5100 \angstrom}$ is the bolometric luminosity at $5100 \angstrom$ in units of $10^{44} \mathrm{erg/s}$ and $M$ is the RM-derived SMBH mass in units of $10^7$ solar masses. The Super-Eddington Accreting Mass Black Hole (SEAMBH) collaboration use $\dot{M}>3$ as a benchmark for the separation between low accretion ``sub-Eddington" sources and the highly accreting ``super-Eddington" sources \citep{Du_2015, HBETA_Du_2016}.

%-----------------------------------------------
\subsection{The Aliasing Problem}
\label{sec: aliasing}

\begin{figure}
    \centering
    \includegraphics[width=0.95\linewidth]{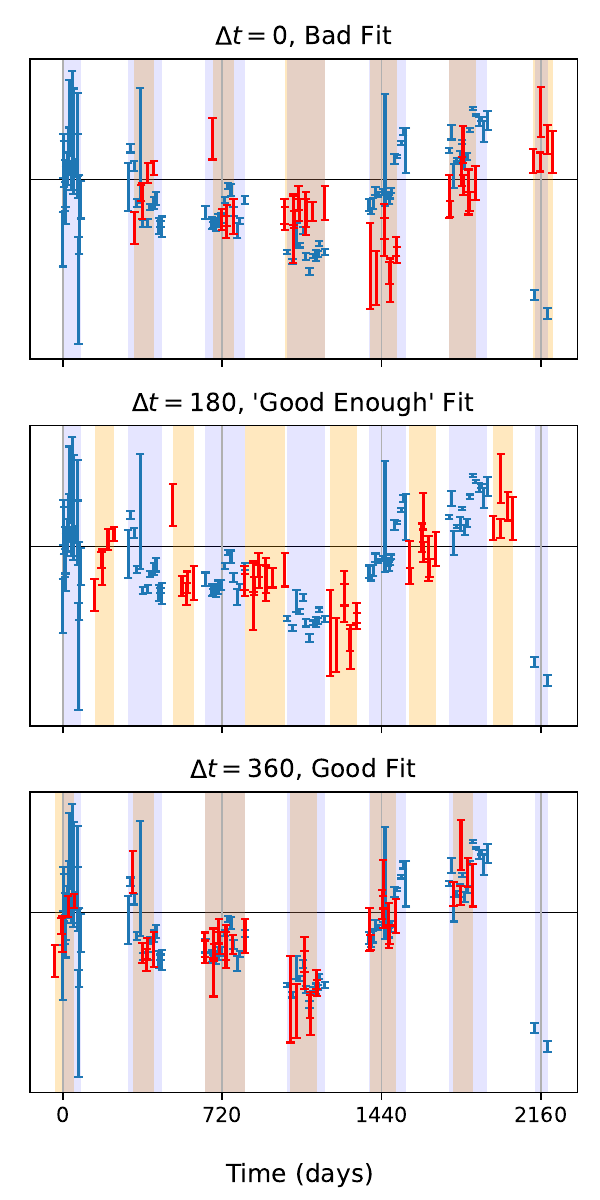}
    \caption{Qualitative demonstration of the source of the aliasing problem for mock RM light curves generated with a true lag of $360\mathrm{d}$, with shaded bands to demonstrate the overlap / gaps in the observations. When observational seasons of our windowing function are of similar or smaller size to the gaps, lags that give no overlap cannot be easily identified as bad fits. This creates local optima in many fitting procedures, inducing `aliasing peaks' in lag recovery distributions every $\approx \! 180 \mathrm{d}$ which can obscure the true lag.}
    \label{fig:aliasing}
\end{figure}

Though the uncertainties in $M_\mathrm{BH}$ within a single source are dominated by the $0.3 - 0.4 $ dex population variability in $f$, lag-recovery presents significant issues with bias and contamination of mass estimates. Cosmological time-dilation means that distant AGN have longer timescales of variability \citep{Lewis_2023}, meaning that RM requires such sources to be observed over a baseline of multiple years to adequately capture variations in their light curves. For the ground-based observations of most distant RM-campaigns, such multi-year observations are necessarily impacted by a seasonal windowing function that imposes $\sim$6-month gaps in observations. These seasonal gaps give rise to the problem of `aliasing' (see Figure~\ref{fig:aliasing}), which can yield spurious lag recoveries at $n+1/2$ yearly gaps \citep[180 days, 420 days, etc.;][]{OzDES-Penton_2021, OzDES-Malik_2022}. 

The dangers posed by aliasing are twofold; firstly, it degrades our ability to detect true lags that fall within seasonal gaps, and more dangerously it can lead to false positives arising within these same gaps \citep{OzDES-Malik_2022}. A common approach is to adopt weighting methods in combination with stringent selection criteria to remove sources with spurious lags. Such approaches reduce contamination from false positives, but at the cost of a drastically reduced sample, with typical acceptance ratios being of order $\approx\!10\%$ in the \ozdes sample. \ozdes selections are based on agreement between competing methods (e.g. \javelin \citep{JAVELIN-Zu_2010} and \PyCCF \citep{PyCCF-Mouyan_2018}; discussed in detail in Section~\ref{sec: methods}), the emergence of a single well constrained lag, and arguments about the physical reasonability of the overall lag posterior distribution (e.g. whether negative lags are properly excluded). More involved approaches rely on characterising the false positive rate (FPR) through simulation of sources \citep[e.g.][]{OzDES-Yu_2023, OzDES-Penton_2025}. \footnote{\nopagebreak Recently, a new lag recovery code has been presented in the form of \texttt{LITMUS},\citep{McDougall_2025_LITMUS} which identifies false positives in a fully Bayesian way, but this method has not been used for any lag recoveries used in this paper.}

%-----------------------------------------------
\section{Data}
\label{sec: observations}

\subsection{\ozdes}
\label{sec: data}

The \ozdes project \citep{OzDES-DR0-Yuan_2015,OZDES-DR1-Childress_2017,OZDES-DR2-Lidman_2020} was a parallel project of the Dark Energy Survey (DES), providing spectroscopic follow-up of sources imaged photometrically by DES. DES itself performed a 6-year imaging survey on the CTIO Blanco 4-metre Telescope in Chile. In addition to a wide-field survey, DES repeatedly imaged ten fields to search for supernovae \citep[field coordinates given in Table 2 of][]{OzDES-Spec-Smith2018}.  These supernova fields had weekly imaging (during each 6 month season) for 5 years from 2013-2018 and a sixth year with lower cadence in 2019 \citep[for details see][]{OZDES-DR2-Lidman_2020}, using \decam in $g, r, i$ and $z$ bands \citep{Flaugher_2015}.  Primarily, these fields were used to obtain supernova light curves, but they also provide an extensive data set for monitoring the variability of other sources, such as AGN.  For ease of access we provide the photometry relevant for AGN reverberation mapping with this paper \red{[link to be provided upon acceptance]}.  The raw images and catalogues from the whole DES survey are available at \url{https://des.ncsa.illinois.edu/releases/dr2}.

To complement the DES photometry, \ozdes obtained monthly spectroscopy using the AAOmega spectrograph fed by the Two Degree Field (2dF) fibre positioner \citep{Lewis_2002_2df} on the Anglo-Australian Telescope \citep[AAT; for details see][]{OZDES-DR2-Lidman_2020}. The 2dF instrument has $\sim$400 optical fibres that can be positioned on targets within a 2-degree diameter field of view, and about a quarter of those were placed on AGN in each exposure. 
Using a spectral resolution of $R=1400$ to $1700$ and a wavelength range of $3700\angstrom$\ to $8800 \angstrom$\ \ozdes monitored a total of $735$ AGN over six years, gathering between $18$ and $25$ epochs on each AGN.  \ozdes typically observed for five months between August and January, perforce leaving large seasonal gaps in the time series data while the fields passed near to the sun. The spectra are publicly available from \url{https://docs.datacentral.org.au/ozdes/overview/dr2/}. 

\subsection{Light Curve \& Spectral Calibration}\label{sec: methods-spec}

For reliable reverberation mapping, a multi-stage calibration procedure of the photometric and spectroscopic measurement is necessary. In \ozdes, we make use of the calibration procedures outlined in \citet{OzDES-Hoormann_2019}, with the exception of line flux and width measurements for the \mgii line for which we use the procedure of \citet{OzDES-Yu_2023}. In this section we briefly review this calibration pipeline, but direct the reader to the original papers for more detail.

\decam measurements provide fluxes at an irregular but approximately weekly cadence in the $g, r, i, z$ filters, though in \ozdes RM we make use of only those in the $g$, $r$ and $i$ bands. In the \citet{OzDES-Hoormann_2019} pipeline, spectroscopic epochs are discarded if they are of low quality \citep[require quality flag 4 or greater, see][]{OzDES-DR0-Yuan_2015} or lack a co-temporal calibrating magnitude. Epochs are then discarded as outliers if they differ by more than $0.2$ magnitudes from the source's mean in the same year and same band pass, with the remaining measurements being averaged if they occur on the same night. An additional filter-dependent calibration uncertainty is also added to the variance as per \citet{Burke_2017}.

\begin{figure}
    \centering
    \includegraphics[width=0.95\linewidth]{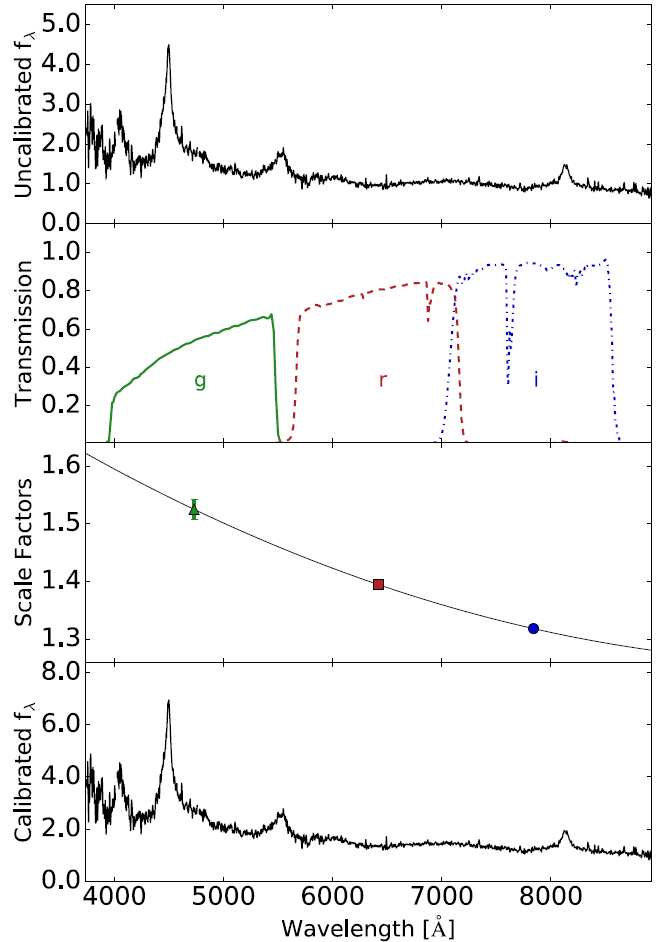}
    \caption{Demonstration of the spectral warping procedure from \citet{OzDES-Hoormann_2019}. The top panel shows a smoothed version of the spectrum of AGN DES J022828.19-040044.30. The second panel shows the $gri$ filter transmission functions, while the third shows the wavelength-dependent transmission coefficients, found by integrating the spectrum with these filters, and the quadratic fit between them, each in units of $10^{-16} \text{erg}\; {\rm s}^{-1} \text{cm}^{-2} \angstrom^{-1} \text{counts}^{-1}$. The bottom panel shows the spectrum after correcting by these scale factors to produce a fully calibrated spectrum in units of $10^{-16} \text{erg}\; {\rm s}^{-1} \text{cm}^{-2} \angstrom^{-1}$.
    }
    \label{fig:JanieCalib}
\end{figure}

Each spectral epoch is calibrated, both converting from photon counts to flux values and correcting for systematic distortions of the spectrum, by comparing the spectra to the  $g$, $r$ and $i$ photometric measurements. Synthetic photometric measurements are made by integrating over the product of the spectrum and band-pass, and a ratio  taken between synthetic and measured brightness. Describing this factor as a function of wavelength, a quadratic is fit across the three measured ratios, acting as a measure of the relative throughput. The entire spectrum was warped to remove this bias (e.g. Figure~\ref{fig:JanieCalib}). After warping, spectra are co-added if they share the same night. 

Once the spectrum is calibrated, line fluxes and widths are calculated by isolating a window of rest-frame wavelengths in which the reverberating line occurs. By linearly interpolating between the flux at the window boundaries, the local continuum is subtracted and the total flux, line dispersion and line full width half maximum estimated. Uncertainties in this value are found by Monte Carlo variation of the window boundaries and the flux measurements within bounds of uncertainty. The flux calibration uncertainties from this procedure is typically on the order of $5-10\%$ across most of the visible wavelength range (See Section~2.2 and Figure~2 of \citet{OzDES-Hoormann_2019}).

Following the approach of \citet{HBETA_Du_2016}, we apply a correction to the $5100 \angstrom$ monochromatic luminosities of all \hbeta sources by way of the empirical scaling relationship provided by Equation~1 in \citet{shen_2011_stellarcontamination}. A more thorough approach would be to decompose the spectra into host galaxy starlight and AGN activity, but is not crucial in this paper as the \ozdes \hbeta sample's small size means that it offers only weak constraining power to the $R-L$ parameters (see Section~\ref{sec: res-RL-Hbeta}).
 
For the \mgii sources we add an additional step to account for the iron emission lines that flank and coincide with the \mgii emission line, which, if not accounted for, can contaminate measurements of line width. In their analysis of the \ozdes \mgii lags, \citet{OzDES-Yu_2023} account for this by fitting iron spectral templates \citep{Template_Tzuzuki_2006, Template_Salviander_2007, Template_Vestergaard_2001} in a Bayesian fashion using the \python Monte Carlo Markov Chain (MCMC) package \emcee. Spectral fits for the emission lines were visually inspected and poorly fitted epochs were removed before lag-fitting with \javelin.

Following the line width pipeline of \citet{OzDES-Hoormann_2019}, the measurements presented here are made using the spectrum averaged over all epochs and use line dispersion as a measure of the line width. All fluxes are converted to luminosities assuming a $\Lambda$CDM cosmology with parameters $H_0=70$~\kmsMpc, $\Omega_m = 0.3$ and $\Omega_\Lambda = 0.7$.

%-----------------------------------------------
\subsection{External data sets}
\label{sec: ext_data}

In addition to presenting our recovered lags, we make use of a number of sources from prior RM works, both for comparison and to improve the constraining power of the $R-L$ relationship. A summary of these sources are listed in Table~\ref{tab: ext_data} The bulk of these lags are those recovered from the Sloan Digital Sky Survey \citep{SDSS-Shen_2023}, a contemporary to \ozdes at a similar `industrial scale' and redshift range. In Section~\ref{sec: SDSS-JAV cuts}, we describe how the \sdss lags use dissimilar selection criteria to \ozdes, including the use of different software as their primary detection method. For the sake of homogeneity in this paper, we adopt the \javelin lags for \sdss and apply an analogue of our own selection criteria. 

\begin{table}
\rowcolors{2}{white}{white}
    \centering
    \begin{tabular}{c|c|c}
        \hline
        \multicolumn{3}{c}{\hbeta Sources}\\\hline
        \thead{Source} & \thead{No. AGN} & \thead{Redshift Range} \\\hline
        \ozdes  &   $8$   &   $0.127 - 0.332$\\\hline
        \sdss   &   $26$  &   $0.289 - 1.003$\\\hline 
        Bentz Collection & $48$& $0.002 - 0.292$\\\hline
        SEAMBH  &   $40$  &   $0.017 - 0.400$\\\hline
        LAMP    &   $16$  &   $0.029 - 0.078$\\\hline
        Misc    &   $6$   &   $0.017 - 0.327$\\\hline 
        \multicolumn{3}{c}{\mgii Sources}\\\hline
        \ozdes  &   $25$  &   $0.840 - 1.860$\\\hline
        \sdss   &   $25$  &   $0.360 - 2.149$\\\hline 
        Zajacek &   $6$   &   $0.003 - 1.890$\\\hline 
        \multicolumn{3}{c}{\civ Sources}\\\hline
        \ozdes  &   $29$  &   $1.922 - 3.451$\\\hline 
        \sdss   &   $15$  &   $1.675 - 2.453$\\\hline 
        Kaspi   &   $17$  &   $0.001 - 3.368$\\\hline 
    \end{tabular}
    \caption{A summary of the number of sources and redshift ranges for the \ozdes, \sdss and other literature data used for constraining $R-L$ relationships in this work. Note that the number of sources here is the number used, and not the total number of published lags from that reference.}
    \label{tab: ext_data}
\end{table}

For low redshift lags associated with the \hbeta line, we also supplement our data with the external lags used in the analysis of \citet{OzDES-Malik_2023}. These include lags from: the Lick AGN Monitoring Project (LAMP) \citep{HBETA_U_2022}, a low redshift ($z<0.08$) study of statistically diverse AGN; the SEAMBH survey, which targets highly accreting high luminosity sources known to produce lower reverberation lags \citep{HBETA_Du_2016, HBETA_Du_2018, HBETA_Hu_2021}; and a collection of low redshift sources collated by \citet{HBETA_Bentz_2013}, coupled with similar sources from \citet{HBETA_Fausnaugh_2017, HBETA_Bentz_2009, HBETA_Bentz_2013, HBETA_Bentz_2014,HBETA_Bentz_2016a,HBETA_Bentz_2016b} and \citet{HBETA_Bentz_2023}. Also included are a number of additional sources from \citet{HBETA_Zhang_2019, HBETA_Li_2021, HBETA_Lu_2016, HBETA_Pei_2014, HBETA_Rakshit_2019}. Sources prior to 2019 and not associated with the \sdss collaboration are drawn from a convenient collation by \citet{HBETA_Martinez-Aldama_2019}.

The combination of a larger sample size and good signal-to-noise yields a larger set of lags associated with the \mgii line in the \ozdes intermediate redshift sample than are recovered in the low redshift \hbeta sample \citep{OzDES-Yu_2021, OzDES-Yu_2023}. In total, $25$ \ozdes \mgii lags are available, which we supplement and contrast with the set of \sdss \mgii lags, a smaller set of lags from \citet{MgII_Zajacek_2021} and a wider collection of sources presented by \citet{MgII_Zajacek_2020}. This collection is comprised of lags collated from \citet{MgII_Lira_2018, MgII_Metzroth_2006, MgII_Czerny_2019} and earlier \sdss lags from \citet{MgII_Shen_2016, MgII_Shen_2019}. We use the entire set of lags as listed by \citet{MgII_Zajacek_2020} when analysing the collection by itself, but when using these data in combination with other surveys we defer to the \sdss measurements when the same source has been reported twice.

For the \civ lags, we compare the \ozdes results presented in \citet{OzDES-Penton_2025} with the latest \sdss \civ lags from \citet{SDSS-Shen_2023} and a diverse collection of earlier sources, which we label as as `The Kaspi Collection'. It is worth noting that the Kaspi Collection includes \sdss lags from \citet{SDSS-Grier_2017, CIV_Grier_2019, MgII_Shen_2019}, as well as earlier \ozdes lags from \citet{OzDES-Hoormann_2019}. As we already analyse more recent and complete \sdss results independently, we do not include these sources in our analysis of the Kaspi Collection.

The sources of the Kaspi Collection can be broadly separated into two groups: high redshift, high luminosity AGN with the \civ line measured in the optical band by \citet{MgII_Lira_2018} and \citet{CIV_Kaspi_2021}, and a number of nearby low luminosity sources from  \citet{CIV_Peterson_2005}, \citet{MgII_Metzroth_2006} and \citet{CIV_Rosa_2015} in which the \civ line is observed in the observer-frame UV. We separate these two groups and label them as `Kaspi High-Z' and `Kaspi Low-Z'. As we discuss further in Section~\ref{sec: res-RL-CIV}, these two sub-samples have different statistical properties, and treating them as a single group runs afoul of tensions when constraining the $R-L$ relationship.

When considering the velocity in equation~\ref{eq:RM_mass}, there are two decisions about how this is interpreted from the quasar spectra: whether to use the dispersion / variance of the line profile (disp) or its full-width half maximum (FWHM), and whether to make these measurements from the spectrum as time-averaged over all epochs (mean-spec) or by its root-mean-squared over all epochs (RMS-spec). Different examples of prior RM work occupy all four quadrants of this decision tree. Note that the disp/FWHM decision represents a different choice of physical measurement, and so leads to a correspondingly different virial factor, while the mean-spec/rms-spec decision is purely about which gives the clearer measurement.

The dispersion approach tends to be the preferred approach in more modern surveys \citep[e.g.][]{SDSS-Shen_2023, HBETA_Rakshit_2019, HBETA_Zhang_2019}, while both mean-spec \citep[e.g.][]{HBETA_Li_2021} and RMS-spec \citep[e.g.][]{Lu_2019} approaches are still in active use. The RMS spectrum is preferred for high SNR sources to isolate the time-varying component of the signals \citep[e.g.][]{HBETA_Bentz_2014, HBETA_Pei_2014}, but many surveys will measure line widths in all, or at least multiple, ways \citep[e.g.][]{SDSS-Shen_2023, HBETA_Fausnaugh_2017}. 

In this work, as with past \ozdes RM papers, we use the dispersion-based definition of line width, as measured on the mean spectrum, as our measure of velocity rather than the FWHM, and default to this measurement for literature measurements where available in the source paper. In making the decision between mean-spec and RMS-spec, we defer to the preferred method of the source paper, e.g. mean spectrum for sources from \citet{HBETA_Bentz_2013}, RMS spectra for sources from \citet{SDSS-Shen_2023} etc. For source-papers in which only the FWHM line widths are provided, we follow the lead of \citet{CIV_Kaspi_2021} and use ${\rm disp}={\rm FWHM}/2.35$ based on an assumption of a Gaussian line profile.

%-----------------------------------------------

\section{Methods}
\label{sec: methods}

%-----------------------------------------------

\subsection{Lag Recovery}

The central element of reverberation mapping is identifying and constraining the delay between two light curves, a task that is complicated by the stochastic variability in the underlying signal, and a host of obstacles arising from the windowing function. These issues have led to a wide range of competing techniques and software for computing the lag in AGN-RM. These can be broadly sorted into two categories: the first containing non-parametric techniques that assume no underlying statistical behaviour and fit to only the light curves, and the second being those that model the AGN variability as a Gaussian process.

Modelling the AGN light curve as a Gaussian process is physically motivated, as they are known to show consistent patterns in their variability \citep{MacLeod_2010, Zu_2013,Kozlowski_2016}, but the associated fitting procedures are computationally expensive and fraught with numerical and statistical obstacles \citep{OzDES-Penton_2021, OzDES-Malik_2022, Read_2019}. As a result, non-parametric methods like \PyCCF \citep{PyCCF-Mouyan_2018} and \pyroa \citep{pyroa-ferus_2021} are still in use as more robust, if less precise, alternatives to full Bayesian modelling. The \sdss team uses \pyroa for their `primary' lag recovery method, while \ozdes use a combination of the GP-based \javelin and the non-parametric \PyCCF for validation and source selection cuts. For the interested reader, in Appendix~\ref{app:methods} we provide an introduction to these fitting methods, their statistical basis, and numerical shortfalls.

\subsection{Post-Recovery Quality Cuts}
The stochastic nature of AGN light curves means that non-physical false positive lags are a possibility even in the presence of arbitrarily good measurements. Coupled with matters of measurement noise, calibration error and, in multi-year surveys such as \ozdes, the impacts of aliasing from the seasonal windowing function, the contamination rate from such false positives can outstrip physically meaningful lag recoveries by near to an order of magnitude if unaccounted for. 

Bespoke analysis can be used to assess reliability of a handful of sources at a time, but this approach is untenable in industrial-scale studies like \ozdes. Instead, it is necessary to use a general, widely applicable criteria for selecting and post-processing the lag recoveries of the sample sources at large to remove spurious results \citep{OzDES-Penton_2021, OzDES-Malik_2022}.

The methods for suppressing the impacts of aliasing through quality cuts varies significantly between different studies. In this section we outline the quality cuts used in each of the previous \ozdes RM papers as well their underlying motivation, and present also a set of selection criteria to apply to the \sdss RM sample of \citet{SDSS-Shen_2023} aimed at drawing from their data a set of sources with selection effects analogous to those of \ozdes.

%=============================================
\subsubsection{Quality Cuts for \hbeta Lags}
\label{sec: methods_Hbetacut}

\ozdes tracked $78$ AGN sources in the redshift range for \hbeta line visibility, with only $5$ sources passing the quality cuts outlined in \citet{OzDES-Malik_2023}. The low luminosities for these sources give lags in the $20-200$ day range, meaning only a small fraction of the brightest sources are expected to be heavily impacted by aliasing, which becomes more severe when the true lag is `off season', i.e. at $\Delta t \approx 180 \mathrm{d}, 540 \mathrm{d}$ etc. \citep{OzDES-Malik_2022}. The \ozdes quality cuts for \hbeta are that each source must have:

\begin{enumerate}
    \item Uncertainties from the \javelin recovered lag are less than $\text{max}(30 \text{d}, \Delta t_{\text{JAV}})$,
	\item Lags from \PyCCF and \javelin agree to within $2 \sigma$ of the \javelin uncertainties,
	\item The maximum correlation from \PyCCF ($r$ in equation~\ref{eq: cross-correlation}) gives $r_{\rm max}>0.6$, and
	\item The \PyCCF false positive rate (p-value) is $<0.05$.
\end{enumerate}

The first cut selects only well constrained lags, the second leverages the robustness of the Interpolated Cross-Correlation Function (ICCF) to account for spuriously constrained \javelin false positives, and the last two cuts ensure that we keep only lags that do not favour the null hypothesis of there being no relation between the two light curves.

%=============================================
\subsubsection{Quality Cuts for \mgii Lags}
\label{sec: methods_MgIIcut}

The majority of \ozdes AGN sources are observed in the redshift range that allow for visibility of the \mgii line, with $453$ of its $753$ target AGN being in the $z\in[0.65, 1.92]$ redshift range where the \mgii line is visible in range of \ozdes data. These \mgii sources exist over a wide redshift range, and the range of observed lags significantly overlaps with the $n+1/2$ year `danger-zones' for aliasing effects. These high redshift AGN lags require strict and principled quality cuts to avoid contamination of the $R-L$ relationship. To this end, \citet{OzDES-Yu_2023} applies two distinct sets of selection criteria, one based on the posterior distribution weighting approach of \citet{CIV_Grier_2019}, and another based on the simulation based false positive rate estimation of \citet{OzDES-Penton_2021}.

In the \sdss approach, the marginalised lag posterior distribution $P(\Delta t)$, as recovered by \pyroa, \javelin or \PyCCF, is attenuated by a weighting function that down-weights lags that correspond to the poorly constrained seasonal gaps. This weighting function is the convolution of two components. The first term, based on the fraction of observations that overlap between the two light curves after shifting by some lag, down-weights `off-season' lags in seasonal gaps where this overlap is small. The second term is the auto-correlation function of the continuum signal (ACF$_{\rm Cont}$), which smooths and widens the `on-season' peaks of the weighting function to account for the continuous nature of the light curve. The entire posterior is then convolved with Gaussian smoothing kernel of width $15$ days (equation~\ref{eq: SDSS-weighting}):
\begin{equation}
    P^\prime(\Delta t) = \left( P(\Delta t) \times 
    \left[ 
        \left(\frac{N(\Delta t)}{N(0)} \right) ^ 2 \circledast {\rm ACF}_{\mathrm{Cont}}(\Delta t) 
    \right] \right)
    \circledast \mathcal{N} \left(\frac{\Delta t}{15{\rm d}} \right) ,
    \label{eq: SDSS-weighting}
\end{equation}
where $\circledast$ represents the convolution operator. The square bracketed terms represent the weighting function while the convolved normal distribution represents the final smoothing.

This ad-hoc approach suppresses lags that are in danger of being the result of aliasing, retaining only those that give an extremely high likelihood. Different quality cuts have been used in past analysis based on this weighting, including discarding sources in which the weighting reduces the posterior evidence by too large of a fraction \citep[e.g.][]{CIV_Grier_2019}. 

In \citet{OzDES-Yu_2023}, the \ozdes analysis reports lags from the un-weighted distribution as produced by \javelin, but uses the weighted distribution to identify the `primary' peak as being the highest likelihood mode in the weighted-distribution $P^\prime(\Delta t)$. Sources are retained only if the unweighted distribution has:
\begin{enumerate}
	\item More than $60$ \% of its evidence (posterior distribution integral) contained within the primary peak, 
	\item The width of this peak, as measured between the 16\textsuperscript{th} and 84\textsuperscript{th} percentiles, be less than $110$ days, and
	\item This peak be in agreement with the unweighted \PyCCF distribution to within $2 \sigma$.
\end{enumerate}

\citet{OzDES-Yu_2023} also includes quality cuts modelled after the more rigorous simulation-work of \citet{OzDES-Penton_2021}, in which simulations of mock DRW signals are used to characterise false positive rates, and quality cuts tuned to remove suspicious sources. These cuts, designed specifically for unweighted \javelin posterior distributions and \ozdes-like observations, require:

\begin{enumerate}
	\item That the standard deviation of the \javelin posterior distribution be less than $110$ days,
	\item That the separation of the \javelin median lag \& peak likelihood lag be within $110$ days of one another, and   
	\item That the lags from \PyCCF and \javelin agree to within $110$ days. 
\end{enumerate}

Of the $453$ available \mgii sources, $25$ pass both sets of quality cuts and produce reliable lags, $\approx\! 5.5 \%$ of the initial sample, with an estimated false-positive rate in this post-cut sample of $\approx \! 4 \%$ (i.e.\ $1$ false positive out of the $25$). 

%=============================================
\subsubsection{Quality Cuts for \civ Lags}
\label{sec: methods_CIVcut}

\ozdes tracked $305$ high redshift AGN with \civ lines visible for reverberation mapping. These distant sources represent the earliest and most luminous AGN in the \ozdes RM sample, and, owing to the anti-correlation of AGN luminosity and optical variability \citep{MacLeod_2010}, the sources with the weakest AGN variability. These sources are also observed with the shortest rest-frame time window due to higher time dilation at these higher redshifts. In combination, lag recovery for the \civ sample is a difficult process with an outsized risk of false positives if approached naively. For \civ RM of the \ozdes sample, \citet{OzDES-Penton_2021} provide a series of quality cuts to maximise the reliability of $R-L$ constraints and minimise the rate of false positives, tuned on mock simulations of \ozdes-like data. Under these cuts, sources are sorted into quality levels of `bronze', `silver', `gold', or are rejected completely based on the properties of their marginalised lag posterior distribution and the lag recovery results from the ICCF method (examples of gold and bronze \civ sources can be seen in Figure~\ref{fig: CIV_example_curves)}. These cuts are:
\begin{enumerate}
    \item That the \javelin posterior rule out $\Delta t = 0$ to within $3 \sigma$,
    \item That the lag of \javelin and ICCF lags agree to within $100 \mathrm{d}$,
    \item That the \javelin lag posterior median and maximum posterior value be similar (within $110 \mathrm{d}$ for bronze quality, $80 \mathrm{d}$ for silver and $65 \mathrm{d}$ for gold),
    \item That the \javelin posterior is strongly constrained to a single peak, with a high fraction ($33\%, 45\%$ and $65\%$ for bronze, silver and gold, respectively) of posterior density / MCMC chain samples falling within a single mode.
\end{enumerate}

After these cuts, \citet{OzDES-Penton_2025} recovered $29$ \civ lags, with $6$ each at the gold and silver confidence level and $17$ at bronze. This analysis included a re-examination of the two \civ lags from \citet{OzDES-Hoormann_2019} with more years of observations.

\subsubsection{SDSS Sub-Sampling Criteria}
\label{sec: SDSS-JAV cuts}

In the final data release for the Sloan Digital Sky Survey's 7-yr RM campaign, \citet{SDSS-Shen_2023} constrain and select lags in a fundamentally different way to \ozdes. Rather than using simulation-based estimates of the false positive rate, \sdss instead allow each source to fit for both positive and negative lags, and then apply the aliasing mitigation weighting strategy outlined in Section~\ref{sec: methods_MgIIcut}. Under the rationale that any lag recovered at $\Delta t<0$ must necessarily be a false positive, they use this as an estimate of the population level false positive rate, using this to tune their selection criteria. To this end, they make use of only one quality cut: that the ICCF $r$ value (equation~\ref{eq: cross-correlation}) must be $>0.4$ as evaluated at the lag recovered by their chosen primary lag recovery method \pyroa. This is similar to the selection criteria of the \ozdes \hbeta analysis of \citet{OzDES-Malik_2023}, which requires $r>0.6$ at the lag as recovered by \javelin.

By contrast to \sdss, \ozdes treats this lag search range as a true prior, examining only the physically reasonable positive lag domain and applying stringent cuts to minimise the effect of aliasing on each individual source. This produces a more restricted sample with fundamentally different statistical properties, which can be seen in the tension between the \sdss and \ozdes $R-L$ constraints. To make joint fits to the $R-L$ relation using both the \sdss and \ozdes data sets, we required a set of lags with relatively homogeneous selection properties. To that end, we derive an alternate set of lags from the published \sdss data of \citet{SDSS-Shen_2023} that differ from their sample in two ways: firstly, in that we use the lags as recovered from \javelin instead of their choice of \pyroa, and secondly that we apply a number of quality cuts to this alternate sample to bring it more in line with the rationale of the \ozdes selections.

Our sub-sampling criteria for the \sdss sources are:
\begin{enumerate}
    \item That recovered lags must be positive, $\Delta t>0$ at $1\sigma$.
    \item \label{enum: subsample2} That results should be robust across methods, such that \javelin, ICCF and \pyroa all agree in peak lag to within $2\sigma$ of the \javelin uncertainty, 
    \item That the fit should be statistically significant, as measured by setting $r>0.6$ evaluated at the \javelin modal lag,
    \item That the anti-aliasing down-weighting weighting scheme must not overly distort the lag probability distribution. We require that the posterior evidence rejected (fraction of MCMC chain samples in their terminology) by the \sdss anti-aliasing weighting must be $<60 \%$.
\end{enumerate}

For step~\ref{enum: subsample2}, in cases where the \javelin lag peak is strongly constrained, we replace the \javelin confidence interval in this test with $30$ days for \hbeta or $110$ days for \civ and \mgii, following a similar reasoning to the cuts of \citet{OzDES-Penton_2021}. After applying these cuts, we retain a sub-sample with $26$ \hbeta lags, $25$ \mgii lags and $15$ \civ lags from the \sdss sample. \javelin's MCMC sampler is known to exhibit strong aliasing artefacts in seasonal signals \citep[see][]{McDougall_2025_LITMUS}, meaning that its use by itself can give misleading results. However, \javelin's use in concert with specifically tuned quality cuts produces a set of reliable lags with a low FPR, as validated by the simulations of \citet{OzDES-Penton_2021}. We use \sdss's \javelin lags rather than their \pyroa lags, as the \ozdes quality cuts are tuned specifically for \javelin's systematics.

The motivation of this sub-sampling is to sacrifice much of the \sdss sample's statistical power by cutting many sources, including discarding what is likely a large fraction of true positives, to yield a smaller sample that has \ozdes's high reliability on a per-source basis. This \sdss-\javelin subsample is broadly similar to the published \sdss sample for the \hbeta and \civ lags, but prefers a markedly for lower scatter in the \mgii $R-L$ relationship. For the \mgii lags, these selection criteria drastically reduce the scatter of the lags about their best-fit $R-L$ relationship, and marginally decrease its preferred slope (see Figure~\ref{fig:RL_params_all_lines} for a comparison).

%----------------------------------------------------
\section{\ozdes Reverberation Mapping Results}
\label{sec: results-RM}
In this section we summarise the reverberation mapping results across past \ozdes papers, including the reverberation lags and the resulting estimated SMBH masses.

\begin{figure*}
    \centering
    \includegraphics[width=0.75\linewidth]{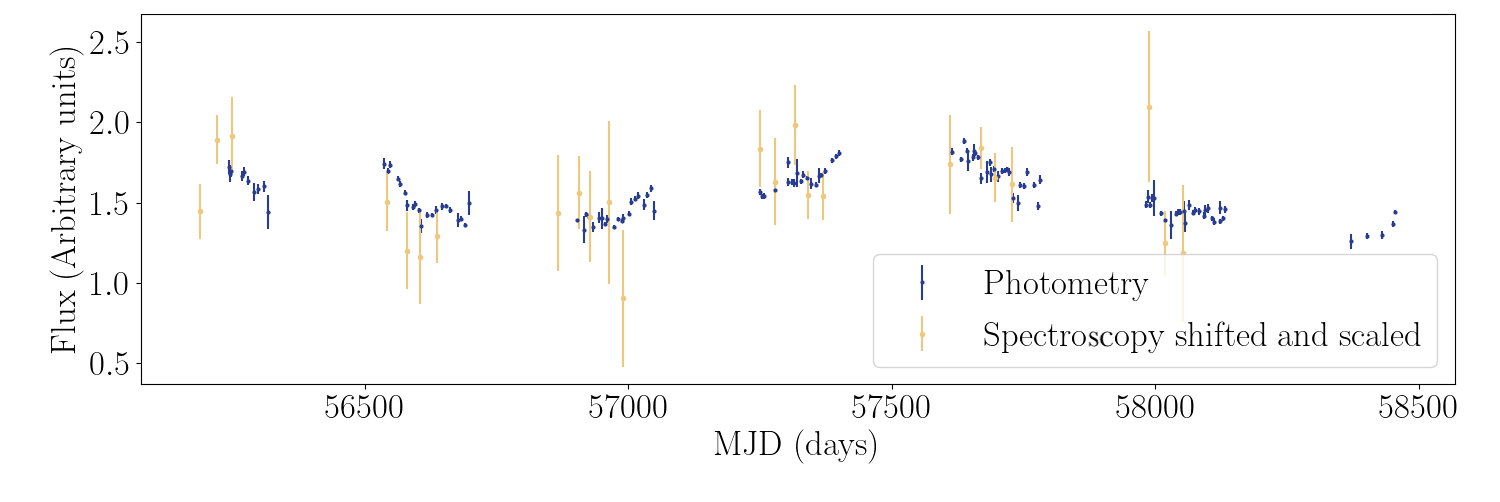}\\
    
    \includegraphics[width=0.75\linewidth]{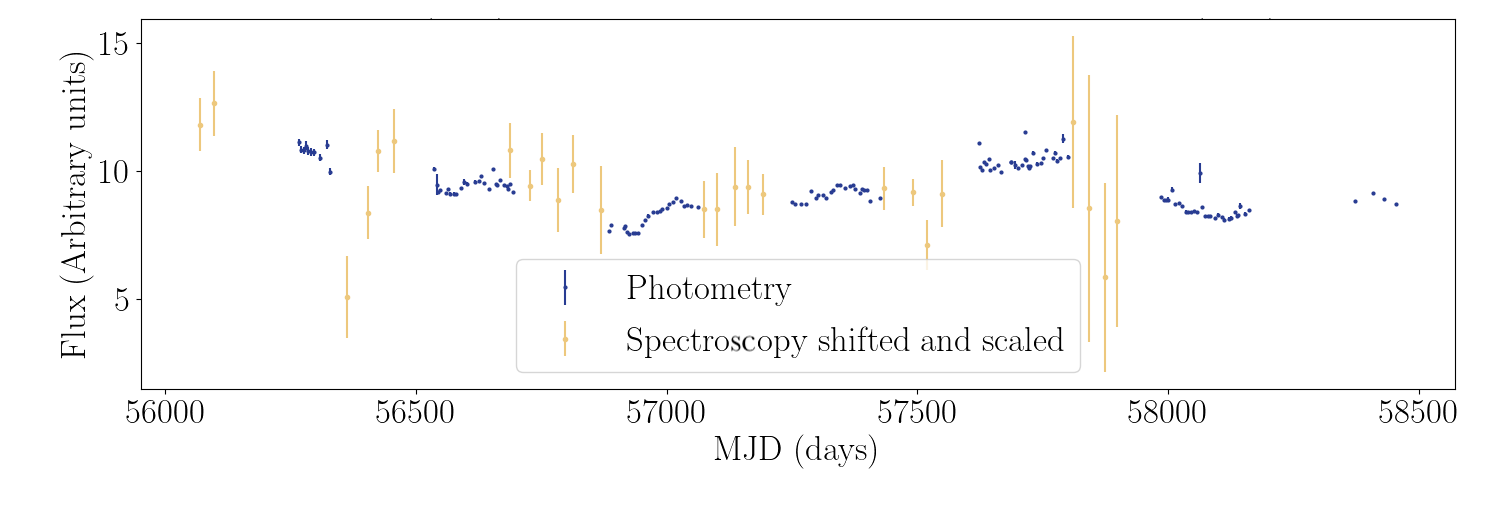}
    \caption{Examples of the lag and scale-corrected light curves (maximum a posteriori estimate) for two \ozdes \civ sources: DES~J022620.86-045946.48(top, gold quality recovery) and DES~J032703.62-274425.27 (bottom, bronze quality recovery). These are adapted from \citet{OzDES-Penton_2025}, the initial paper for these lag recoveries. For a list of the criteria used to classify sources into these grades, see Section~\ref{sec: methods_CIVcut}.}
    \label{fig: CIV_example_curves}
\end{figure*}

%--------------------------
\vspace{-5mm}
\subsection{Lag measurements}

\label{sec: res-lags}
In total, the \ozdes reverberation mapping program has produced $62$ high quality AGN lags, consisting of $8$ \hbeta lags \citep{OzDES-Malik_2023}, $25$ \mgii lags \citep{OzDES-Yu_2021, OzDES-Yu_2023} and $29$ \civ lags \citep{OzDES-Penton_2025}. For each of these sources, we calculate the mass using equation~\ref{eq:RM_mass} and accretion rate using equation~\ref{eq: accretion}, where the $5100 \angstrom$ monochromatic luminosity uses bolometric corrections of \citet{Runnoe_2012} (Table 2 in their paper) and we use the virial factor of $\log_{10}(f) = 0.62 \pm 0.07$ from \citet{SDSS-Shen_2023}, as well as their $0.31$ dex inherent scatter. Rest-frame lags range from $16^{+2.7}_{-2.7} \; \dayu$\footnote{$\dayu$ being days, here and elsewhere in the paper.} to $335^{+2.5}_{-16.0} \; \dayu$, sitting within the range of lags recovered by other works. We find that the distribution of lags over redshift is reasonably consistent with existing reverberation mapping results (Figure~\ref{fig:All_population_plots}, upper left). We note that the distribution of lags changes abruptly as one moves between lines, most notably between the \mgii and \civ lines, indicating different lags, and thus radii, for different ions. This is quantified in Section~\ref{sec: res-bolom}. These RM results, as well as the redshifts and luminosities for the relevant sources, are listed in Table~\ref{tab:Results_All_Hbeta} for our \hbeta sources, Table~\ref{tab:Results_All_MgII} for \mgii sources, and Table~\ref{tab:Results_All_CIV} for \civ sources.

%--------------------------
\vspace{-5mm}
\subsection{Black hole masses from direct RM measurements}
\label{sec: res-masses}

In total, we present a compilation of the $62$ SMBH masses from \ozdes reverberation mapping, ranging from $25.3_{-6.8}^{+8.5}\times 10^6 \msol$  to $4.0_{-1.7}^{+1.8} \times 10^9 \msol$. The relevant measurements are listed, along with estimates of the corresponding accretion rates, in Tables \ref{tab:Results_All_Hbeta}, \ref{tab:Results_All_MgII} and \ref{tab:Results_All_CIV}. Accretion rates are calculated per equation~\ref{eq: accretion}, correcting luminosities to the $L_{5100 \angstrom}$ using the bolometric corrections of \citet{Runnoe_2012}. The calculation uses full error propagation by varying the lag, luminosity and line width within their respective measurement uncertainties. Results are reported as the media, with uncertainties drawn from the 16\textsuperscript{th} and 84\textsuperscript{th} percentiles.

\begin{figure*}
    \centering
    \includegraphics[width=0.45\linewidth]{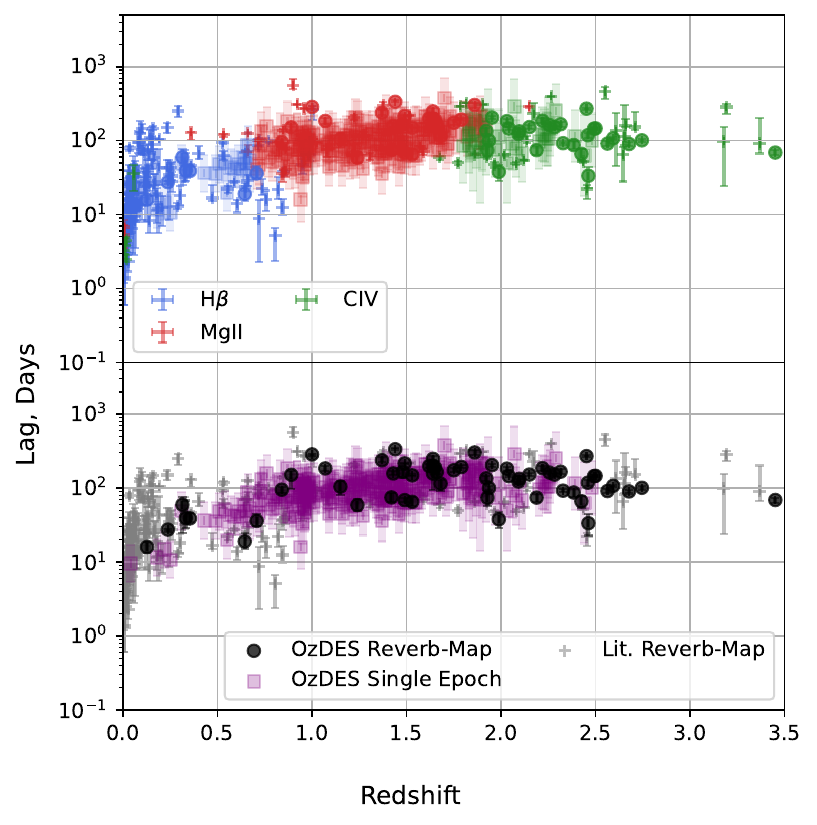}
    \includegraphics[width=0.45\linewidth]{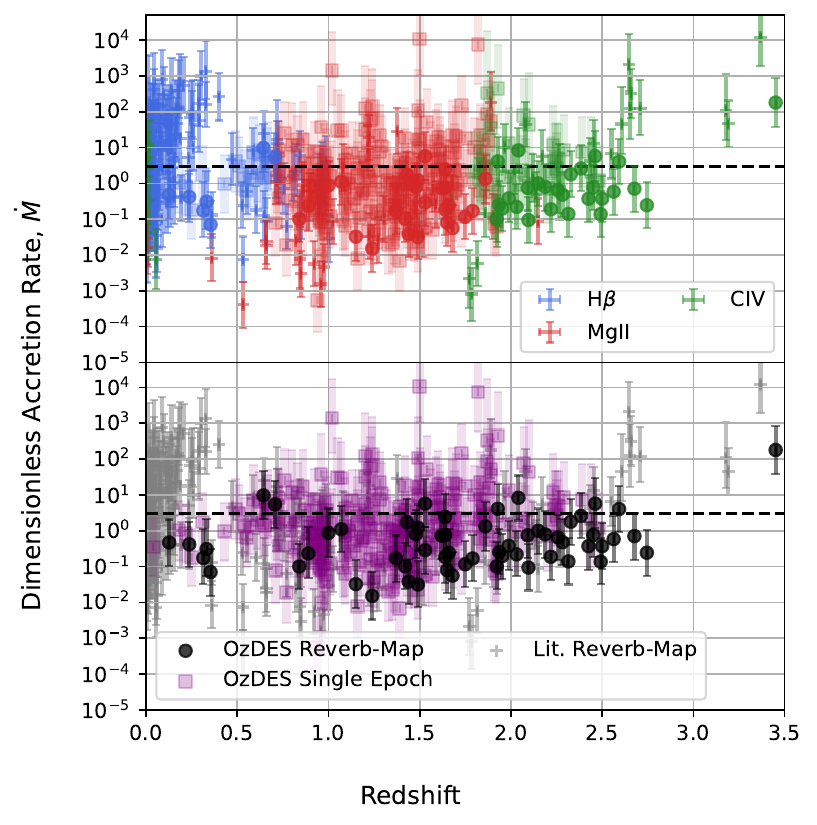}\\
    \includegraphics[width=0.45\linewidth]{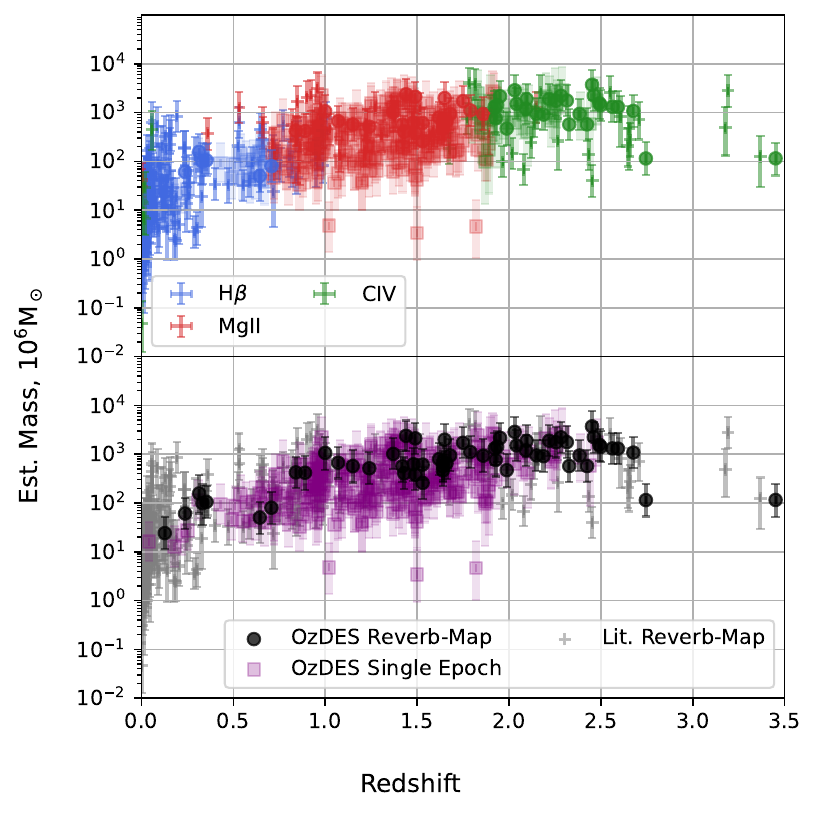}
    \includegraphics[width=0.45\linewidth]{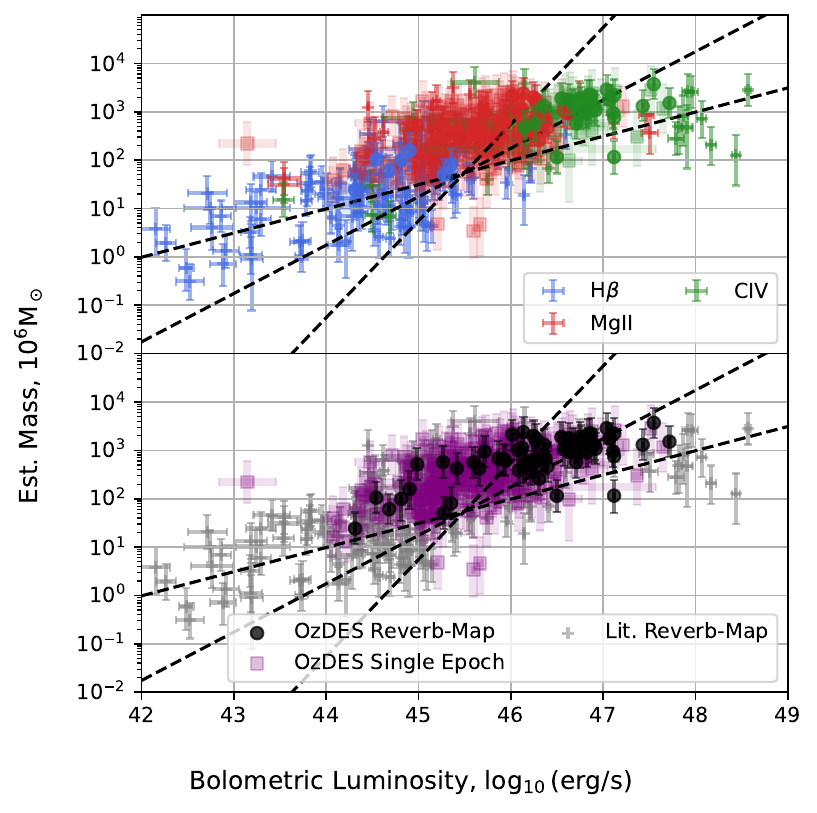}
    \caption{
    A summary of all \ozdes reverberation mapping (circles) and single epoch findings (squares) as well as comparison with literature RM results (plus signs). The top row plots measured and estimated lags (left, estimates from the R-L relationship) and accretion rates (right) against redshift, while the bottom row shows SBMH mass plotted against redshift (left) and bolometric luminosity (right). In each plot, the top panel colours sources by emission line, while the bottom colours by data source. On the mass vs luminosity plot, we also overlay power laws of index $0.5, 1.0$ and $2.0$ as a way to illustrate the slope of the relation.}
    \label{fig:All_population_plots}
\end{figure*}

We broadly see an increase in mass with redshift (Figure~\ref{fig:All_population_plots} lower left), which is expected, as the AGN at higher redshifts are selected to be more luminous \citep[see, for example, Figure 1 in][]{OzDES-Malik_2024}, and this selection effect will dominate any evolutionary effects. We similarly see the expected general increase of more luminous AGN being more massive, but are unable to infer a clear power-law relationship (Figure~\ref{fig:All_population_plots} lower right). 

It should be noted also that the estimated accretion rates for high redshift \civ sources can be extremely high (Figure~\ref{fig:All_population_plots} upper right), with most direct RM samples sitting above the $\dot{M}=3$ threshold used by \citet{Du_2015} as the division between low and high accretion sources. In Section~\ref{sec: res-RL-CIV} we note an apparent flattening of the \civ $R-L$ curve at high luminosity, and, given that SEAMBH associate high accretion rates with low lags \citep{HBETA_U_2022}, this offers a possible physical explanation. Within the \ozdes sample, there is a weak trend for more luminous, higher-redshift sources to accrete more strongly (black data points, Figure~\ref{fig:All_population_plots} upper right).

%----------------------------------------------------

\begin{figure*}
    \centering
    \includegraphics[width=\linewidth]{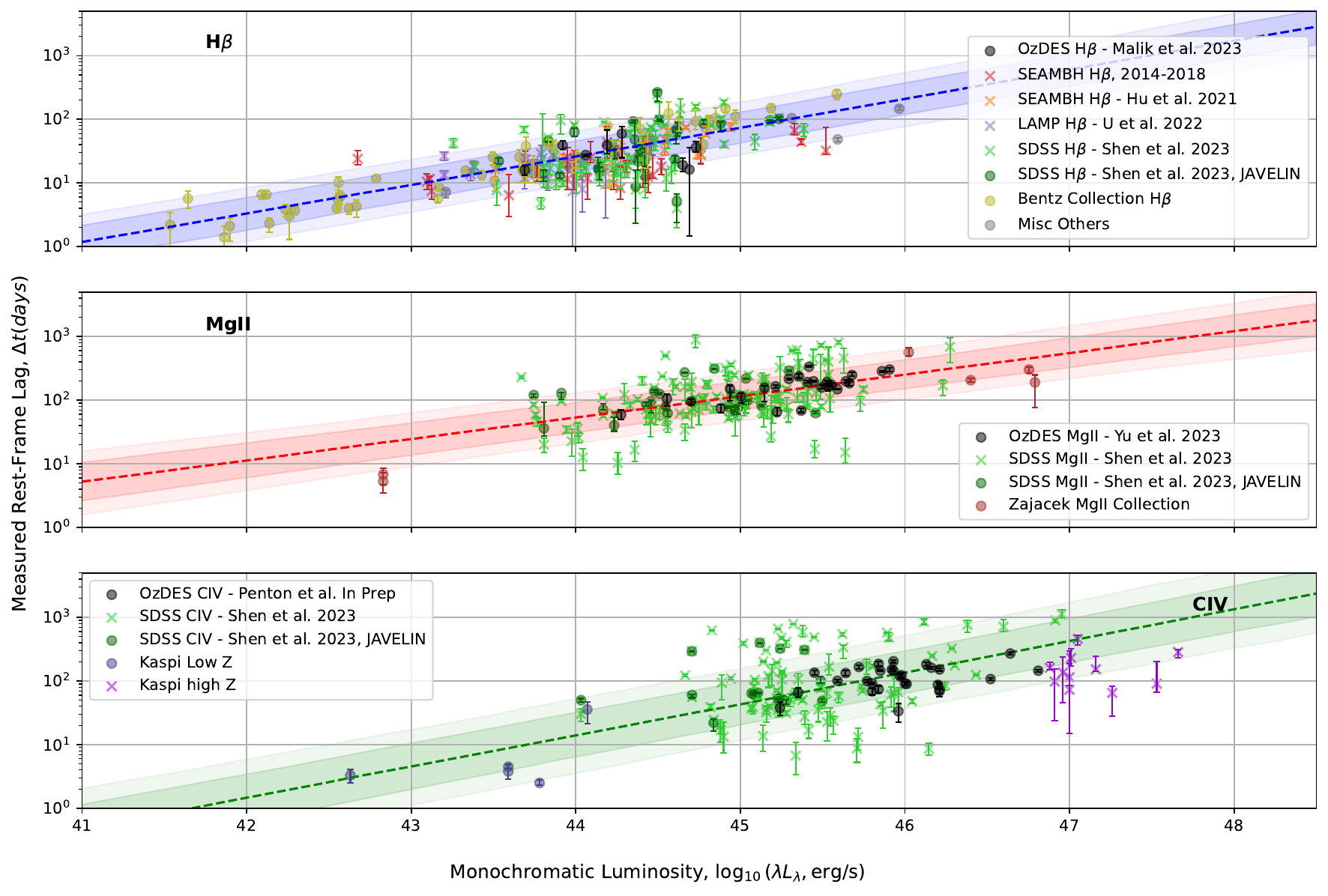}
    \caption{Rest-frame lags and monochromatic luminosities for all data sources from all lines, shown on a log-log scale to show the linear trend that forms the basis for the $R-L$ relationship. Sources marked with a circle contribute to the constraint of the shown $R-L$ relationship, while sources marked with a cross do not. Sub-plots from top to bottom are for \hbeta, \mgii and \civ lags with their `best fit' $R-L$ relationships overlaid (see Table \ref{tab:R-L_Summary}). The monochromatic luminosity is measured at $5100 \angstrom$, $3000 \angstrom$ and $1350 \angstrom$ from top to bottom.}
    \label{fig:RL-all-data}
\end{figure*}

%--------------------------
% R-L CONTOURS

    \begin{figure*}
        \centering

        \subfigure[]{
            \includegraphics[width=0.39\linewidth]{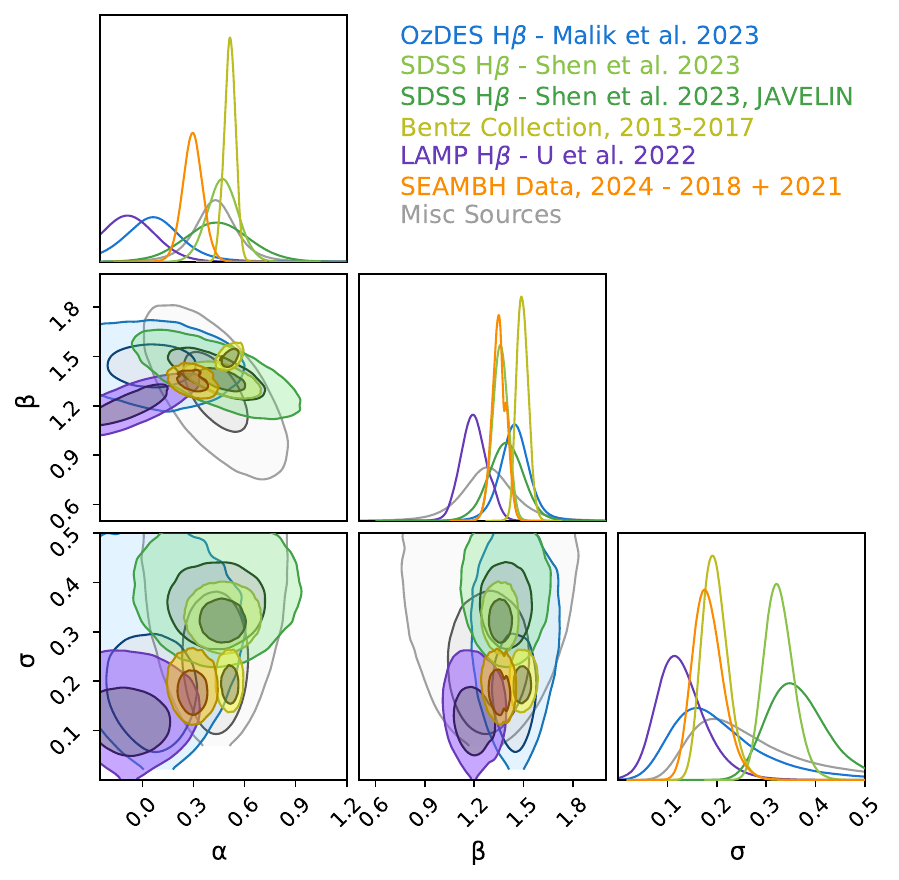}
        }\subfigure[]{
            \includegraphics[width=0.39\linewidth]{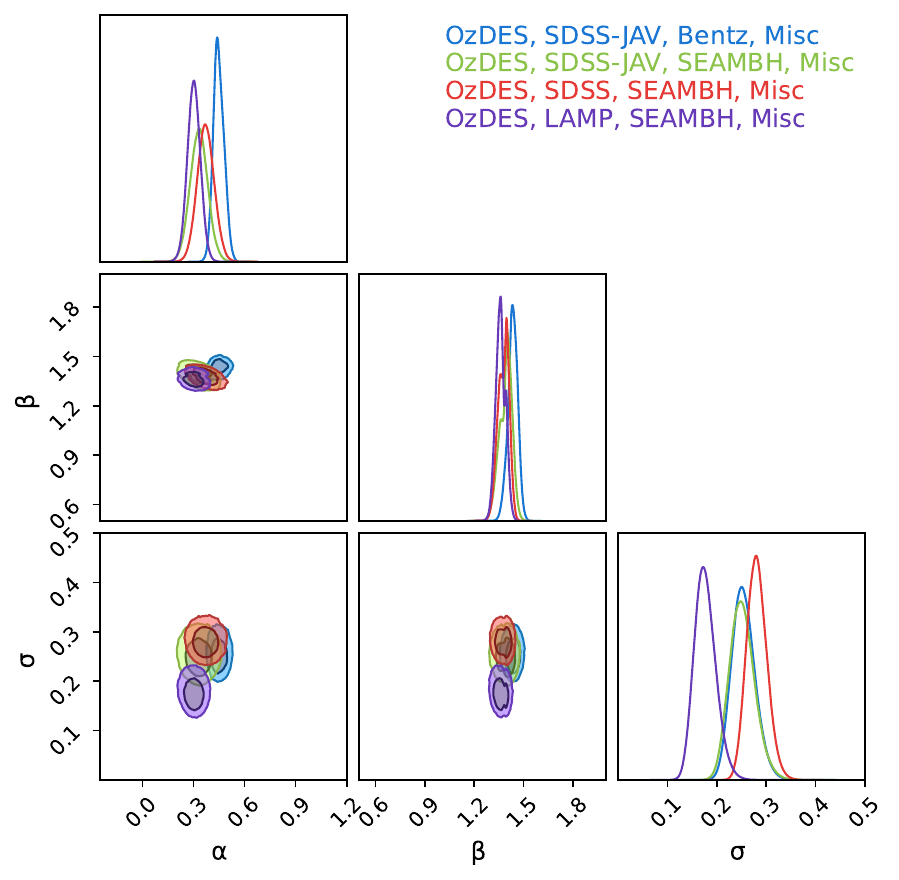}
        }\\
        \subfigure[]{
            \includegraphics[width=0.39\linewidth]{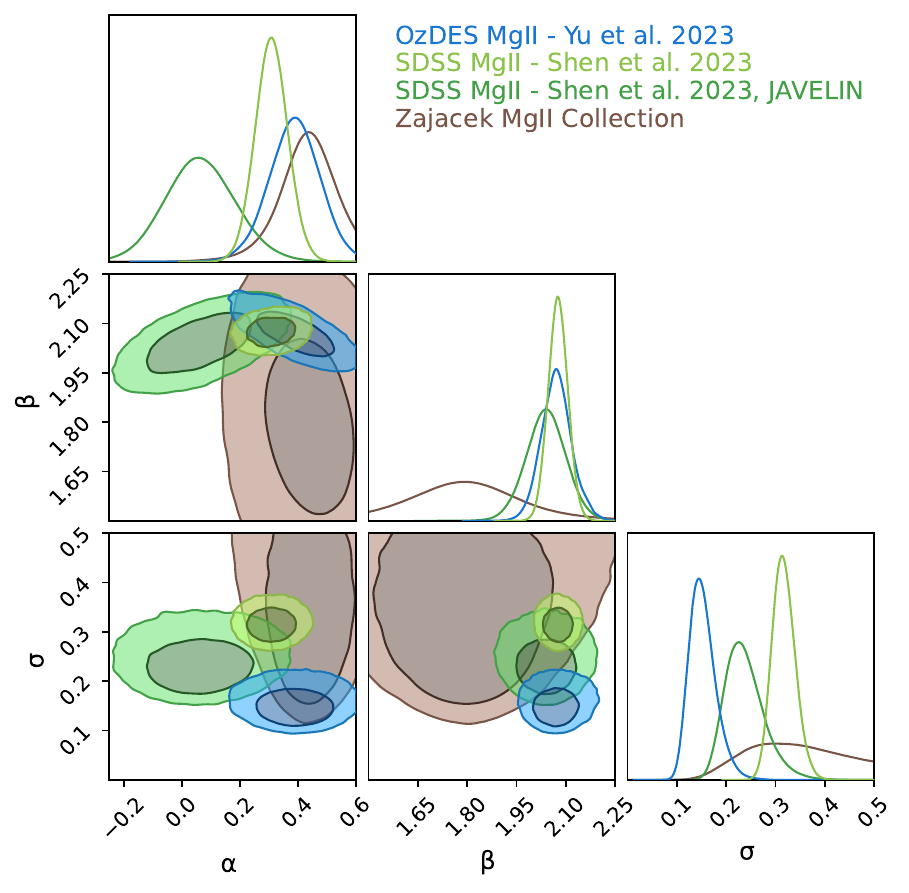}
        }\subfigure[]{
            \includegraphics[width=0.39\linewidth]{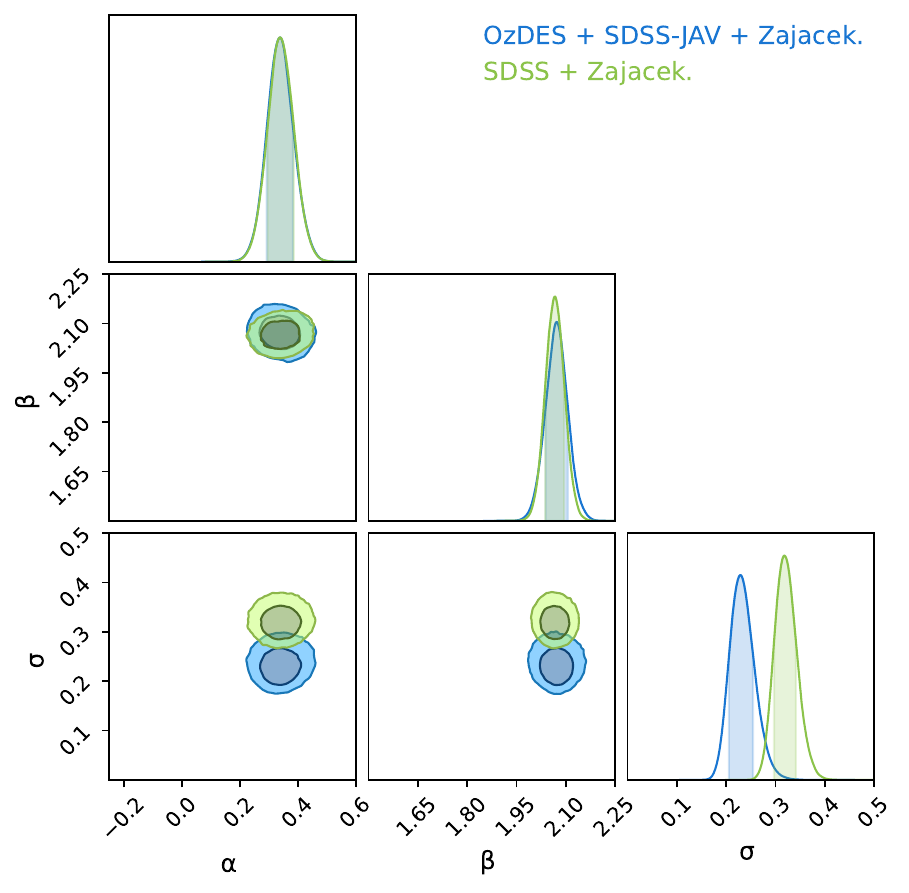}
        }\\
        \subfigure[]{
            \includegraphics[width=0.39\linewidth]{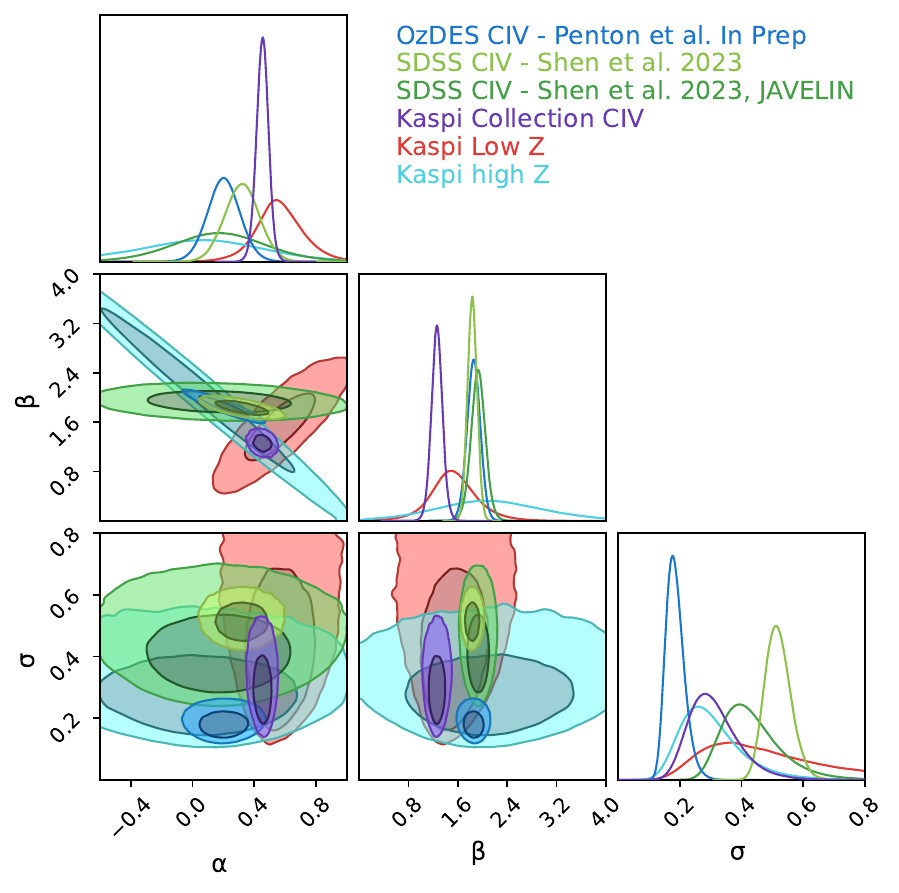}
        }\subfigure[]{
            \includegraphics[width=0.39\linewidth]{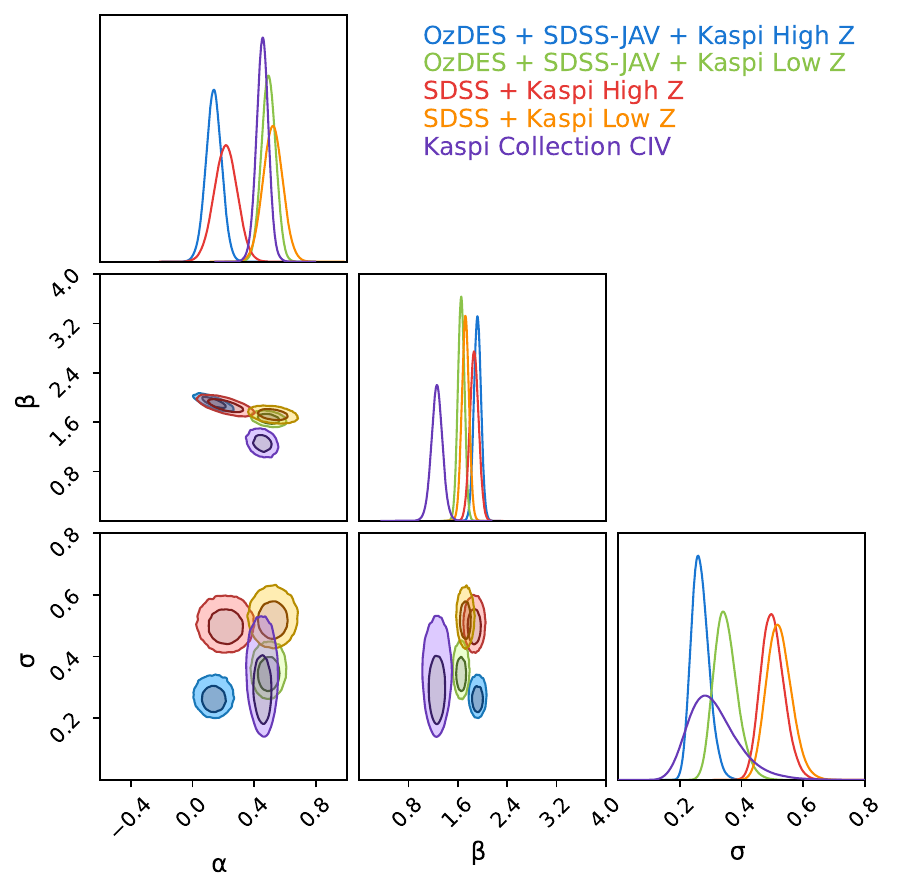}
        }
        \caption{Constraints on $R-L$ relationship parameters for (from top to bottom) \hbeta, \mgii and \civ. The left column shows constraints for each individual survey / data source, while the right column shows constraints for mutually consistent data sources, as listed in Table \ref{tab:hbeta_consistency}. To minimise covariance between $\alpha$ and $\beta$, fitting is performed using units of $10^{44} \mathrm{erg/s}$ for \hbeta and units of $10^{45} \mathrm{erg/s}$ in equation~\ref{eq: R-L model} for \mgii and \civ.
        }
        \label{fig:RL_params_all_lines}
    \end{figure*}

%----------------------------------------------------
\section{Comparison of New Results With Past Surveys \& Tuning of $R-L$ Relationship}
\label{sec: literature_and_RL}
Here we examine the $R-L$ relationship using the full array of lags available at the close of this latest generation RM surveys. In addition to lags from \ozdes and the final \sdss release of \citet{SDSS-Shen_2023}, we use a wide array of past \hbeta, \mgii and \civ lags as outlined in Section~\ref{sec: ext_data}. In Section~\ref{sec: res-RL}, we use a Bayesian approach to constrain the parameters of the $R-L$ relationship for all three reverberating emission lines, and to identify where the results of different surveys are in tension. For each line, we then select a `primary' set of data for constraining an $R-L$ relationship that is most representative of the high redshift results of \ozdes and \sdss, with the aim of demonstrating that such relations, marginalised over the footprint of such industrial scale surveys, are now within reach.

We investigate in Section~\ref{sec: res-accretion} the possibility of using line width or accretion rate to reduce scatter in the single epoch model, and finally we use then use the $R-L$ relationship as a probe of the relative sizes of the \hbeta, \mgii and \civ emission regions of the BLR in Section~\ref{sec: res-bolom}

%=============================================

%-----------------------------------------------
\subsection{Comparison of $R-L$ Parameter Constraints}
\label{sec: res-RL}

In this section examine what constraints can be places on the $R-L$ relationship for all lines using \ozdes data alone, and also using \ozdes data in combination with data from other sources. Though \ozdes and \sdss have similar physical target populations, the same can not be said for the wealth of prior sources. Further still, different surveys have different lag measurement methods and quality cut criteria, meaning we cannot assume them all to trace the same population in the $R-L$ plane. We resolve some of this issue by using the \sdss data products that most closely match the published \ozdes population (see Section~\ref{sec: SDSS-JAV cuts}), but for other surveys we take a high level approach of only allowing data sets to be used in combination if there is some reasonable combination of $R-L$ parameters that satisfy all surveys in the grouping simultaneously. The rationale here is that every survey will have complex multi-layered selection effects, but we are here concerned only if these impose enough of a difference in the $R-L$ plane that they are significantly in tension with one another. A more complex treatment would require modelling of every survey, their selection effects and pipelines and the AGN population, while the approach taken here is a natural extension of our comparison of the surveys in the $R-L$ plane. In Table \ref{tab:R-L_Summary} we list the $R-L$ parameters for our chosen `primary' data sets, while the following sub-sections elaborate on alternate $R-L$ relationships with different combinations of literature data and quality cuts.

For consistency, we refit the $R-L$ relationship with the same model for all data, while making use of the published lags and monochromatic luminosities from previous papers.\footnote{\ozdes luminosities differ slightly from previous papers due to recalibration.} We model the $R-L$ relationship as being linear in log-log space, with slope $\alpha$, offset $\beta$, and an intrinsic scatter $\sigma$ such that the a given luminosity $\lambda L_{\lambda,i}$ predicts a lag $\Delta t_{\mathrm{RL},i}$ by:
\begin{equation}
    \label{eq: R-L model}
    \log_{10}{\Delta t_{\mathrm{RL},i}} = \alpha (\log_{10}(\lambda L_{\lambda,i}) - \log_{10}(\lambda L_0)) + \beta \pm \sigma.
\end{equation}
Here, $\log_{10}(\lambda L_0))$ represents the choice of luminosity units, typically $\lambda L_0=10^{44}$ or $10^{45}$ erg/s, and is chosen to minimise the covariance between offset $\beta$ and slope $\alpha$ in the posterior distribution. $\beta$ varies with the choice of time units for the lag, and we here follow the convention of all lags being in days.

The observed RM lag, $\Delta t_{\mathrm{RM}}$, has an additional scatter from observational uncertainty. We describe this uncertainty in log-space, i.e. $\log_{10}(\Delta t_{\mathrm{RM},i}) = \log_{10}(\Delta t_{\mathrm{RL},i}) \pm E_{\mathrm{log},i}$. Under this modelling, the resulting likelihood for fit parameters $\theta={\alpha,\beta,\sigma}$ is:
\begin{equation}
\label{eq: R-L Likelihood}
    \mathcal{L} = \prod_i{ \frac{1}
    {\sqrt{2\pi (E_{\log,i}^2+\sigma^2)}}
    \exp{ 
        \frac{
            -(\log_{10}{\Delta t_{\mathrm{RM},i}} - \log_{10}{\Delta t_{\mathrm{RL},i}})^2
        }
        {
            2 (E_{\log,i}^2+\sigma^2)
        }
    }
    }.
\end{equation}
We approximate the measurement uncertainty, $E_{\mathrm{log}}$, as an asymmetric `split Gaussian' with uncertainties found from the positive and negative uncertainties of the lag measurement. I.e. if a lag measurement $\Delta t_{\mathrm{RM},i}$ is constrained within upper and lower bounds $\Delta t_{\mathrm{+RM},i}$ and $\Delta t_{\mathrm{-RM},i}$, then the log-space measurement uncertainty is:
\begin{equation}
\label{eq: R-L likelihood error}
    E_{\log,i} = 
    \begin{cases}
        \log_{10}\left( \frac{\Delta t_{\mathrm{+RM,},i}}{\Delta t_{\mathrm{RM},i}}\right) , \quad
            \Delta t_{\mathrm{RL},i}>\Delta t_{\mathrm{RM},i} \\
        \log_{10}\left( \frac{\Delta t_{\mathrm{RM},i}}{\Delta t_{\mathrm{-RM},i}}\right), \quad
            \Delta t_{\mathrm{RL},i}<\Delta t_{\mathrm{RM},i}\\
    \end{cases}.
\end{equation}
For similar consistency reasons, we use the same procedure to re-fit the $R-L$ relation of literature sources and combinations thereof. We treat all published lags and their uncertainties as describing the cumulative summary statistics of the lag posterior distribution (i.e. median and 16\textsuperscript{th}, 84\textsuperscript{th} percentiles).

All fits are performed using the Bayesian analysis tool \numpyro \citep{numpyro}, specifically its No U-Turn (NUTS) sampler, a Hamiltonian Monte Carlo (HMC) method that is well suited to unimodal distributions. We note that this fitting is relatively insensitive to the choice of sampler, as it is unimodal and produces approximately Gaussian contours (e.g. Figure~\ref{fig:RL_params_all_lines}).

\begin{table*}
\rowcolors{2}{white}{white}
    \centering
    \begin{tabular}{c|c|c|c|c|c}
        \hline
        
         \thead{Line}   &   \thead{Luminosity\\Unit} & \thead{Slope \\ ($\alpha$)} & \thead{Offset \\($\beta$)} & \thead{Scatter\\ ($\sigma$)}  & \makecell{Slope - Offset \\ Fit Correlation \\ ($\phi$)}\\[12pt]
        
         \hline
         \hbeta & $\log_{10}\left(\lambda L_{5100 \angstrom}\right) - 44$  
         & 
             $0.44^{+0.04}_{-0.02} $ 
            & $1.43^{+0.04}_{-0.02} $
            & $0.25^{+0.02}_{-0.02} $ 
            & $\quad0.14$\ \\[4pt]
         \hline
         \mgii & $\log_{10}\left(\lambda L_{3000 \angstrom}\right) - 45$  
         & 
            $0.34^{+0.04}_{-0.05} $ 
            & $2.07^{+0.03}_{-0.03} $ 
            & $0.23^{+0.03}_{-0.02} $ 
            & $-0.13$\\[4pt]
         \hline
         \makecell{\civ} & $\log_{10}\left(\lambda L_{1350 \angstrom}\right) - 45$  
         & 
            $0.47^{+0.05}_{-0.04} $ 
            & $1.65^{+0.05}_{-0.06} $ 
            & $0.36^{+0.04}_{-0.06} $ 
            & $-0.36$\\[4pt]
         \hline
         \multicolumn{6}{c}{$\log_{10}(\Delta t) = \beta + \alpha \mathrm{Luminosity \; Unit} \pm \sigma$}\\
         \hline
    \end{tabular}
    
    \caption{
        Our final $R-L$ Relationships for \hbeta, \mgii and \civ using a combination of multiple data sets for each. Monochromatic luminosities are measured in the rest-frame in units of $\text{erg/s}$, and resulting radii are in units of $\log_{10}$ light-days.
    }
    \label{tab:R-L_Summary}
\end{table*}

%=============================================
\subsubsection{\hbeta $R-L$ Relationship}
\label{sec: res-RL-Hbeta}

\ozdes is designed primarily for higher redshifts, and so captures only a small number of sources in the \hbeta redshift range. Coupled with the high rejection rate, this yields only $8$ high quality lag recoveries for the \ozdes \hbeta $R-L$ relationship \citep{OzDES-Malik_2023}. In past \ozdes analyses, the limited number of sources has been bolstered by lags from other RM projects, namely the 5-year \sdss lags \citep{SDSS-Grier_2017}, a wide range of low redshift lags collated by \citet{HBETA_Bentz_2009} as well as later additions to this `Bentz Collection' \citep{HBETA_Bentz_2013, HBETA_Bentz_2014, HBETA_Bentz_2016a, HBETA_Bentz_2016b, HBETA_Bentz_2023}, lags from the SEAMBH project, lags from the LAMP collaboration \citep{HBETA_U_2022}, and a small number of lags from miscellaneous sources \citep{HBETA_Li_2021, HBETA_Lu_2016, HBETA_Zhang_2019, HBETA_Rakshit_2019}. We make use of these same sources, in addition to the final \sdss lags of \citet{SDSS-Shen_2023}, but re-fit each data set individually to more fairly measure tensions between them. Considering surveys to be in tension if they differ by more than $2\sigma$, the groups of surveys with mutually consistent constraints are listed in Table \ref{tab:hbeta_consistency}. In total, $8$ \ozdes sources and $26$ \sdss sources are used.

Figure~\ref{fig:RL_params_all_lines} (a) shows $R-L$ relationship parameters for data from \ozdes, SDSS, the Bentz Collection, the SEAMBH Survey, the LAMP survey, and a number of other miscellaneous sources. Shown also are the contours for \sdss lags as recovered from \javelin instead of \pyroa, and selected according to the criteria outlined in Section~\ref{sec: SDSS-JAV cuts}. The constraining power is highest for surveys with a high source-count, like \sdss \& SEAMBH, or a wide range of luminosities as in the Bentz Collection. 

Due to the low sample count, the \ozdes and miscellaneous sources are poorly constrained and are consistent with all other surveys. The largest tension exists between the slope and offset of the Bentz Collection and SEAMBH data, with this tension being the deciding factor about which surveys are mutually consistent. The SEAMBH data are known to produce consistently lower lags from their highly accreting AGN, and this is borne out by the lower values for $\alpha$ and $\beta$ in collections including SEAMBH sources. 

The \sdss results are consistent with these SEAMBH values in slope and scatter, but show systematically higher scatter than any other survey with strong constraints. This lower offset and higher scatter puts the \sdss data in tension with the low redshift Bentz Collection. Limiting the \sdss lags to only those that pass our cuts in Section~\ref{sec: SDSS-JAV cuts} (the \sdss-\javelin contour), the constraints become weak enough to overcome this inconsistency, but the preferred values for $\alpha$, $\beta$ and $\sigma$ do not significantly change. The LAMP survey lags sit the lowest of any data set, and their comparatively narrow range of luminosities mean they do not strongly constrain the $R-L$ slope by themselves.

Figure~\ref{fig:RL_params_all_lines} (b) shows the constraints on $R-L$ relationship parameters when combining all datasets that are not in tension with one another. Four such groups are possible: 
\begin{enumerate}
    \item \ozdes combined with the Bentz Collection, miscellaneous sources and the sub-sampled \sdss data,
    \item As for 1, but with SEAMBH lags in place of Bentz,
    \item As for 2 but using the full \sdss sample,
    \item \ozdes data combined with miscellaneous sources, the SEAMBH lags and the lags from the LAMP survey.
\end{enumerate}
Where we use a `primary' \hbeta grouping elsewhere in the paper, it refers to the first of these groupings. We make this choice of grouping as it combines the data of \ozdes and \sdss with a broad set of literature lags, while excluding the lags of the SEAMBH survey that are, by the survey's design, deliberately intended to be atypical of the broader AGN population.

\newcommand{\Tt}{\cellcolor{green!25}T}
\newcommand{\Tf}{\cellcolor{red!25}F}

\begin{table}
\rowcolors{2}{white}{white}
    \centering
    \small{
    \begin{tabular}{| p{0.2\linewidth} || c | c | c | c|c | c | c|}
        \hline
        \textbf{Source} & \rotatebox{90}{OzDES} & \rotatebox{90}{SDSS} & \rotatebox{90}{SDSS-JAV  } & \rotatebox{90}{LAMP} & \rotatebox{90}{Bentz} & \rotatebox{90}{SEAMBH} & \rotatebox{90}{Misc}\\[5pt]
        \hline
        \hline
        \ozdes & - & \Tt & \Tt & \Tt & \Tt & \Tt & \Tt\\
        \hline
       \sdss& \Tt & - & - & \Tf & \Tf & \Tt & \Tt\\
        \hline
        SDSS-JAV & \Tt & - & - & \Tf & \Tt & \Tt & \Tt\\
        \hline
        LAMP & \Tt & \Tf & \Tf & - & \Tf & \Tt & \Tt\\
        \hline
        Bentz & \Tt & \Tf & \Tt & \Tf & - & \Tf & \Tt\\
        \hline
        SEAMBH & \Tt & \Tt & \Tt & \Tt & \Tf & - & \Tt\\
        \hline
        Misc & \Tt & \Tt & \Tt & \Tt & \Tt & \Tt & -\\
        \hline
    \end{tabular}
    }
    \caption{Consistency between data sources for \hbeta $R-L$ relation parameters. If two data sources are statistically consistent (T), their recovered parameters for slope, offset and scatter are consistent to within $2\sigma$. Otherwise they are visibly in tension (F). No result is listed for the main \sdss results and the sub-sampled \javelin \sdss results, as they are drawn from the same survey. These tensions yield four distinct sub-groups of mutually consistent data sources.}
    \label{tab:hbeta_consistency}
\end{table}

%=============================================
\subsubsection{MgII $R-L$ Relationship}
\label{sec: res-RL-MgII}

Until this recent generation of high redshift surveys, the bulk of reverberation mapping lags have been measured using \hbeta, with a far more sparse set of lags in \mgii. Rather than dividing these few sources by survey, we compare the \ozdes and \sdss results to the convenient collection by \citet{MgII_Zajacek_2020}, consisting of $5$ lags from \citet{MgII_Lira_2018}, \citet{MgII_Czerny_2019} and \citet{MgII_Metzroth_2006}, and a recent lag from \citet{MgII_Zajacek_2021}. We exclude the $6$ lags from \citet{MgII_Shen_2016} and \citet{MgII_Shen_2019}, as they are already included in the \sdss results sample. From the \ozdes and \sdss data releases we use $25$ sources each.

Comparing $R-L$ relationship parameters for \sdss and \ozdes, we find that they broadly agree in terms of the slope and offset of the fit, but that \sdss lags exhibit significantly higher scatter by roughly $0.2$ dex. The low number of sources in the Zajacek Collection means that its fit is poorly constrained, and so is not in tension with \ozdes or \sdss. Restricting the \sdss results to the SDSS-JAV sample drastically reduces the number and constraining power, but also decreases the scatter significantly. This decrease in scatter removes the tension with \ozdes, allowing these two data sets to be combined.

This yields two mutually consistent groupings, making use of the full \sdss data set and excluding \ozdes or including \ozdes but restricting the \sdss data. Owing to the complementary correlations of the data sets, both choices have near identical results for the main properties of the $R-L$ relation, but with a slightly higher scatter for the \sdss-based results (Figure~\ref{fig:RL_params_all_lines} (c)). We proceed with the \ozdes + \sdss-\javelin + Zajacek fit as our primary \mgii $R-L$ relation as it is derived from the most diverse set of sources while still showing very low scatter about the mean relationship, suggestive of a lower false positive rate.

%=============================================
\subsubsection{$CIV$ $R-L$ Relationship}
\label{sec: res-RL-CIV}
Like \mgii, there are only a small number of \civ lags outside of large-scale high redshift surveys. For these supplemental lags, we draw sources from the collation by \citet{CIV_Kaspi_2021}, including $6$ low redshift AGN in which the $1350 \angstrom$ \civ line is observed in the observer-frame ultra-violet range from \citet{CIV_Peterson_2005, CIV_Rosa_2015} and \citet{MgII_Metzroth_2006}, as well as $6$ high redshift sources from \citet{MgII_Lira_2018} and $3$ high redshift sources new to the Kaspi paper itself. As with the collation of existing \mgii lags, we exclude the Kaspi sources from \citet{CIV_Grier_2019} and \citet{OzDES-Hoormann_2019} as they are already included in the \sdss and \ozdes data.\footnote{Interestingly, only one of the two sources from \citet{OzDES-Hoormann_2019} makes it into the final OzDES sample.  In that paper we used the down-weighting of seasonal gaps method employed by SDSS (equation~\ref{eq: SDSS-weighting}), however when that is removed, the confidence in those sources drop and only one of them passes (at bronze level) the simulation-based criteria of \citet{OzDES-Penton_2021}.} We use $29$ high quality lags from \ozdes and $15$ from \sdss.

Unlike the \hbeta and \mgii $R-L$ relationship fits, it is much harder to find consistent agreement between different data sources. The lags from prior works sit systematically lower than the trends set by either \ozdes or \sdss, and there is consistently greater scatter than exhibited for either \hbeta or \mgii. Some of this discrepancy can be attributed to selection effects (see below); some may be due to \civ\ being contaminated by a non-virial wind component \citep{Denney_2012}.  We attempt to account for selection effects in this section, but leave more detailed study of contamination for future work. 

The high redshift surveys of \sdss and \ozdes are highly consistent with one another except for the markedly higher scatter of the \sdss data, much as is seen for \hbeta and \mgii. The \ozdes-like SDSS-JAV sample has lower scatter, removing the tension with the \ozdes sample, though partially through a loss of constraining power from the reduction in source count.

The literature lags sit noticeably lower than either \sdss or \ozdes lags, even at high redshift. Though it is not statistically objectionable to combine both high redshift surveys with either the Kaspi Collection's high redshift lags or its low redshift anchor, this gap means that it is not reasonable to do both at once. This allows for five possible groupings if taking the standard approach to the fitting of the $R-L$ relationship:
\begin{enumerate}
    \item \label{enum: civ_group_1} The high redshift literature lags with the combined \ozdes and SDSS-JAV sample, 
    \item \label{enum: civ_group_2} As above but with the low redshift anchor literature sources,
    \item \label{enum: civ_group_3} As per 1. but with the full \sdss and no \ozdes sources,
    \item \label{enum: civ_group_4} As above but with the low redshift anchor literature sources,
    \item \label{enum: civ_group_5} The low and high redshift sources of the Kaspi Collection with no new sources.
\end{enumerate}

Though still exhibiting a high scatter, the above groupings that fit for the entire luminosity range (groups \ref{enum: civ_group_1}, \ref{enum: civ_group_3} and \ref{enum: civ_group_5}) give slopes roughly consistent with those of \hbeta and \mgii, while the flattened branch of the high-luminosity only groupings (\ref{enum: civ_group_2} and ~\ref{enum: civ_group_4}) have flat relationships between reverberation lag and AGN luminosity. This apparent disagreement between sources means we are unable to choose a primary data set for fitting the \civ $R-L$ relationship.

\begin{figure}
    \centering
    \includegraphics[width=\linewidth]{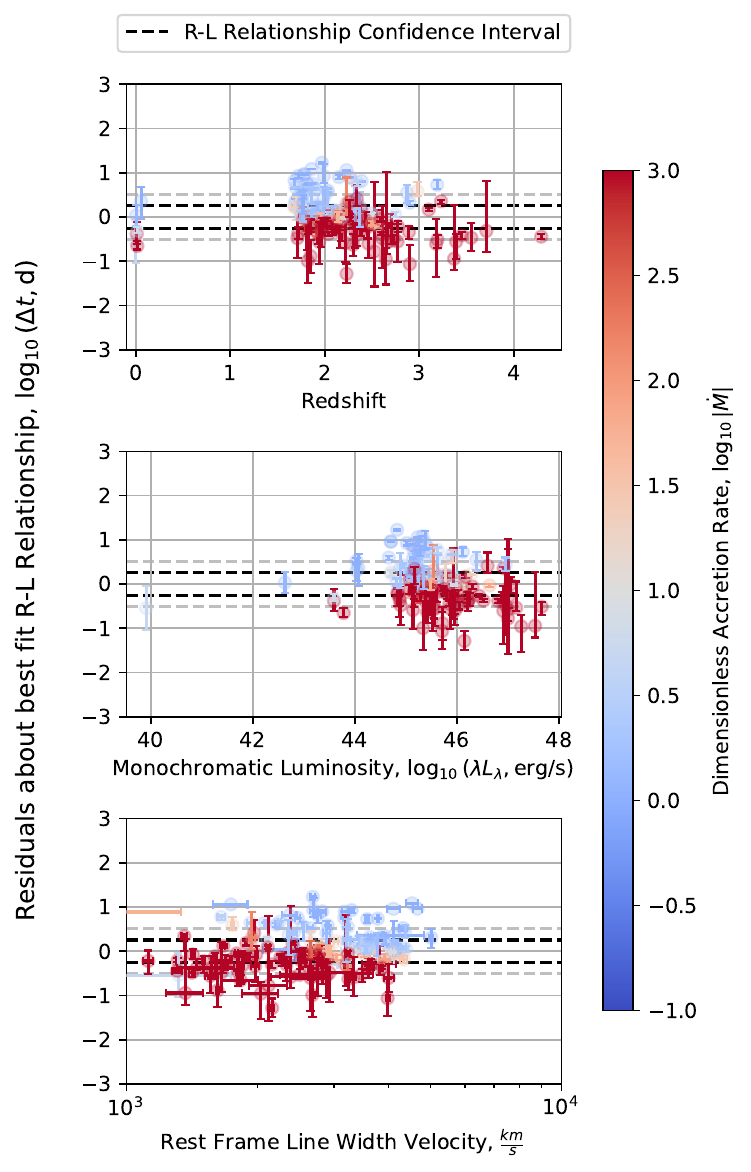}
    \caption{Residuals of \civ lag measurements for all RM sources about the best fit high redshift $R-L$ relationship, coloured by accretion rate $\log_{10}(\dot{M})$. There is a clear trend of the low redshift / luminosity sources (most of which are drawn from the Low-Z Kaspi sample) sitting below the fit. Though the high-accretion rate sources sit below the $R-L$ fit, there is no independent correlation with either accretion rate or emission line velocity dispersion (bottom panel, discussed further in Section~\ref{sec: res-accretion}).}
    \label{fig:Residuals-CIV}
\end{figure}

The tension between the slope of the high and low redshift is manifest in an apparent `levelling out' of the $R-L$ relation at high redshift (Figure~\ref{fig:RL-all-data}, bottom panel). This is best illustrated when examining the relationship between the residuals and the `high-$z$' fit (our `grouping~\ref{enum: civ_group_1})'). As shown in Figure~\ref{fig:Residuals-CIV}, high luminosity / redshift sources give unmistakably and consistently higher lags than the lower  luminosity / redshift data. 

%----------------------------------

\label{sec: selection_effects}
This nonlinear $R-L$ relationship for \civ may have a physical basis, such as distant AGN having higher accretion rates, which have been associated with lower lags \citep[e.g.][]{Du_2015, HBETA_Hu_2021}. A simpler explanation lies in describing this as a selection effect: more luminous AGN have longer rest-frame lags, stretched further at high redshift by time dilation, and are therefore more likely to have lags that exceed the length of the survey. This induces a selection effect in which high redshift / luminosity sources are biased downwards from the true population distribution as the finite survey length imposes a ceiling on what observer-frame lags can be recovered. \civ sources are often the most distant, and consequently the most luminous and associated with the longest observer-frame lags, and so their observations are the most impacted by this effect. 

Coupled with recovery rates and quality cuts, the exact effects of this under-sampling of high lag sources is difficult to characterise, as it interacts with the entire observational and analysis pipeline. A complete treatment would require modelling the underlying source population, survey selection, and the entire analysis chain including quality cuts.  Though not intractable, this is an involved task, particularly when trying to unify multiple surveys and studies.

\citet{OzDES-Penton_2025} estimates that, for the $6$-year baseline of \ozdes, a lag-recovery selection effect becomes significant for \civ sources at observer-frame lags of $\approx 1000  \; \dayu$ and continues until lags above $1500 \; \dayu$ are almost impossible to recover. We make the coarse approximation of treating this $1250 \; \dayu$ limit as a hard cutoff, altering our Bayesian model so that that \ozdes cannot detect lags above this limit. This is tantamount to altering the normal distribution of equation~\ref{eq: R-L Likelihood} to be a truncated normal distribution with a cutoff at $\Delta t_i = \frac{1250 \; \dayu}{1+z_i}$. To a rough first order approximation, we scale this limit for each survey proportional to the length of their spectroscopic campaign as compared to \ozdes, giving nominal lag `cutoffs' in Table~\ref{tab: truncation_table}. 

\begin{table}
\rowcolors{2}{white}{white}
    \centering
    \begin{tabular}{c|c|c|c}
        \hline
        Survey & \makecell{Highest Observer \\ Frame Lag \\ (days)} & \makecell{Spectroscopic \\ Baseline \\ (yrs)} & \makecell{Nominal \\ Lag Cutoff \\ (days)} \\[\tablespacing]
        \hline
        \ozdes &   $934.1^{+31.1}_{-31.1} $ & $6$ & $1250$ \\[\tablespacing]
        \hline
        SDSS-\javelin &   $1298.6^{+14.7}_{-12.0}$ & $7$ & $1460$ \\[\tablespacing]
        \hline
        Kaspi High-Z &    $1629.9^{+252.1}_{-326.7}$ & $9.5$ & $1980$ \\[\tablespacing]
        \hline
    \end{tabular}
    \caption{Maximum observer-frame lags for each \civ survey, along with their estimated maximum recoverable lag cutoff.}
    \label{tab: truncation_table}
\end{table}

\begin{figure}
    \centering
    \includegraphics[width=0.8\linewidth]{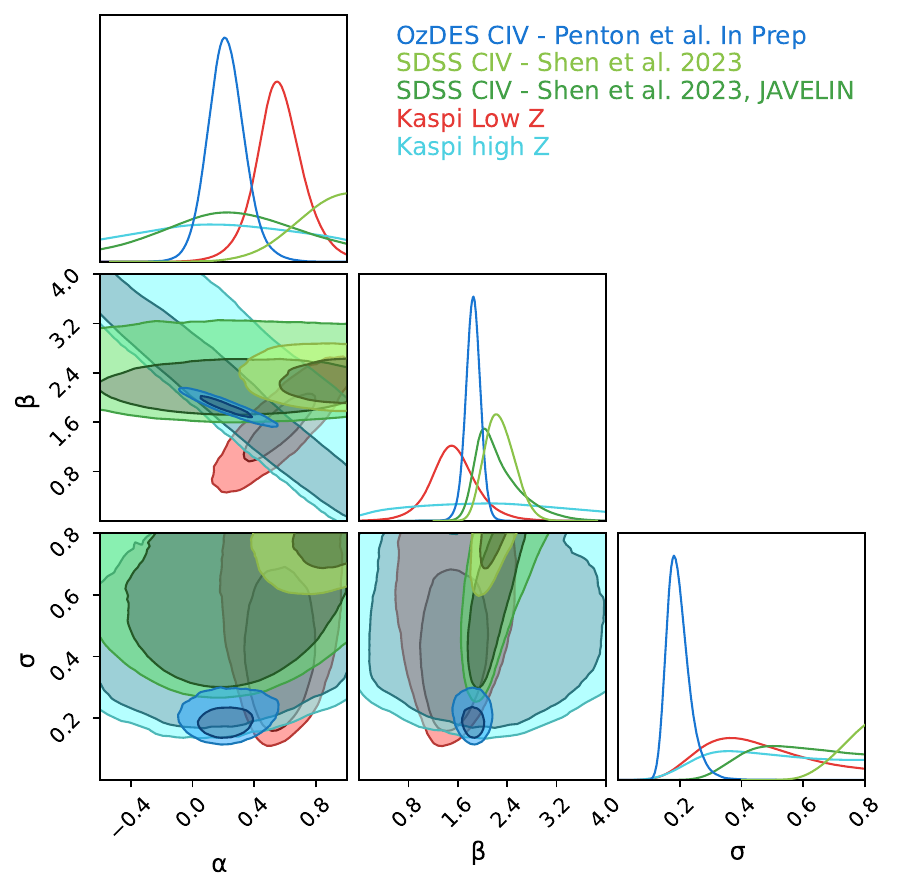}
    \includegraphics[width=0.8\linewidth]{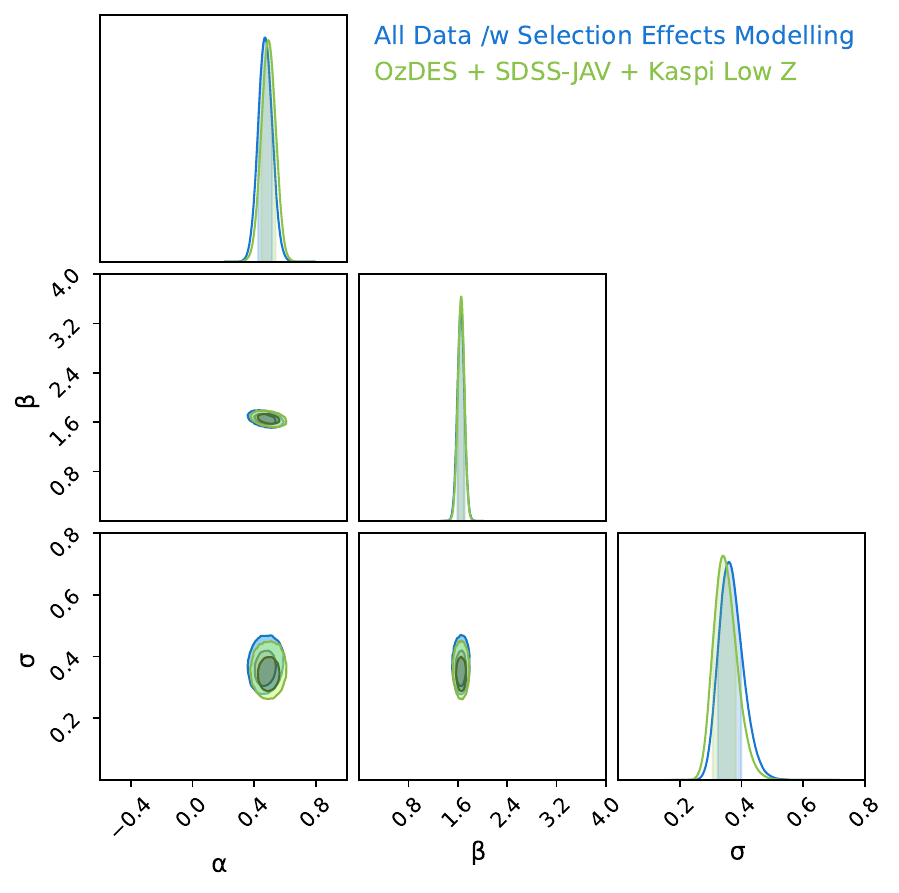}
    \caption{Constraints on the \civ $R-L$ relationship parameters after accounting for a maximum observable lag due to survey lengths. The top panel shows constraints for each individual data set, and the bottom panel compares the combined data. These are based on the same data groupings as used in Figure~\ref{fig:RL_params_all_lines}, but with a model incorporating the cutoffs in Table~\ref{tab: truncation_table}.}
    \label{fig: RL-trunc}
\end{figure}

This model significantly relaxes the constraints from each survey, introducing the freedom for an un-observed fraction of sources to exist above the lag cutoff. This relaxes the tension between the surveys, creating a single parameter-space region of overlap between all four data sources (Figure~\ref{fig: RL-trunc}, top panel).  The high luminosity literature sources, drawn from the high-$z$ Kaspi sample, are dominated by this observational window effect, loosening their constraining power such that this fit matches the constraints of the more generic $R-L$ fit achieved when excluding the high-$z$ literature lags all together (Figure~\ref{fig: RL-trunc}, bottom panel). It is this fit, with all \civ data accounted for, that we use as our primary $R-L$ fit for single-epoch mass estimates in Section~\ref{sec: single_epoch}.

%-----------------------------------------------

\subsection{Accretion Rate \& Line Width as Predictors of RM Lag}
\label{sec: res-accretion}

It is matter falling on to the accretion disk circling the SMBH that drives AGN activity, and there is evidence of a separation in behaviour between low and high accretion AGN. Highly accreting `super-Eddington' AGN are associated with lower lags than their `sub-Eddington' counterparts at the same luminosity \citep{Du_2015}. It is also evident that high accretion sources tend to sit below our best fit $R-L$ relation (e.g. Figure~\ref{fig:Residuals-CIV}), while low accretion sources sit above. We are then motivated to see if observations other than luminosity are predictors of AGN lag as a means to reduce the scatter of our single epoch methods, improve the predictive constraints of our single epoch methods and as a probe of the underlying AGN behaviour.

To examine this, we examine a new functional form for the single epoch lag prediction in which we add emission line width as a variable in the $R-L$ relation, i.e. an $R-L-V$ relation: 
\begin{equation}
    \log_{10}(\Delta t) = \alpha \log_{10}(\lambda L_{\lambda}) + \gamma  \log_{10}(\braket{\sigma_v^2})+ \beta,
    \label{eq:R-L_relationship-vel}
\end{equation}
where $\gamma$, the slope of log-lag against log velocity dispersion, represents the predictive power of line-width independent of luminosity.

Our estimates for mass and accretion rate (Equation~\ref{eq:RM_mass} and Equation~\ref{eq: accretion}, respectively) both follow power laws, making them linear for the logarithmic space that we fit our single epoch relations in. In this way, fitting for line width and luminosity also covers the fitting for accretion rate, as it depends entirely on these two other measurements.

Performing this $R-L-V$ fit for all lines, using our primary data groupings from Section~\ref{sec: res-RL}, we find no significant improvement in predictive power (Table \ref{tab:R-L-V-summary}), and constraints on the lag-velocity dependence are broadly consistent with zero, i.e. no dependence of lag on line width. For \mgii we see a slight preference towards $\gamma>0$, but this is only weakly constrained and there is still no marked decrease in scatter about the model.

\begin{table}
\rowcolors{2}{white}{white}
    \centering
    \begin{tabular}{|c||c|c|c|}
    \hline
        \makecell{Line \\ Type} & \makecell{$\gamma$} & \makecell{$\sigma$ for \\$R-L$} &  \makecell{$\sigma$ for \\$R-L-V$} \\[\tablespacing] \hline
         \hbeta 
            & $-0.04^{+0.13}_{-0.13}$  
            & $0.25^{+0.02}_{-0.02}$ 
            & $0.25^{+0.03}_{-0.02}$ 
        \\[\tablespacing]
         \hline
         \mgii 
            & $0.59^{+0.30}_{-0.23}$  
            & $0.23^{+0.02}_{-0.02}$ 
            & $0.22^{+0.03}_{-0.02}$ 
        \\[\tablespacing]
         \hline
         \civ 
             & $0.04^{+0.38}_{-0.34}$  
             & $0.34^{+0.04}_{-0.03}$ 
             & $0.35^{+0.04}_{-0.04}$ 
         \\[\tablespacing]
         \hline
    \end{tabular}
    \caption{Constraints on line-width velocity as a supplementary predictor of lag, and comparison of the scatter about this model compared to the luminosity-only $R-L$ relationship. In all cases we fail to see a reduction in the scatter. Similarly, all lines are consistent with zero  log luminosity / velocity ($\gamma=0$ in equation \ref{eq:R-L_relationship-vel}), i.e. no lag-velocity dependence, though with \mgii preferring a positive relation.}
    \label{tab:R-L-V-summary}
\end{table}

The apparent trend between accretion rate and the residuals about our best fit $R-L$ relationships is a projection of the luminosity dependence onto the log-accretion rate axis, rather than an independent axis of variability. For a fixed luminosity, higher than average lags will give higher masses, which will give lower accretion rates, producing an apparent anti-correlation between the two.

\begin{figure}
    \centering
    \includegraphics[width=1.0\linewidth]{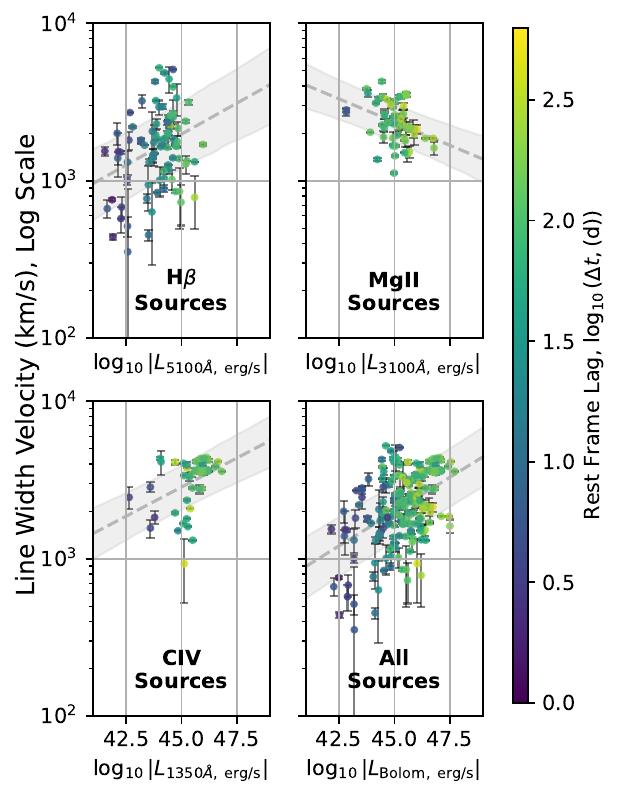}
    \caption{Luminosity / velocity scatter plot for our `primary' data-sets, coloured by rest frame lag. Under-laid are fits for the correlation between lag and velocity, modelled as a power law. The colouring shows how lag evolves strongly over the luminosity axis, demonstrating it is a better predictor of lag compared to velocity.}
    \label{fig:Luminosity-Velocity}
\end{figure}

There is the possibility that the lack of predictive power comes from a strong correlation between luminosity and velocity dispersion in our sample: if we only probe a narrow band of $L-V$ space, the two parameters become degenerate in our fitting. We can rule this out by examining the correlation between luminosity and velocity (Figure~\ref{fig:Luminosity-Velocity}). We find that this correlation is weak, and is not consistent between lines. Table \ref{tab:lum-vel-summary} notes an extremely shallow scaling index between luminosity and velocity dispersion for all lines, indicating that our sampling of the two parameters is sufficiently uncorrelated.

We note that this lack of predictive power in velocity is in contrast to the findings of the SEAMBH collaboration, \citep{HBETA_Du_2016, HBETA_Du_2018, HBETA_Hu_2021} who find that higher accretion rates (derived from higher velocities) is associated with lower lags. Our results here do not necessarily contradict their findings, this analysis is for our primary data group which does not include their data specifically because their highly accreting sources produce statistically lower \hbeta lags (see Section~\ref{sec: res-RL-Hbeta}), and we use a different functional form for the accretion dependence in place of their use of a critical  accretion rate threshold. Our fit only demonstrates that line width / accretion rate does not serve to explain the residuals about our best fits. It is also worth noting that different data sources have non-uniform approaches to measuring these line widths, and unlike the lag parameter we make no attempt to account for these differences in analysis.

\begin{table}
\rowcolors{2}{white}{white}
    \centering
    \begin{tabular}{|c||c|c|}
    \hline
        \makecell{Line \\ Type} & \makecell{Log-Vel \\ / Log-Lum \\ Slope} & \makecell{Scatter \\ (Dex)} \\[\tablespacing] \hline
         \hbeta 
             & $0.08^{+0.02}_{-0.03}$  
             & $0.22^{+0.02}_{-0.02}$ 
         \\[\tablespacing]
         \hline
         \mgii 
             & $-0.06^{+0.02}_{-0.02}$  
             & $0.12^{+0.01}_{-0.01}$ 
         \\[\tablespacing]
         \hline
         \civ 
             & $-0.07^{+0.02}_{-0.03}$  
             & $0.14^{+0.02}_{-0.02}$
         \\[\tablespacing]
         \hline
         All Sources 
             & $0.09^{+0.01}_{-0.01}$  
             & $0.19^{+0.01}_{-0.01}$
         \\[\tablespacing]
         \hline
    \end{tabular}
    \caption{Constraints on the scaling index and scatter between AGN luminosity and velocity dispersion for all emission lines, per equation~\ref{eq:R-L_relationship-vel}. Constraints are for sources in our `primary' data sets as outlined in Section~\ref{sec: res-RL}.}
    \label{tab:lum-vel-summary}
\end{table}

%-----------------------------------------------

\section{Probing BLR Stratification by Comparisons of Bolometric $R-L$ Relations}
\label{sec: res-bolom}

In cases where there are lags for multiple lines from a single source \citep[e.g.][]{MgII_Metzroth_2006, MgII_Lira_2018}, it has been observed that different RM lines produce different lags, indicating a stratification in the structure of the BLR, with different lines emitted from distinct regions. By comparing these lags, we can probe the relative size and overall geometry of the broad line region. Though a small subset of the \ozdes target AGN do exhibit multiple lines in the optical range, these data are of insufficient number and quality to meaningfully measure the BLR stratification. Here we instead leverage the full set of available RM results by making use of the $R-L$ relationships themselves (as fit in Section~\ref{sec: res-RL}) as a probe of the BLR geometry.

For each line, in Section~\ref{sec: res-RL} we derived $R-L$ relationships relative to the monochromatic luminosity evaluated at a line-specific wavelength conveniently similar to that of the reverberating line itself. Though each line's $R-L$ relationship refers to a different luminosity, simple bolometric corrections allow us to reposition these relationships onto the same luminosity axis. Doing so allows us to compare the expected lags, and so typical radius of emission, for the different ions that produce reverberating emission lines. Here we use a variation of the $R-L$ relationship fitting in Section~\ref{sec: res-RL} to examine implications of the radii of the \hbeta, \mgii and \civ emission regions.

In this analysis, we assume a fixed proportionality between monochromatic and bolometric luminosity:
\begin{equation}
    L_{\mathrm{bol}} = C_\lambda \times\lambda L_\lambda.
    \label{eq:bolometric_correction}
\end{equation}
We use the bolometric corrections of \citet{Runnoe_2012}, namely that $C_{5100 \angstrom} = 8.1\pm0.4$, $C_{3000 \angstrom} = 5.2\pm0.2$, and $C_{1450 \angstrom} = 4.2\pm0.1$. We use the correction for $1450 \angstrom$ for \civ, though note that the luminosity is defined at $1350 \angstrom$. As we take the logarithm of these values, small variations should have little impact. The resulting bolometric luminosities span the range $\log_{10}(L_\mathrm{bol})\in[40.53, 47.51]$.

These corrections result in a horizontal shift of the relations that were derived in Section~\ref{sec: res-RL} and presented in Table~\ref{tab:R-L_Summary}. We note that this is the simplest possible bolometric model, applying only a simple translation of the $R-L$ relationship, while more expressive alternatives exist that describe the bolometric correction as a luminosity dependent power law would also impact the relationship's slope. We note that it is a coarse assumption to treat these corrections as being equally applicable across all redshift ranges, and stress that this is a simplified first pass of this stratification measurement.

By correcting each source to its bolometric luminosity, we can repeat the $R-L$ fitting procedure of Section~\ref{sec: res-RL} and more meaningfully compare the relations for \hbeta, \mgii and \civ. For all lines, we use the same data sets as the final $R-L$ fits in Table \ref{tab:R-L_Summary}, including our treatment of selection effects for \civ as discussed in Section~\ref{sec: selection_effects}.

Performing this fit with luminosity units of $10^{45} \mathrm{erg/s}$ (Figure~\ref{fig:Bolometric_Relations} (b)), we find that the \mgii and \hbeta relations are very similar, though the \mgii lags sit at a slightly higher offset and prefer a slightly shallower slope. By contrast, the fit for \civ sits lower, but with a slope and scatter that is broadly similar to \hbeta. This is in keeping with prior studies of sources with multiple lines visible simultaneously, which found that \hbeta and \mgii were produced by roughly co-spatial regions of the BLR \citep[i.e. yielded similar lags for the same source, for an example see][]{MgII_Shen_2019}, while \civ lags tend to be smaller by a factor of $~2$, though this ratio is poorly constrained. \citep{MgII_Lira_2018, Kaspi_2007}.

We can constrain these relative radii for the different emission lines by adopting an adjusted $R-L$ relationship with dimensionless scaling factors `$S_\lambda$' for \mgii and \civ (equation~\ref{eq:bolometric_scaling}):
\begin{equation}
    \log_{10}(\Delta t_\lambda \times S_\lambda) = \alpha (\log_{10}(L_\mathrm{bol}) - \log_{10}(L_0)) + \beta
    \label{eq:bolometric_scaling}.
\end{equation}
Here, units for lag and luminosity are as in equation \ref{eq: R-L model}. This allows each line's $R-L$ relation to have a different vertical offset, equivalent to the assumption that the ratio of these lags is the same across different AGN. \footnote{A further model was trialled in which each line was free to fit its own inherent scatter, but this did not significantly change the resulting parameter constraints for slope or scale factors compared to when a single intrinsic scatter is fit.}
 
By fitting this combined model we arrive at a bolometric $R-L$ relationship with a slope of $\alpha = 0.44 ^ {+0.02} _ {-0.02}$. The constraints on this model give two interesting results:
\begin{enumerate}
    \item Constraints on $S_{\mgii}$ suggest \mgii produces longer lags than \hbeta, indicating that it may be emitted from a radius larger than \hbeta by a factor of $\approx~\frac{1}{0.55}\approx1.82$. \citet{OzDES-Malik_2024} who did a stacked RM analysis (see their Table 1), in which they found that the average \mgii\ lags in the OzDES sample were significantly longer than those for \hbeta\ at comparable luminosities.
    \item We find that the \civ radius is consistent with begin co-spatial with the \hbeta radius, though it prefers to be smaller by a factor of $\approx{1}{1.21}=0.83$
\end{enumerate}

\begin{figure*}
    \centering
    \includegraphics[width=0.9\linewidth]{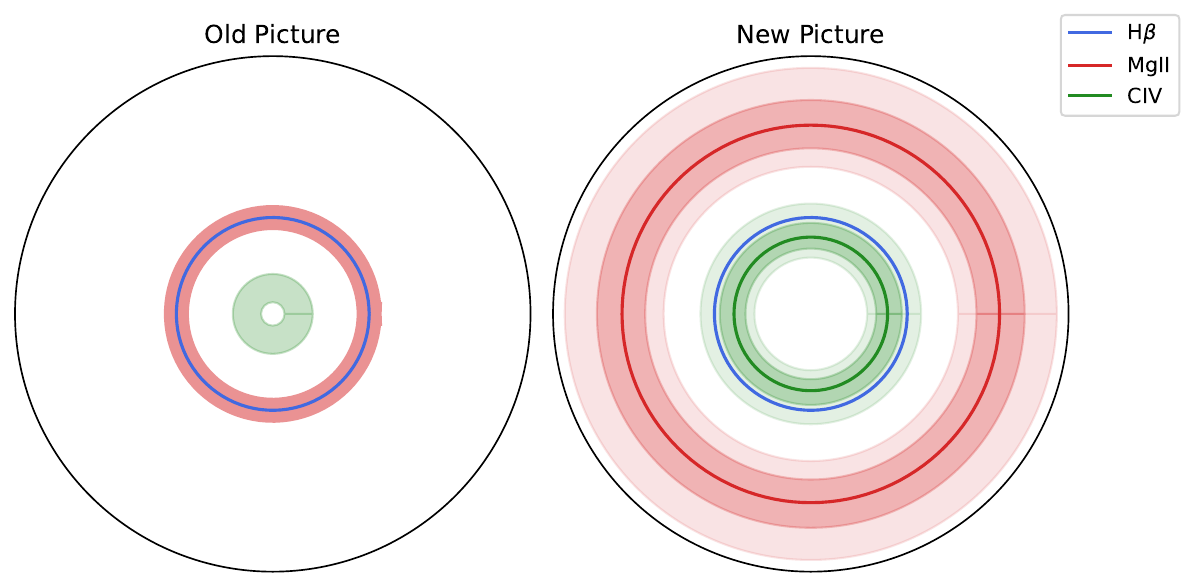}
    \caption{A sketch of the linear scales of the \mgii and \civ emission regions in the BLR relative to \hbeta, comparing the previously understood average stratification (left panel) with the new picture suggested from our relative lag scaling (right panel). For the left panel, the \hbeta and \mgii regions are roughly the same size per \citet{MgII_Shen_2019}, while the \civ region is $~2-4$ times smaller per \citet{MgII_Lira_2018, Kaspi_2007}. For the right panel, solid lines represent the nominal values in Table~\ref{tab: RL-bolomparams} while shaded regions indicate bounds of uncertainty. For the left panel, the shading shows the rough bounds of the scale factors.}
    \label{fig:bolom_scale_picture}
\end{figure*}

\begin{landscape}
    \begin{figure}
        \centering

        \subfigure[The {\em bolometric} $R-L$ relationships for each line after marginalising over the uncertainty in scale factors. The combined fit is the dashed line.]{
            \includegraphics[width=0.90\linewidth]{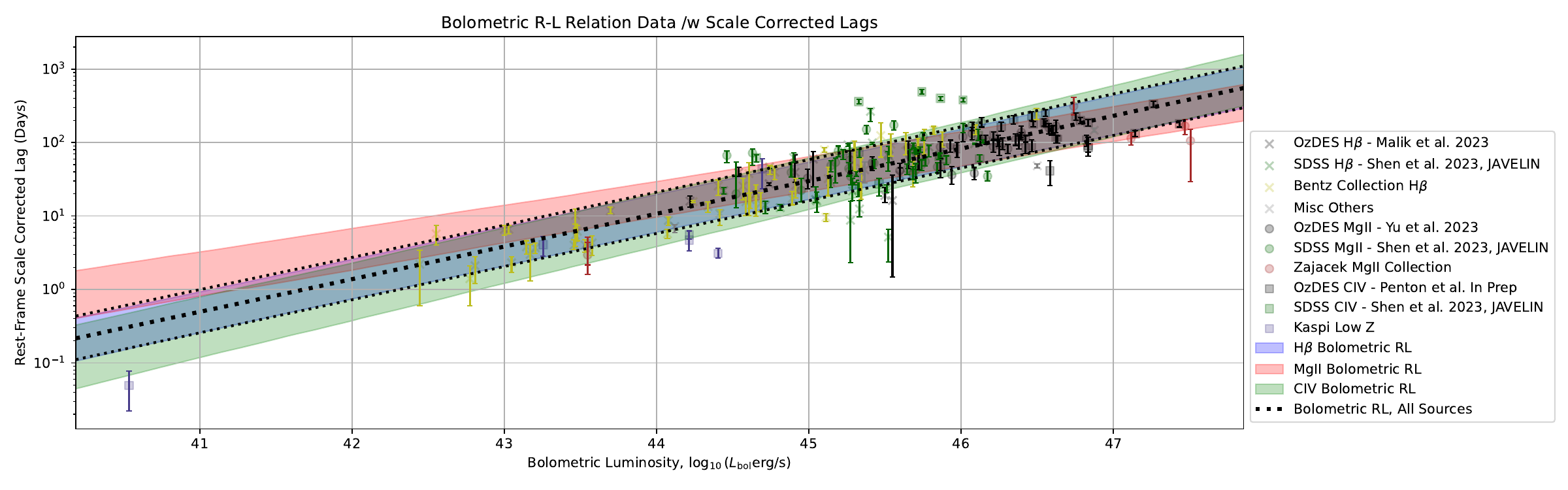}
        }\\
        \subfigure[Best fit bolometric $R-L$ relation parameters for all three emission lines.]{
            \includegraphics[width=0.33\linewidth]{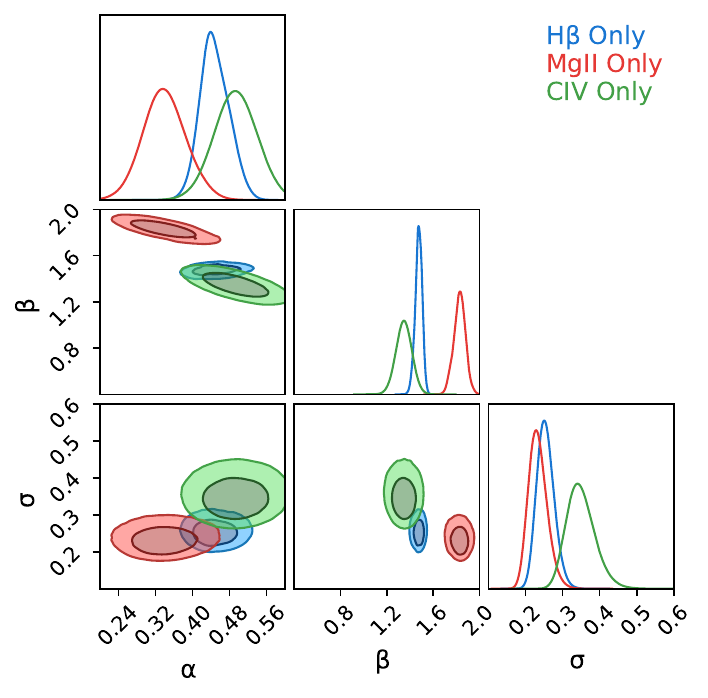}
        }\subfigure[Best simultaneous multi-line bolometric $R-L$ fit.]{
            \includegraphics[width=0.33\linewidth]{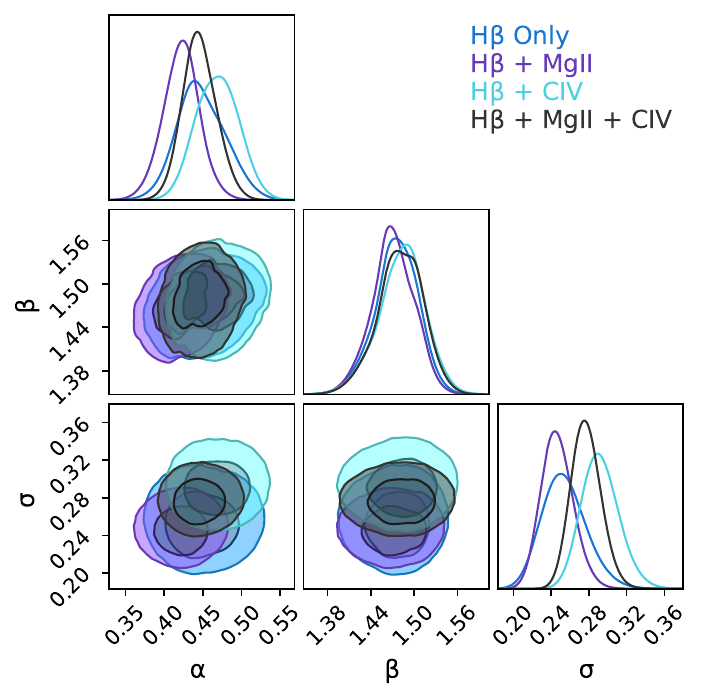}
        }\subfigure[Constraints on the scale factors between lines.]{
            \includegraphics[width=0.25\linewidth]{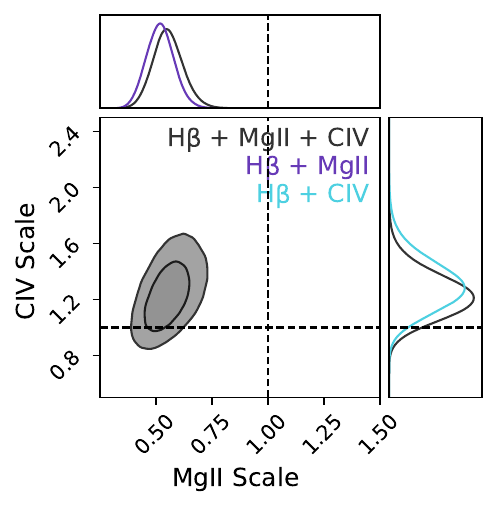}
        }
        
        \caption{We constrain the relative sizes of the BLR by assuming a single bolometric $R-L$ relation, as shown in panel (a). 
    (b) shows the $R-L$ parameters for the primary data sets from Section~\ref{sec: res-RL} after converting to bolometric luminosities with the factors of \citet{Runnoe_2012}. 
    (c) shows the results of simultaneously fitting data from multiple lines to find a single bolometric $R-L$ relation.  To achieve this we allow \civ and \mgii lags to occur at different radii to the \hbeta lags and combine their data sets, i.e. allow for vertical offset between the $R-L$ relations for different lines. Note that figures (b) and (c) have different axis scaling. Figure (d) shows the best fit scaling needed to bring the lines to a common bolometric $R-L$ relation. The case of the regions overlapping with that of \hbeta, i.e. $C_{\mathrm{MgII}}\approx1$ and $C_{\mathrm{CIV}}\approx1$, are marked with dashed lines. A larger scale factor indicates a smaller emission radius relative to the \hbeta region. }
        \label{fig:Bolometric_Relations}
    \end{figure}
\end{landscape}

The constraints on the $R-L$ relation as well as the relative scaling factors are summarised in Table~\ref{tab: RL-bolomparams}, with a sketch of the implications for the BLR geometry shown in Figure~\ref{fig:bolom_scale_picture}. These findings are in line with those of \citet{SDSS-Shen_2023}, who similarly found \mgii lags to be systematically longer than those for \hbeta and \civ lags to be systematically shorter.

\begin{table*}
\rowcolors{2}{white}{white}
    \centering
    \begin{tabular}{||c||c|c|c|c||c|c||}
    \hline
             Data &  
             \makecell{Slope \\($\alpha$)} & 
             \makecell{Offset \\($\beta$)} & 
             \makecell{Scatter \\($\sigma$)} & 
             \makecell{Slope - Offset \\ Fit Correlation \\ ($\phi$)} &
             \makecell{$S_\mathrm{MgII}$\\ 
             $\left( \frac{R_{\mathrm{H}\beta}}    {R_{\mathrm{MgII}}} \right)$
             } &
             \makecell{$S_\mathrm{CIV}$ \\ 
                $\left( \frac{R_{\mathrm{H}\beta}}{R_{\mathrm{CIV}}} \right)$
             }
         \\[\tablespacing] \hline
            \hbeta 
            & $0.44 ^ {+0.04} _ {-0.02}$
            & $1.47 ^ {+0.04} _ {-0.02}$
            & $0.25 ^ {+0.02} _ {-0.02}$
            & $0.23$
            & - 
            & - 
        \\[\tablespacing] \hline
            \hbeta + \mgii 
            & $0.42 ^ {+0.02} _ {-0.02}$
            & $1.47 ^ {+0.04} _ {-0.03}$
            & $0.24 ^ {+0.02} _ {-0.02}$
            & $0.50$
            & $0.52 ^ {+0.06} _ {-0.06}$
            & - 
        \\[\tablespacing] \hline
            \hbeta + \civ
            
        	  & $0.47 ^ {+0.03} _ {-0.03}$
        	  & $1.6493 ^ {+0.03} _ {-0.04}$
        	& $0.29 ^ {+0.02} _ {-0.02}$
            & $0.50$
            &  - 
            & $1.28 ^ {+0.20} _ {-0.17}$
        \\ [\tablespacing] \hline
            \hbeta + \mgii + \civ 
            & $0.44 ^ {+0.02} _ {-0.02}$
            & $1.48 ^ {+0.04} _ {-0.02}$
            & $0.28 ^ {+0.02} _ {-0.01}$
            & $0.49$
            & $0.55 ^ {+0.07} _ {-0.06}$
            & $1.21 ^ {+0.17} _ {-0.16}$
        \\[\tablespacing] \hline
        
         \multicolumn{7}{||c||}{$\log_{10}(\Delta t) = \beta + \alpha 
         \left( 
            \log_{10}(L_{\mathrm{bol}}
         \right)-45) 
         \pm \sigma$}\\
         \hline
        
    \end{tabular}
    \caption{Parameter constraints for bolometric $R-L$ relationship. All slopes and offsets are fit for equation~\ref{eq:bolometric_scaling} with units of $10^{45} \mathrm{erg/s}$.}
    \label{tab: RL-bolomparams}
\end{table*}
It is worth noting that the relative radius of \hbeta and \mgii are strongly dependent on the assumptions of AGN spectral properties. Bolometric corrections are entirely degenerate with relative radius parameter $S_\lambda$, and the corrections of \citet{Runnoe_2012} are derived from only low redshift AGN. There is also considerable diversity amongst bolometric corrections from different sources \citep[e.g.][]{Runnoe_2012, Netzer_2019, Richards_2006}, though there is broad agreement on the relative corrections between \hbeta and \mgii, i.e. that $\log_{10}(L_{3000 \angstrom} / L_{5100 \angstrom})\approx 1.55$.

%--------------------------
% BOLOMETRIC CONTOURS

%--------------------------

\section{Black Hole Masses From Single-Epoch Measurements (New Results)}

\label{sec: single_epoch}

We apply the $R-L$ relationships derived in Section~\ref{sec: res-RL} and summarised in Table \ref{tab:R-L_Summary}, to derive black hole masses for OzDES AGN that do not have time-series spectra.  We use the luminosity of the AGN to estimate their lag, and combine that with a measure of the line width at a single epoch to estimate their mass. 
The results are subject to a set of quality cuts to avoid contamination of poor quality spectra. We require that all single-epoch sources:
\begin{enumerate}
    \item Have a clearly visible line profile by visual inspection,
    \item Have a mass that excludes a nonphysical zero/negative mass at a $1\sigma$ level
\end{enumerate}

After these cuts, we derive the masses of an additional $246$ OzDES AGN using these single-epoch estimates, consisting of $19$ \hbeta sources, $198$ \mgii sources and $29$ \civ sources. This single epoch sample results in masses ranging from $13.0^{+21.1}_{-8.0}\times 10^6 \msol$ to $2.8^{+5.8}_{-1.9}\times 10^9 \msol$. For the \civ sources, we make use of our `selection effect adjusted' $R-L$ relationship (see Section~\ref{sec: res-RL} for details).

As with the RM sources in Section~\ref{sec: res-masses}, we also use equation~\ref{eq: accretion} to estimate the accretion rates of these sources, using the bolometric corrections of \citet{Runnoe_2012} to convert monochromatic luminosities to the required units. All single epoch sources, including their masses, accretion rates, single epoch radii and measured velocity dispersions, are listed in Table \ref{tab:single_epoch}. These masses are included, along with the RM and literature measurements, in Figure~\ref{fig:All_population_plots}.

Masses are again calculated by varying all measurements within their uncertainties, with added variation for the single epoch lag within the uncertainties of our $R-L$ parameters. As with all other calculations, values are quoted as the distribution median, and uncertainties quoted using from 16\textsuperscript{th} and 84\textsuperscript{th} percentiles of the marginalised posterior distribution for each parameter. To avoid extrapolating $R-L$ relationships too far, monochromatic log-luminosities must be $>41\,\mathrm{erg\,s}^{-1}$ for \hbeta sources,  $>43.5\,\mathrm{erg\,s}^{-1}$ for \mgii sources and $>44.5\,\mathrm{erg\,s}^{-1}$ for \civ sources. The luminosity limit is somewhat arbitrary, so we include the single \mgii source outside this range in Table \ref{tab:single_epoch}, marked with a \textdagger. 

\begin{figure*}
    \centering
    \includegraphics[width=\linewidth]{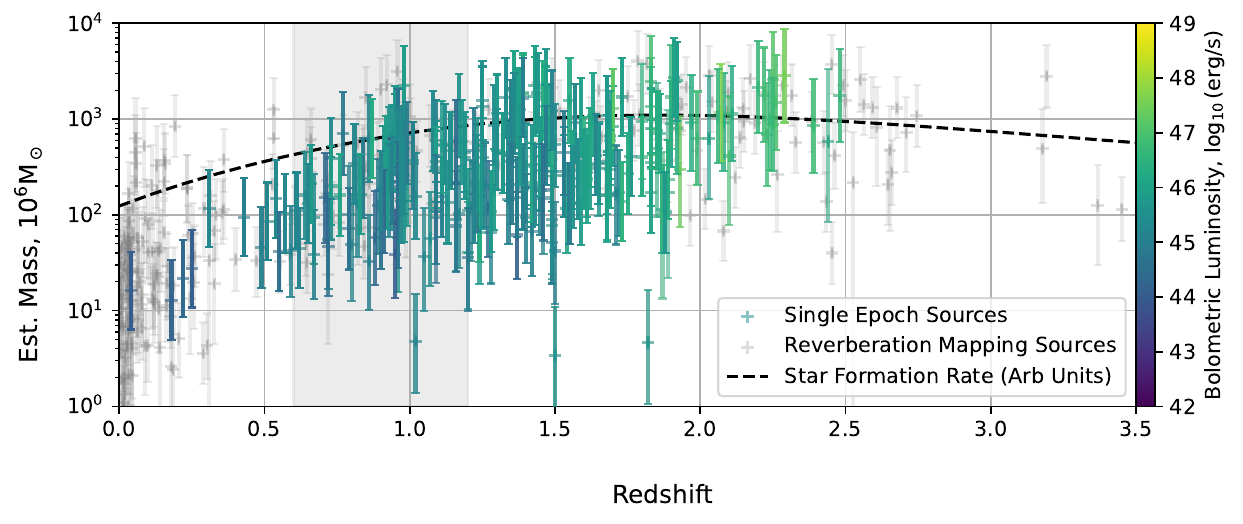} 
    \caption{Single epoch masses plotted against redshift for all single epoch sources, with shading showing the limits of the redshift bins in figure \ref{fig: mass_density}. Shown for comparison with a black dotted line is an estimate of star formation rate vs redshift using the functional form and parameters of \citet{Madau_2014} (equation $15$ in their paper). The opacity of the error bars scale inversely proportional to their width. Shown underneath in low opacity grey are the mass estimates from RM sources, including both our own and those from the existing works.}
    \label{fig: SE_masses}
\end{figure*}

\begin{figure}
    \centering
    \includegraphics[width=0.95\linewidth]{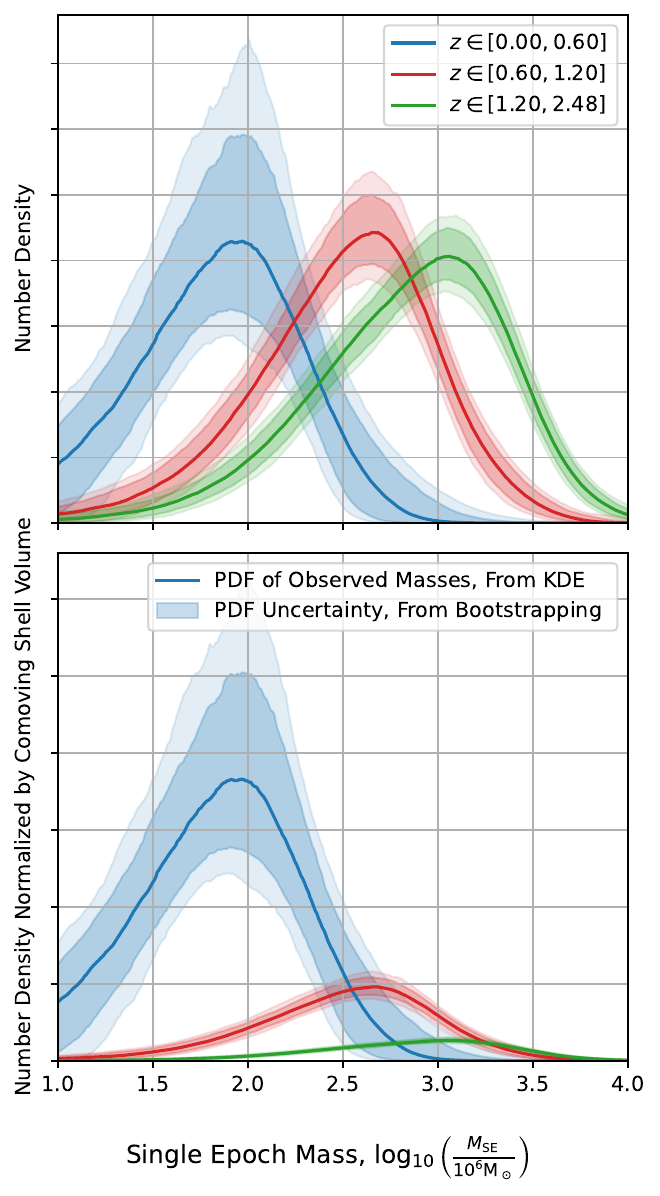} \\\vspace{-0.1cm}
    \caption{Kernel Density estimates of SMBH mass density at varying redshifts for the single epoch sources in Table \ref{tab:single_epoch}. The top panel shows the density of observed sources, while the bottom panel have their normalisation corrected by a factor of co-moving shell density such that they act as estimates of number-density per co-moving volume. The shaded regions represent uncertainties on the density acquired from bootstrapping and varying the data within measurement uncertainties.}
    \label{fig: mass_density}
\end{figure}

We find that the ceiling on the AGN SMBH masses has lowered towards more recent times, with no SMBH above $10^9 M_\odot$ nearer than $z=0.5$ in our results. If we limit ourselves to only the DES single-epoch data of Table \ref{tab:single_epoch} we acquire a single homogeneous data set that avoids the observation baseline dependent effects discussed in Section~\ref{sec: res-RL-CIV}. Even for this data set, we still see a significant evolution of the mass-distribution of sources as a function of redshift (Figure~\ref{fig: SE_masses}), with nearby sources showing a consistent shift towards lower masses. As \ozdes is a magnitude limited survey, some of the lack of distant low-mass observations can be explained by low completeness, dim objects being harder to observe at greater distances, but we can still confirm that the high-mass tail of the mass-density distribution, which should be subject only to count-based statistical effects, also decreases for nearby times to a degree that cannot be explained by statistical uncertainty or by a decrease in the number of available sources (see Figure~\ref{fig: mass_density}). This is not a shocking result, as it aligns with the well established trend of `cosmic downsizing' by which massive SMBH become less active towards more recent cosmic history \citep{Barger_2005_downsizing, Vestergaard_2009_downsizing, Kelly_2010_downsizing, Fanidakis_2011_downsizing}.

%-------------------------------------
\section{Discussion and Conclusions}
\label{sec: discussion}

This work marks the close of the \ozdes reverberation mapping program, and, in conjunction with the final RM data release of the Sloan Digital Sky Survey by \citet{SDSS-Shen_2023}, the end of the first generation of industrial scale high redshift reverberation mapping. \ozdes has measured lags for $62$ AGN, including $8$, $25$ and $29$ measurements from the \hbeta, \mgii and \civ reverberating lines, respectively, and we have in this work presented an additional $246$ SMBH masses from single epoch methods. These $308$ SMBH masses represent a significant contribution to the number of high redshift black hole mass measurements. The SMBHs range in mass from $2.3 \times 10^6 M_\odot$ to $4.0 \times 10^9 M_\odot$, putting our most massive measurement at the higher range of presently observed SMBHs.

Supplementing \ozdes data with existing lags that are consistent \ozdes's high-redshift sources, we provide significantly improved constraints on the radius-luminosity relationship for sub-Eddington \hbeta and \mgii lags. For all lines, we do not find a strong impact from accretion rate or emission line broadening as a single epoch lag predictor. Applying our \mgii $R-L$ relationship to ULAS J1120+0641, a particularly distant SMBH observed by JWST at a redshift of $z=7.09$ and a luminosity of $\log_{10}(L_{3000\angstrom}=46.47)$ and \mgii line FWHM of $2,500^{+480}_{-320}$ \citep{Banados_2017}, we recover a mass of $M=3.39^{+5.38}_{-2.10} \times 10^8 \msol$, slightly lower, but still consistent with, \citet{Banados_2017}'s estimate of $8\times10^8 \msol$.

Our single epoch relations for \hbeta and \mgii  have a significantly smaller scatter than those of the final \sdss release, $\sim0.25$ dex against their $\sim 0.45$ dex. As these $R-L$ relations are tuned over a large redshift range, they offer a broadly applicable tool for estimating SMBH masses to higher redshifts. We also offer the best constrained \civ $R-L$ fit at high redshift, but we recommend some caution in extrapolating this relation beyond the luminosity range of the available RM-data (i.e. $L_{1350\angstrom}>10^{47} \mathrm{erg/s}$). 

Using our new $R-L$ relations, we make single epoch mass estimates for $246$ AGN up to redshift $z=2.480$. In these estimates we are able to observe a dearth of massive AGN, with the upper range of AGN mass trending downwards at recent cosmic times. We note that measurements we present are subject to a selection function that depends on interaction between the physical properties of the AGN, the varied analysis pipelines of different surveys and the quality cuts that we apply here. These effects are particularly strong at high redshift where lags abut our upper observational limit of $1000 \; \dayu$ observer-frame lags. We make no attempt to model these selection effects, and so defer any interrogation of the black hole population properties to future work.

The similarity of the \mgii and \hbeta $R-L$ relationship in slope and scatter, even with a statistically significant sample such as we now have access to, supports our understanding of their emission regions as being stratified. Of interest is the systematically larger \mgii lags for comparable bolometric luminosity. Though simple arguments of ionisation energy suggest that the higher energy \mgii should precipitate out closer in than \hbeta, some photoionisation models \citep[e.g.][]{Guo_2020_Photoionization} do support \mgii being farther out from the SMBH engine. We note this with the caveat that our analysis is based on the simplest possible modelling of the spectral-energy distribution of AGN. Conversely, our fits indicate that \civ may be emitted from a radius much closer to that of \hbeta than found from simultaneous RM of \hbeta and \civ \citep{MgII_Lira_2018, Kaspi_2007}, though a slightly smaller radius is statistically preferred. The challenges in fitting a single $R-L$ relationship for this line make it difficult to constrain this with any certainty.

The strong windowing effects of the \civ $R-L$ trend highlights the need for further study of this emission line,\footnote{For example, to investigate whether the \civ line is affected by outflows, which could contaminate the line's velocity dispersion measurement.} either through intensive study of low redshift sources or by extending the luminosity envelope of \civ lag recoveries. Nearby, low luminosity sources can be observed in the observer-frame ultraviolet range, \cite[e.g.][]{MgII_Metzroth_2006, MgII_Lira_2018}, and extending such studies to higher redshift or luminosities would help to clarify the origins of the apparent flattening of the \civ $R-L$ curve. For the same reason, there is also motivation to use infrared spectroscopy to observe the high redshift counterparts to the \hbeta and \mgii sample to determine if these systems are homogeneous or show evolution over time. Extending these samples to more distant, more luminous sources will allow us to determine whether this `levelling out' of the $R-L$ relationship is physical, a statistical anomaly, or a selection effect. It will also let us probe the evolution of the geometry of such systems over time, and help delineate between accretion and luminosity based effects. In particular, follow up of sources in the low redshift domain, where the lower impact of time dilation means AGN variability is fast and lags are short, will allow direct comparison between \civ lags to \hbeta lags from $z\in[0,0.8]$ and to \mgii lags from $z\in[0.45, 1.25]$. \citet{MgII_Metzroth_2006} and \citet{MgII_Lira_2018} already do this for some local AGN, but extending to more distant and luminous sources will help fill the low luminosity branch of the \civ relationship.

The major findings of this paper are twofold: firstly that we have enough data as of this generation in BLR RM to make meaningful scaling relationships for \hbeta, \mgii and \civ out to high redshift, and that the large apparent disagreement between surveys can be resolved with even a simple accounting of selection effects and a consistent framework for screening false positives. In the resulting $R-L$ relations the inherent scatter now vastly outstrips statistical uncertainty, meaning the next generation of BLR RM may concern itself with the second order effects of sub-populations, multi-variable lag predictors (e.g. accretion rate), sub-populations and more granular statistical biases. The second finding, and one we make an effort to stress, is that the handling of these selection effects and false positives are not a fully settled issue.

Though BLR RM surveys are conceptually simple, surveys can adopt wildly divergent approaches in the particulars of their lag recovery. Different regimes for fitting lags and different criteria for quantifying the significance of these fits interact with initial survey target selection in a complex and multi-layered way. In this paper we offer $R-L$ relations that are a significant improvement over the existing standard, specifically for high redshift sources. However there remains work to be done in future to improve these constraints, even with existing RM measurements. There are two main avenues of interest for strengthening our RM-derived scaling relations in their current form: firstly, to re-fit lags for all sources across all surveys in a single framework, and secondly to properly model how this framework interacts with surveys to understand what selection functions arise in the luminosity / redshift / lag plane. We suggest that characterising these window functions should include a combination of simulation-based forward modelling in the style of \citep{OzDES-Penton_2021} with modern fitters like \texttt{LITMUS} and \pyroa do not suffer from  \javelin's numerical artefacts when encountering aliasing \citep[see][]{McDougall_2025_LITMUS}. Of particular interest to the \ozdes data would be a re-analysis that discards less false negatives for a larger final sample, and understanding how such cuts affects the apparent population distributions in addition to the overall false positive rate.

\section*{Contribution Statement}
Project conception and coordination: TMD, CL, PM; Analysis, programming, calculations: HM with input from ZY, UM, AP; Writing: HM; Original figures: HM; Editing: TMD, CL, PM, ZY, RS, GL, BT, BP; Data generation and/or curation: all authors.

\section*{Acknowledgements}

HGM, TMD, AP acknowledge support for early stages of this project from an Australian Research Council (ARC) Laureate Fellowship (project number FL180100168) and for later stages from the ARC Centre of Excellence for Gravitational Wave Discovery, OzGrav (CE230100016).

Funding for the DES Projects has been provided by the U.S. Department of Energy, the U.S. National Science Foundation, the Ministry of Science and Education of Spain, 
the Science and Technology Facilities Council of the United Kingdom, the Higher Education Funding Council for England, the National Center for Supercomputing 
Applications at the University of Illinois at Urbana-Champaign, the Kavli Institute of Cosmological Physics at the University of Chicago, 
the Center for Cosmology and Astro-Particle Physics at the Ohio State University,
the Mitchell Institute for Fundamental Physics and Astronomy at Texas A\&M University, Financiadora de Estudos e Projetos, 
Funda{\c c}{\~a}o Carlos Chagas Filho de Amparo {\`a} Pesquisa do Estado do Rio de Janeiro, Conselho Nacional de Desenvolvimento Cient{\'i}fico e Tecnol{\'o}gico and 
the Minist{\'e}rio da Ci{\^e}ncia, Tecnologia e Inova{\c c}{\~a}o, the Deutsche Forschungsgemeinschaft and the Collaborating Institutions in the Dark Energy Survey. 

The Collaborating Institutions are Argonne National Laboratory, the University of California at Santa Cruz, the University of Cambridge, Centro de Investigaciones Energ{\'e}ticas, 
Medioambientales y Tecnol{\'o}gicas-Madrid, the University of Chicago, University College London, the DES-Brazil Consortium, the University of Edinburgh, 
the Eidgen{\"o}ssische Technische Hochschule (ETH) Z{\"u}rich, 
Fermi National Accelerator Laboratory, the University of Illinois at Urbana-Champaign, the Institut de Ci{\`e}ncies de l'Espai (IEEC/CSIC), 
the Institut de F{\'i}sica d'Altes Energies, Lawrence Berkeley National Laboratory, the Ludwig-Maximilians Universit{\"a}t M{\"u}nchen and the associated Excellence Cluster Universe, 
the University of Michigan, NSF NOIRLab, the University of Nottingham, The Ohio State University, the University of Pennsylvania, the University of Portsmouth, 
SLAC National Accelerator Laboratory, Stanford University, the University of Sussex, Texas A\&M University, and the OzDES Membership Consortium.

Based in part on observations at NSF Cerro Tololo Inter-American Observatory at NSF NOIRLab (NOIRLab Prop. ID 2012B-0001; PI: J. Frieman), which is managed by the Association of Universities for Research in Astronomy (AURA) under a cooperative agreement with the National Science Foundation.

The DES data management system is supported by the National Science Foundation under Grant Numbers AST-1138766 and AST-1536171.
The DES participants from Spanish institutions are partially supported by MICINN under grants PID2021-123012, PID2021-128989 PID2022-141079, SEV-2016-0588, CEX2020-001058-M and CEX2020-001007-S, some of which include ERDF funds from the European Union. IFAE is partially funded by the CERCA program of the Generalitat de Catalunya.

We  acknowledge support from the Brazilian Instituto Nacional de Ci\^encia
e Tecnologia (INCT) do e-Universo (CNPq grant 465376/2014-2).

This document was prepared by the DES Collaboration using the resources of the Fermi National Accelerator Laboratory (Fermilab), a U.S. Department of Energy, Office of Science, Office of High Energy Physics HEP User Facility. Fermilab is managed by Fermi Forward Discovery Group, LLC, acting under Contract No. 89243024CSC000002.

Based in part on data acquired at the Anglo-Australian Telescope, under program Ab/2013B/012]. We acknowledge the traditional owners of the land on which the AAT stands, the Gamilaroi people, and pay our respects to elders past and present.

Calculations were made using \python \citep{VanRossum_2009_python} and with the aid of \texttt{numpy} \citep{harris_2020_numpy}. Plots and figures were generated with the aid of \texttt{matplotlib} \citep{Hunter_2007_matplotlib} and \texttt{chainconsumer} \citep{Hinton_2016_chainconsumer}.

We acknowledge and pay respect to the traditional owners of the land on which the University of Queensland and University of Southern Queensland are situated, upon whose unceded, sovereign, ancestral lands we work. We pay respects to their Ancestors and descendants, who continue cultural and spiritual connections to Country.

Based in part on data acquired at the Anglo-Australian Telescope, under program A/2013B/12. We acknowledge the traditional custodians of the land on which the AAT stands, the Gamilaraay people, and pay our respects to elders past and present.

%%%%%%%%%%%%%%%%%%%%%%%%%%%%%%%%%%%%%%%%%%%%%%%%%%
\section*{Data Availability}

With this paper we release a table of the \ozdes RM light curves for the full sample, as well as tabulated forms of the RM and single epoch samples listed in Appendix~\ref{app:methods} (Tables~\ref{tab:Results_All_Hbeta}, \ref{tab:Results_All_MgII}, \ref{tab:Results_All_CIV}), and Appendix~\ref{sec: SE_results} (Table~\ref{tab:single_epoch}).  This is available at \href{https://zenodo.org/records/20120159?token=eyJhbGciOiJIUzUxMiJ9.eyJpZCI6ImEyOWM0NzRkLTgwNjktNDUyNS1iM2UwLTFmNTQ2Y2E2YjkzMyIsImRhdGEiOnt9LCJyYW5kb20iOiJjYTc1Y2IzODM5ZDhjN2E1MmEwOTg3ODRlZjgyMjczZiJ9.ubrvuKXrTumeo83TU4ndyu54RzXdDupGWOytqtr50a6HQqMErXcY_XVrpdeeL0KKMxxs0TKcZmLvJOJvVcTW_A}{this zenodo repository}.

The raw images and catalogues from the whole DES survey are available at \url{https://des.ncsa.illinois.edu/releases/dr2}.

The spectra are publicly available from \url{https://docs.datacentral.org.au/ozdes/overview/dr2/}.

%%%%%%%%%%%%%%%%%%%% REFERENCES %%%%%%%%%%%%%%%%%%
\def \aap{A\&A}
\def \aaps{A\&AS}
\def \ag{Astron. Geophys.}
\def \aj{AJ}
\def \ajp{Am. J. Phys.}
\def \al{Astron. Lett.}
\def \ap{Appl. Phys.}
\def \apj{ApJ}
\def \apjl{ApJ}
\def \apjl{ApJ Lett.}
\def \apjs{ApJS}
\def \apss{Astrophys. and Space Science}
\def \araa{ARA\&A}
\def \arns{Annu. Rev. Nuc. Sci.}
\def \asp{Astron. Soc. Pac.}
\def \azh{Astronomicheskii Zhurnal}
\def \baas{BAAS}
\def \baps{Bull. Am. Phys. Soc.}
\def \bist{Bull. Inf. Sci. Tech.}
\def \ca{Comments on Astrophys.}
\def \cqg{Class. Quantum Gravity}
\def \epjc{Euro. Phys. J. C}
\def \grg{Gen. Relativ. Gravitation}
\def \ijmpd{Int. J. Mod. Phys. D}
\def \jhep{J. High Energy Phys.}
\def \jms{J. Molecular Spectrosc.}
\def \jos{J. Opt. Soc. Am.}
\def \jcap{J. Cosmo. Astropart. Phys.}
\def \josb{J. Opt. Soc. Am. B}
\def \jpcrd{J. Phys. Chem. Ref. Data}
\def \jpcrds{J. Phys. Chem. Ref. Data Suppl.}
\def \jqsrt{J. Quant. Spectrosc. Radiat. Transfer}
\def \jtp{J. Technical Phys.}
\def \met{Metrologia}
\def \mnras{MNRAS}
\def \mpla{Mod. Phys. Lett. A}
\def \nat{Nature}
\def \nature{Nature}
\def \npb{Nucl. Phys. B}
\def \nsrds{Natl. Stand. Rel. Data Ser.}
\def \nw{Naturwiss.}
\def \pasa{Publ. Astron. Soc. Austral.}
\def \pasp{PASP}
\def \pawk{Preuss. Akad. Wiss. K}
\def \plb{Phys. Lett. B}
\def \phd{PhD thesis}
\def \physrep{Physics Reports}
\def \pr{Phys. Rev.}
\def \pra{Phys. Rev. A}
\def \prb{Phys. Rev. B}
\def \prc{Phys. Rev. C}
\def \prd{Phys. Rev. D}
\def \prep{in preparation}
\def \prl{Phys. Rev. Lett.}
\def \prsa{Proc. R. Soc. A}
\def \psc{Phys. Scr.}
\def \ptp{Progress Theor. Phys.}
\def \ptrsla{Phil. Trans. R. Soc. London}
\def \qjras{Quart. J. R. Astron. Soc.}
\def \rmp{Rev. Mod. Phys.}
\def \rpp{Rep. Prog. Phys.}
\def \sovast{Soviet Astronomy}
\def \spu{Sov. Phys. Uspekhi}
\def \ssr{Space Sci. Rev.}
\def \tms{Trans. Math. Software}

\def \cup{Cambridge U. Press}
\def \cupadr{Cambridge U.K.}
\def \pup{Princeton U. Press}
\def \pupadr{Princeton, U.S.A.}

\def \etal{et al.}

\section*{Affiliations}
{\footnotesize
$^{1}$School of Mathematics and Physics, University of Queensland,  Brisbane, QLD 4072, Australia, $^{2}$Kavli Institute for Particle Astrophysics \& Cosmology, P. O. Box 2450, Stanford University, Stanford, CA 94305, USA, $^{3}$Center for Cosmology and Astro-Particle Physics, The Ohio State University, Columbus, OH 43210, USA, $^{4}$Department of Astronomy, The Ohio State University, Columbus, OH 43210, USA, $^{5}$Centre for Gravitational Astrophysics, College of Science, The Australian National University, ACT 2601, Australia, $^{6}$The Research School of Astronomy and Astrophysics, Australian National University, ACT 2601, Australia, $^{7}$Sydney Institute for Astronomy, School of Physics, A28, The University of Sydney, NSW 2006, Australia, $^{8}$School of Mathematical \& Physical Sciences, 12 Wally's Walk, Macquarie University, Macquarie Park, NSW 2113, $^{9}$Fermi National Accelerator Laboratory, P. O. Box 500, Batavia, IL 60510, USA, $^{10}$Physik-Institut, University of Z\'{u}rich, Winterthurerstrasse 190, CH-8057 Z{\'u}rich, Switzerland, $^{11}$Departamento de F\'{\i}sica Te\'orica, Centro de Astropart\'iculas y F\'isica de Altas Energ\'ias (CAPA), Universidad de Zaragoza, 50009 Zaragoza, Spain, $^{12}$Institute of Cosmology and Gravitation, University of Portsmouth, Portsmouth, PO1 3FX, UK, $^{13}$University Observatory, LMU Faculty of Physics, Scheinerstr. 1, 81679 Munich, Germany, $^{14}$Department of Physics \& Astronomy, University College London, Gower Street, London, WC1E 6BT, UK, $^{15}$Instituto de Astrofisica de Canarias, E-38205 La Laguna, Tenerife, Spain, $^{16}$Laborat\'orio Interinstitucional de e-Astronomia - LIneA, Av. Pastor Martin Luther King Jr, 126 Del Castilho, Nova Am\'erica Offices, Torre 3000/sala 817, Brazil, $^{17}$Universidad de La Laguna, Dpto. Astrof{\'i}sica, E-38206 La Laguna, Tenerife, Spain, $^{18}$INAF-Osservatorio Astronomico di Trieste, via G. B. Tiepolo 11, I-34143 Trieste, Italy, $^{19}$Korea Astronomy and Space Science Institute, 776 Daedeok-daero, Yuseong-gu, Daejeon 34055, South Korea, $^{20}$Institut de F\'{\i}sica d'Altes Energies (IFAE), The Barcelona Institute of Science and Technology, Campus UAB, 08193 Bellaterra (Barcelona) Spain, $^{21}$Kapteyn Astronomical Institute, University of Groningen, Landleven 12 (Kapteynborg, 5419), 9747 AD Groningen, The Netherlands, $^{22}$Hamburger Sternwarte, Universit\"{a}t Hamburg, Gojenbergsweg 112, 21029 Hamburg, Germany, $^{23}$Centro de Investigaciones Energ\'eticas, Medioambientales y Tecnol\'ogicas (CIEMAT), Madrid, Spain, $^{24}$California Institute of Technology, 1200 East California Blvd, MC 249-17, Pasadena, CA 91125, USA, $^{25}$Instituto de F\'{i}sica Te\'{o}rica UAM/CSIC, Universidad Aut\'{o}noma de Madrid, 28049 Madrid, Spain, $^{26}$Centre for Astrophysics \& Supercomputing, Swinburne University of Technology, Victoria 3122, Australia, $^{27}$Santa Cruz Institute for Particle Physics, Santa Cruz, CA 95064, USA, $^{28}$Center for Astrophysics $\vert$ Harvard \& Smithsonian, 60 Garden Street, Cambridge, MA 02138, USA, $^{29}$Lawrence Berkeley National Laboratory, 1 Cyclotron Road, Berkeley, CA 94720, USA, $^{30}$Lowell Observatory, 1400 Mars Hill Rd, Flagstaff, AZ 86001, USA, $^{31}$Jet Propulsion Laboratory, California Institute of Technology, 4800 Oak Grove Dr., Pasadena, CA 91109, USA, $^{32}$Department of Physics and Astronomy, University of Pennsylvania, Philadelphia, PA 19104, USA, $^{33}$George P. and Cynthia Woods Mitchell Institute for Fundamental Physics and Astronomy, and Department of Physics and Astronomy, Texas A\&M University, College Station, TX 77843,  USA, $^{34}$Universit\'e Grenoble Alpes, CNRS, LPSC-IN2P3, 38000 Grenoble, France, $^{35}$Center for Astrophysical Surveys, National Center for Supercomputing Applications, 1205 West Clark St., Urbana, IL 61801, USA, $^{36}$Department of Astronomy, University of Illinois at Urbana-Champaign, 1002 W. Green Street, Urbana, IL 61801, USA, $^{37}$Instituci\'o Catalana de Recerca i Estudis Avan\c{c}ats, E-08010 Barcelona, Spain, $^{38}$Department of Astrophysical Sciences, Princeton University, Peyton Hall, Princeton, NJ 08544, USA, $^{39}$School of Mathematics and Physics, University of Surrey, Guildford, Surrey, GU2 7XH, UK, $^{40}$Observat\'orio Nacional, Rua Gal. Jos\'e Cristino 77, Rio de Janeiro, RJ - 20921-400, Brazil, $^{41}$Ruhr University Bochum, Faculty of Physics and Astronomy, Astronomical Institute, German Centre for Cosmological Lensing, 44780 Bochum, Germany, $^{42}$Physics Department, Lancaster University, Lancaster, LA1 4YB, UK, $^{43}$Computer Science and Mathematics Division, Oak Ridge National Laboratory, Oak Ridge, TN 37831, $^{44}$Department of Astronomy and Astrophysics, University of Chicago, Chicago, IL 60637, USA, $^{45}$Cerro Tololo Inter-American Observatory, NSF's National Optical-Infrared Astronomy Research Laboratory, Casilla 603, La Serena, Chile, $^{46}$Berkeley Center for Cosmological Physics, Department of Physics, University of California, Berkeley, CA 94720, US
}
%%%%%%%%%%%%%%%%%%%%%%%%%%%%%%%%%%%%%%%%%%%%%%%%%%

%%%%%%%%%%%%%%%%% APPENDICES %%%%%%%%%%%%%%%%%%%%%

\appendix
%----------------------------------------------------

\section{Lag recovery methods}\label{app:methods}
%-----------------------------------------------

The main two lag recovery methods are the interpolated cross-correlation function and Javelin, which we describe here.

\subsection{\PyCCF}

Introduced to the field of reverberation mapping by \citet{ICCF-Gaskell_1987}, the Interpolated Cross-Correlation Function (ICCF) is a purely non-parametric lag recovery method that relies on the cross-correlation ($r$) of the two light curves as they are shifted against one another in the time domain. The procedure is to linearly interpolate one (or both) of the light curves between observations and measure the Pearson correlation, 
\begin{equation}
    r =\frac{\braket{y_1, y_2}}{\sqrt{\braket{y_1, y_1} \braket{y_2, y_2}}},
    \label{eq: cross-correlation}
\end{equation}
where $y_1$ and $y_2$ represent the sets of photometric and spectroscopic amplitudes respectively, and angled brackets indicate an inner product, approximated by a sum.\footnote{$\braket{y_1,y_2}=\sum_{i=1}^N (y_{1,i}-\bar{y}_1)(y_{2,i}-\bar{y}_2)$, where the overline represents the mean.} The physical lag is then taken to be the delay that maximises this correlation, while the uncertainties are estimated by `bootstrapping': recovering the peak correlation-lag from many realisations generated by randomly sub-sampling the observations and then re-sampling these observations within their measurement uncertainties.

The ICCF method is numerically low cost and robust, and has seen wide use in RM, particularly in low redshift campaigns with high cadence measurements \citep[e.g][]{ HBETA_Barth_2013, HBETA_Rakshit_2019, HBETA_Zhang_2019}. Through simulations, it has been found to agree with more rigorous models like \javelin to within statistical bounds, though with higher reported uncertainties \citep{OZDES-Yu_2019}. The low numerical cost of the ICCF means that it can easily be used as a diagnostic tool for testing the reliability of other more involved methods, as the $r$ test statistic gives a general `goodness of fit' at any lag, and the low cost of evaluation means the entire lag parameter space can be searched exhaustively. \ozdes makes use of \PyCCF \citep{PyCCF-Mouyan_2018}, a vectorised \python-based implementation of ICCF. \citet{OzDES-Malik_2023} estimates a false positive rate / $p$-value for each lag by shuffling the time ordering of measurements, creating mock observations of signals with the same stationary statistics as the observed light curve. These were then used to tune the quality cuts outlined in Section~\ref{sec: methods_Hbetacut}.

ICCF is robust and `light-weight', but has shortfalls in its lack of precision and its inability to adequately characterise the vague constraints of light curves where large gaps exist between observations, such as the six month seasonal gaps of \ozdes and \sdss \citep{OzDES-Malik_2022}. The quality of lag recovery degrades rapidly as measurement cadence or observational season length decrease, as the linear interpolation lends false confidence while failing to capture the stochastic variations that take place between measurements. By contrast, the \sdss team's preferred lag recovery method, \pyroa \citep{pyroa-ferus_2021}, interpolates the light curves by adopting a `rolling average' of the observations within a window of variable width, giving conservatively broad constraints within seasonal gaps compared to \PyCCF's spuriously tight ones. Non-parametric methods like \PyCCF and \pyroa allow a certain degree of flexibility to account for their approximate modelling, and so naturally inherit a degree of robustness against outliers or under-estimated errors \citep{OZDES-Yu_2019}.

%-----------------------------------------------

\subsection{\javelin}
Where non-parametric methods interpolate between observations with as few assumptions about the underlying light curves as possible, the second class of fitting methods leverages our understanding of the underlying stochasticity of AGN variability. Though AGN do not follow a consistent light curve shape like some astrophysical objects \citep[e.g. supernovae, ][]{Guy_2005}, they do exhibit consistent statistical properties in their signals, namely a set power spectrum for the variations in their brightness. Specifically, these variations are known to follow a `damped random walk' (DRW) \citep{Kelly_2009, Kozlowski_2010a, MacLeod_2010}, with only small deviations at very long and short timescales of variation \citep{Zu_2013}. The DRW is an example of a Gaussian process, specifically a first order continuous auto-regressive process, yielding a red-noise like $1/f^2$ scaling at high frequencies. This Gaussian-process-like structure in the stochastic variations creates statistical correlations within and between light curves, allowing us to interpolate between measurements in a physically principled way. 
For the DRW specifically, the auto-correlation is encoded in a covariance matrix, whose elements are defined as:
\begin{equation}
    C_{ij} = \sigma_i \sigma _j \exp(\vert t_i-t_j\vert / \tau_d) + E_i E_j \delta_{ij}
    \label{eq: GP_covariance},
\end{equation}
where $\sigma_{i/j}$ are the variabilities of the light curves that measurements $i$ and $j$ belong to, $E_{i/j}$ are their measurement uncertainties, $t_{i/j}$ are their measurement times after the spectroscopic measurements offset by the lag $\Delta t$, $\tau$ is the timescale of variability in the damped random walk and $\delta_{ij}$ is the Kronecker delta.

The predominant Gaussian process modelling program at time of writing is the \python-based program \javelin \citep{JAVELIN-Zu_2010}, a DRW-based implementation of the method outlined by \citet{Press_Rybicki_1989}. This method models the AGN light curves in a Bayesian way, constructing a unified light curve from the photometric and spectroscopic observations by treating the response as a shifted, scaled and smoothed copy of the driving continuum. This approach gives a closed form likelihood function:
\begin{equation}
    \mathcal{L} =
    \frac{1}
    {\sqrt{\mathrm{det}((2 \pi)^N C^{-1})}}
    \exp \left(- \frac{1}{2} (\vec{y} - \vec{y}_0)^T C^{-1} (\vec{y} - \vec{y}_0)\right)
    \label{eq: GP_likelihood},
\end{equation}
where $C$ is the covariance matrix defined in equation~\ref{eq: GP_covariance} and $\vec{y}-\vec{y_0}$ are the light curve measurements (photometric and spectroscopic) after offsetting by the mean of their respective light curves and $N$ is the total number of observations across all light curves. This likelihood allows the signal parameters, including the lag, to be constrained in a Bayesian framework. This procedure uses all available information simultaneously to constrain the light curve behaviour between measurements, more accurately characterising the uncertainty arising from large seasonal gaps.

Equation~\ref{eq: GP_likelihood} requires the inversion of a large matrix of rank $N$. Though the near-diagonal shape of equation~\ref{eq: GP_covariance} ameliorates this cost somewhat, this is still a computationally expensive process. To reduce the number of evaluations, \javelin uses the MCMC approach using the \python-based package \emcee \citep{emcee-Foreman_Mackey_2013}, which uses the Affine-Invariant Ensemble Sampler \citep{AffineInvariant-Goodman_2010} to direct the bulk of its samples to high likelihood regions of parameter space. The seasonal windowing function of equatorial surveys, and the resulting aliasing problem (see Section~\ref{sec: aliasing} for details) can give rise to multi-modal distributions along the lag axis of parameter-space \citep[e.g.][]{CIV_Grier_2019}, which can result in poor efficiency and inaccurate posterior distributions from \emcee. Gaussian-process-based modelling is also sensitive to outlier data-points or under-estimated measurement uncertainties, which can spuriously over-constrain the light curve and disrupt the lag recovery results. 

Another Gaussian process based lag recovery program similar to \javelin is \cream \citep{Starkey_2015_CREAM}, which broadly adopts the same methodology as \javelin but assumes different Gaussian process statistics and parameterises its Bayesian priors in terms of physical properties (e.g. mass, accretion rate etc). \cream has seen use in earlier \sdss papers, \citep[e.g.][]{CIV_Grier_2019}, and includes an `error scaling' parameter in its generative model to relax the light curve constraints and side-step the impacts of under-estimated uncertainties. Outlier rejection, for both \javelin and \cream, still relies on `by-eye' identification and removal of suspicious measurements \citep[e.g.][]{OzDES-Yu_2023}.

\onecolumn

\section{\ozdes Reverberation Mapping Results}

\begin{longtable}{c|c|c|c|c|c|c}
        \centering

\tableheader{5100}
\hline \hline \\
$\mathrm{DES J002802.42-424913.52}$ & 0.127 & $43.31 \pm 0.01$ & $16_{-2.7}^{+2.7}$ & $1385.00 \pm 3.00$ & $24_{-13}^{+28}$ & $0.5_{-0.4}^{+1.6}$ \\ [\tablespacing]
$\mathrm{DES J024347.34-005354.84}$ & 0.237 & $43.83 \pm 0.01$ & $27.5_{-1.6}^{+0.8}$ & $1645.00 \pm 6.00$ & $61_{-31}^{+64}$ & $0.4_{-0.3}^{+1.4}$ \\ [\tablespacing]
$\mathrm{DES J034028.46-292902.41}$ & 0.310 & $44.64 \pm 0.00$ & $16_{-15}^{+20}$ & $1716.00 \pm 4.00$ & $33_{-35}^{+68}$ & $20_{-10}^{+240}$ \\ [\tablespacing]
$\mathrm{DES J022249.67-051453.01}$ & 0.314 & $44.12 \pm 0.01$ & $60_{-35}^{+16}$ & $1867.00 \pm 5.00$ & $160_{-100}^{+210}$ & $0.17_{-0.14}^{+0.94}$ \\ [\tablespacing]
$\mathrm{DES J003954.13-440509.97}$ & 0.332 & $44.00 \pm 0.01$ & $39_{-11}^{+29}$ & $1838.00 \pm 9.00$ & $100_{-60}^{+140}$ & $0.3_{-0.3}^{+1.9}$ \\ [\tablespacing]
$\mathrm{DES J022330.16-054758.06}$ & 0.354 & $43.63 \pm 0.01$ & $38.9_{-4.6}^{+6.9}$ & $1831.00 \pm 9.00$ & $110_{-60}^{+120}$ & $0.07_{-0.05}^{+0.24}$ \\ [\tablespacing]
$\mathrm{DES J002904.43-425243.04}$ & 0.644 & $44.59 \pm 0.01$ & $19_{-3.8}^{+5.3}$ & $1836.00 \pm 4.00$ & $50_{-27}^{+57}$ & $10_{-8}^{+36}$ \\ [\tablespacing]
$\mathrm{DES J022617.85-043108.99}$ & 0.707 & $44.69 \pm 0.01$ & $36_{-5.3}^{+9}$ & $1671.00 \pm 7.00$ & $80_{-43}^{+91}$ & $5_{-4}^{+19}$ \\ [\tablespacing]

        \caption{The 8 \ozdes results for \hbeta, as listed in \citet{OzDES-Malik_2023}. Masses and dimensionless accretion rates are calculated as per equation~\ref{eq:RM_mass} and \ref{eq: accretion}, with full error propagation.}
        \label{tab:Results_All_Hbeta}
    \end{longtable}

    \begin{longtable}{c|c|c|c|c|c|c}
        \centering

\tableheader{3000}
\hline \hline \\
$\mathrm{DES J033328.93-275641.21}$ & 0.840 & $44.70 \pm 0.01$ & $95.1_{-9.2}^{+8.7}$ & $2344.2 \pm 66.9$ & $420_{-220}^{+450}$ & $0.1_{-0.08}^{+0.33}$ \\ [\tablespacing]
$\mathrm{DES J024831.08-005025.60}$ & 0.890 & $44.94 \pm 0.01$ & $151_{-54}^{+19}$ & $1868 \pm 163$ & $420_{-230}^{+500}$ & $0.2_{-0.2}^{+1}$ \\ [\tablespacing]
$\mathrm{DES J002933.85-435240.69}$ & 1.000 & $45.86 \pm 0.01$ & $286_{-15}^{+7}$ & $2147 \pm 275$ & $1100_{-600}^{+1200}$ & $0.9_{-0.7}^{+3.3}$ \\ [\tablespacing]
$\mathrm{DES J003710.86-444048.11}$ & 1.070 & $45.66 \pm 0.00$ & $185_{-15}^{+16}$ & $2121.3 \pm 81.9$ & $670_{-350}^{+730}$ & $1.1_{-0.8}^{+3.7}$ \\ [\tablespacing]
$\mathrm{DES J033836.19-295113.50}$ & 1.150 & $44.55 \pm 0.02$ & $105_{-24}^{+21}$ & $2644 \pm 189$ & $570_{-310}^{+670}$ & $0.03_{-0.03}^{+0.12}$ \\ [\tablespacing]
$\mathrm{DES J033211.42-284323.99}$ & 1.240 & $44.27 \pm 0.01$ & $59_{-9}^{+12}$ & $3313.5 \pm 95.3$ & $510_{-270}^{+580}$ & $0.015_{-0.012}^{+0.054}$ \\ [\tablespacing]
$\mathrm{DES J003922.97-430230.41}$ & 1.370 & $45.36 \pm 0.01$ & $238_{-14}^{+22}$ & $2291 \pm 176$ & $1000_{-500}^{+1100}$ & $0.17_{-0.13}^{+0.57}$ \\
\end{longtable}

    %\onecolumn
    \begin{longtable}{c|c|c|c|c|c|c}
        \centering
\tableheader{1350}
\hline \hline \\
$\mathrm{DES J022327.85-040119.16}$ & 1.922 & $45.45 \pm 0.08$ & $136_{-11}^{+11}$ & $3428 \pm 145$ & $1300_{-700}^{+1400}$ & $0.1_{-0.08}^{+0.37}$ \\ [\tablespacing]
$\mathrm{DES J022514.39-044700.14}$ & 1.928 & $46.21 \pm 0.01$ & $74_{-17}^{+17}$ & $3601.0 \pm 27.0$ & $760_{-410}^{+870}$ & $4_{-3}^{+16}$ \\ [\tablespacing]
$\mathrm{DES J022537.03-050109.34}$ & 1.936 & $45.77 \pm 0.02$ & $100.1_{-5.8}^{+5.8}$ & $4190.0 \pm 40.0$ & $1400_{-700}^{+1500}$ & $0.26_{-0.2}^{+0.87}$ \\ [\tablespacing]
$\mathrm{DES J033401.79-265054.28}$ & 1.952 & $45.93 \pm 0.02$ & $204.6_{-9.8}^{+9.8}$ & $3604.0 \pm 67.0$ & $2200_{-1100}^{+2300}$ & $0.2_{-0.15}^{+0.63}$ \\ [\tablespacing]
$\mathrm{DES J025159.70-005159.89}$ & 1.988 & $45.24 \pm 0.05$ & $37.8_{-9}^{+9}$ & $3956 \pm 159$ & $470_{-260}^{+550}$ & $0.4_{-0.3}^{+1.5}$ \\ [\tablespacing]
$\mathrm{DES J032703.62-274425.27}$ & 2.031 & $46.13 \pm 0.01$ & $183_{-12}^{+12}$ & $4395.0 \pm 43.0$ & $2900_{-1500}^{+3000}$ & $0.22_{-0.17}^{+0.76}$ \\ [\tablespacing]
$\mathrm{DES J002959.21-434835.24}$ & 2.041 & $46.81 \pm 0.01$ & $145.4_{-5.9}^{+5.9}$ & $3587.0 \pm 13.0$ & $1500_{-800}^{+1600}$ & $8_{-6}^{+27}$ \\
\end{longtable}

%-----------------------------------------------

\section{\ozdes Single Epoch Results}
\label{sec: SE_results}

\begin{longtable}{l|c|c|c|c|c|c|c}
Source ID  &  Redshift &  Line & \makecell{Monochromatic \\ Luminosity \\ $\log_{10}(\lambda L_\lambda \mathrm{erg/s})$} & \makecell{Single \\ Epoch \\ Lag \\ (Days)}  & \makecell{   Velocity \\ Dispersion \\ $\sigma_V $ \\ (km/s)  }&  \makecell{Single \\ Epoch \\ Mass \\($10^6 \msol$) }&  \makecell{Dimensionless \\ Accretion \\ Rate \\ $\dot{M}$}  \\[12pt] \hline 
$\mathrm{DES15E1arx}$ & 0.040 & \hbeta & $43.45 \pm 0.01$ & $15_{-7}^{+12}$ & $1444.0 \pm 27.0$ & $16_{-10}^{+25}$ & $0.4_{-0.3}^{+1.9}$ \\ [\tablespacing]
$\mathrm{DES14S2aog}$ & 0.180 & \hbeta & $43.59 \pm 0.04$ & $18_{-8}^{+14}$ & $1182 \pm 130$ & $13_{-8}^{+21}$ & $1.0_{-0.9}^{+5.9}$ \\ [\tablespacing]
$\mathrm{OzDES-DR2\_04692}$ & 0.220 & \hbeta & $43.76 \pm 0.02$ & $21_{-9}^{+17}$ & $1346.0 \pm 70.0$ & $22_{-13}^{+33}$ & $0.8_{-0.7}^{+4.7}$ \\ [\tablespacing]
$\mathrm{OzDES-DR2\_03800}$ & 0.250 & \hbeta & $43.53 \pm 0.05$ & $17_{-7}^{+13}$ & $1767.0 \pm 52.0$ & $27_{-17}^{+41}$ & $0.18_{-0.15}^{+0.96}$ \\ [\tablespacing]
$\mathrm{SVA1\_COADD-2939653045}$ & 0.310 & \hbeta & $44.65 \pm 0.01$ & $52_{-23}^{+42}$ & $1710.00 \pm 4.00$ & $120_{-70}^{+180}$ & $1.8_{-1.5}^{+9.5}$ \\ [\tablespacing]
$\mathrm{OzDES-DR2\_04665}$ & 0.430 & \hbeta & $44.41 \pm 0.08$ & $41_{-18}^{+32}$ & $1786.0 \pm 40.0$ & $90_{-60}^{+150}$ & $1.0_{-0.8}^{+5.1}$ \\ [\tablespacing]
$\mathrm{OzDES-DR2\_04637}$ & 0.490 & \hbeta & $44.41 \pm 0.08$ & $41_{-18}^{+31}$ & $1258 \pm 155$ & $45_{-28}^{+75}$ & $4_{-4}^{+24}$ \\
\end{longtable}

%-----------------------------------------------

\twocolumn
\bibliographystyle{apj}
\bibliography{bib_main,bib_Hbetasources,bib_MgIIsources,bib_CIVsources,bib_OzDES}

@article{CIV_Kaspi_2021,
    doi = {10.3847/1538-4357/ac00aa},
    url = {https://doi.org/10.3847/1538-4357/ac00aa},
    year = {2021},
    month = {jul},
    publisher = {The American Astronomical Society},
    volume = {915},
    number = {2},
    pages = {129},
    author = {Kaspi, Shai and Brandt, W. N. and Maoz, Dan and Netzer, Hagai and Schneider, Donald P. and Shemmer, Ohad and Grier, C. J.},
    title = {Taking a Long Look: A Two-decade Reverberation Mapping Study of High-luminosity Quasars},
    journal = {The Astrophysical Journal},
}

@article{CIV_Grier_2019,
   title={The Sloan Digital Sky Survey Reverberation Mapping Project: Initial C iv Lag Results from Four Years of Data},
   volume={887},
   ISSN={1538-4357},
   url={http://dx.doi.org/10.3847/1538-4357/ab4ea5},
   DOI={10.3847/1538-4357/ab4ea5},
   number={1},
   journal={The Astrophysical Journal},
   publisher={American Astronomical Society},
   author={Grier, C. J. and Shen, Yue and Horne, Keith and Brandt, W. N. and Trump, J. R. and Hall, P. B. and Kinemuchi, K. and Starkey, David and Schneider, D. P. and Ho, Luis C. and Homayouni, Y. and Li, Jennifer I-Hsiu and McGreer, Ian D. and Peterson, B. M. and Bizyaev, Dmitry and Chen, Yuguang and Dawson, K. S. and Eftekharzadeh, Sarah and Guo, Yucheng and Jia, Siyao and Jiang, Linhua and Kneib, Jean-Paul and Li, Feng and Li, Zefeng and Nie, Jundan and Oravetz, Audrey and Oravetz, Daniel and Pan, Kaike and Petitjean, Patrick and Ponder, Kara A. and Rogerson, Jesse and Vivek, M. and Zhang, Tianmeng and Zou, Hu},
   year={2019},
   month=dec, pages={38} }

@article{CIV_Peterson_2005,
   title={Multiwavelength Monitoring of the Dwarf Seyfert 1 Galaxy NGC 4395. I. A Reverberation‐based Measurement of the Black Hole Mass},
   volume={632},
   ISSN={1538-4357},
   url={http://dx.doi.org/10.1086/444494},
   DOI={10.1086/444494},
   number={2},
   journal={The Astrophysical Journal},
   publisher={American Astronomical Society},
   author={Peterson, Bradley M. and Bentz, Misty C. and Desroches, Louis‐Benoit and Filippenko, Alexei V. and Ho, Luis C. and Kaspi, Shai and Laor, Ari and Maoz, Dan and Moran, Edward C. and Pogge, Richard W. and Quillen, Alice C.},
   year={2005},
   month=oct, pages={799–808} }

@article{CIV_Rosa_2015,
   title={SPACE TELESCOPE AND OPTICAL REVERBERATION MAPPING PROJECT. I. ULTRAVIOLET OBSERVATIONS OF THE SEYFERT 1 GALAXY NGC 5548 WITH THE COSMIC ORIGINS SPECTROGRAPH ONHUBBLE SPACE TELESCOPE},
   volume={806},
   ISSN={1538-4357},
   url={http://dx.doi.org/10.1088/0004-637X/806/1/128},
   DOI={10.1088/0004-637x/806/1/128},
   number={1},
   journal={The Astrophysical Journal},
   publisher={American Astronomical Society},
   author={Rosa, G. De and Peterson, B. M. and Ely, J. and Kriss, G. A. and Crenshaw, D. M. and Horne, Keith and Korista, K. T. and Netzer, H. and Pogge, R. W. and Arévalo, P. and Barth, A. J. and Bentz, M. C. and Brandt, W. N. and Breeveld, A. A. and Brewer, B. J. and Dalla Bontà, E. and Lorenzo-Cáceres, A. De and Denney, K. D. and Dietrich, M. and Edelson, R. and Evans, P. A. and Fausnaugh, M. M. and Gehrels, N. and Gelbord, J. M. and Goad, M. R. and Grier, C. J. and Grupe, D. and Hall, P. B. and Kaastra, J. and Kelly, B. C. and Kennea, J. A. and Kochanek, C. S. and Lira, P. and Mathur, S. and McHardy, I. M. and Nousek, J. A. and Pancoast, A. and Papadakis, I. and Pei, L. and Schimoia, J. S. and Siegel, M. and Starkey, D. and Treu, T. and Uttley, P. and Vaughan, S. and Vestergaard, M. and Villforth, C. and Yan, H. and Young, S. and Zu, Y.},
   year={2015},
   month=jun, pages={128} }

@article{HBETA_Bentz_2013,
   title={THE LOW-LUMINOSITY END OF THE RADIUS-LUMINOSITY RELATIONSHIP FOR ACTIVE GALACTIC NUCLEI},
   volume={767},
   ISSN={1538-4357},
   url={http://dx.doi.org/10.1088/0004-637X/767/2/149},
   DOI={10.1088/0004-637x/767/2/149},
   number={2},
   journal={The Astrophysical Journal},
   publisher={American Astronomical Society},
   author={Bentz, Misty C. and Denney, Kelly D. and Grier, Catherine J. and Barth, Aaron J. and Peterson, Bradley M. and Vestergaard, Marianne and Bennert, Vardha N. and Canalizo, Gabriela and De Rosa, Gisella and Filippenko, Alexei V. and Gates, Elinor L. and Greene, Jenny E. and Li, Weidong and Malkan, Matthew A. and Pogge, Richard W. and Stern, Daniel and Treu, Tommaso and Woo, Jong-Hak},
   year={2013},
   month=apr, pages={149} }

@article{Du_2015,
   title={SUPERMASSIVE BLACK HOLES WITH HIGH ACCRETION RATES IN ACTIVE GALACTIC NUCLEI. IV. H$\beta$TIME LAGS AND IMPLICATIONS FOR SUPER-EDDINGTON ACCRETION},
   volume={806},
   ISSN={1538-4357},
   url={http://dx.doi.org/10.1088/0004-637X/806/1/22},
   DOI={10.1088/0004-637x/806/1/22},
   number={1},
   journal={The Astrophysical Journal},
   publisher={American Astronomical Society},
   author={Du, Pu and Hu, Chen and Lu, Kai-Xing and Huang, Ying-Ke and Cheng, Cheng and Qiu, Jie and Li, Yan-Rong and Zhang, Yang-Wei and Fan, Xu-Liang and Bai, Jin-Ming and Bian, Wei-Hao and Yuan, Ye-Fei and Kaspi, Shai and Ho, Luis C. and Netzer, Hagai and Wang, Jian-Min},
   year={2015},
   month=jun, pages={22} }

@article{HBETA_Du_2016,
   title={SUPERMASSIVE BLACK HOLES WITH HIGH ACCRETION RATES IN ACTIVE GALACTIC NUCLEI. V. A NEW SIZE–LUMINOSITY SCALING RELATION FOR THE BROAD-LINE REGION},
   volume={825},
   ISSN={1538-4357},
   url={http://dx.doi.org/10.3847/0004-637X/825/2/126},
   DOI={10.3847/0004-637x/825/2/126},
   number={2},
   journal={The Astrophysical Journal},
   publisher={American Astronomical Society},
   author={Du, Pu and Lu, Kai-Xing and Zhang, Zhi-Xiang and Huang, Ying-Ke and Wang, Kai and Hu, Chen and Qiu, Jie and Li, Yan-Rong and Fan, Xu-Liang and Fang, Xiang-Er and Bai, Jin-Ming and Bian, Wei-Hao and Yuan, Ye-Fei and Ho, Luis C. and Wang, Jian-Min},
   year={2016},
   month=jul, pages={126} }

@article{HBETA_Du_2018,
   title={Supermassive Black Holes with High Accretion Rates in Active Galactic Nuclei. IX. 10 New Observations of Reverberation Mapping and Shortened H$\beta$-Lags},
   volume={856},
   ISSN={1538-4357},
   url={http://dx.doi.org/10.3847/1538-4357/aaae6b},
   DOI={10.3847/1538-4357/aaae6b},
   number={1},
   journal={The Astrophysical Journal},
   publisher={American Astronomical Society},
   author={Du, Pu and Zhang, Zhi-Xiang and Wang, Kai and Huang, Ying-Ke and Zhang, Yue and Lu, Kai-Xing and Hu, Chen and Li, Yan-Rong and Bai, Jin-Ming and Bian, Wei-Hao and Yuan, Ye-Fei and Ho, Luis C. and Wang, Jian-Min},
   year={2018},
   month=mar, pages={6} }

@ARTICLE{HBETA_Hu_2021,
       author = {{Hu}, Chen and {Li}, Sha-Sha and {Yang}, Sen and {Yang}, Zi-Xu and {Guo}, Wei-Jian and {Bao}, Dong-Wei and {Jiang}, Bo-Wei and {Du}, Pu and {Li}, Yan-Rong and {Xiao}, Ming and {Songsheng}, Yu-Yang and {Yu}, Zhe and {Bai}, Jin-Ming and {Ho}, Luis C. and {Brotherton}, Michael S. and {Aceituno}, Jes{\'u}s and {Winkler}, Hartmut and {Wang}, Jian-Min and {Seambh Collaboration}},
        title = "{Supermassive Black Holes with High Accretion Rates in Active Galactic Nuclei. XII. Reverberation Mapping Results for 15 PG Quasars from a Long-duration High-cadence Campaign}",
      journal = {\apjs},
         year = 2021,
        month = mar,
       volume = {253},
       number = {1},
          eid = {20},
        pages = {20},
          doi = {10.3847/1538-4365/abd774},
       adsurl = {https://ui.adsabs.harvard.edu/abs/2021ApJS..253...20H}
}

@article{HBETA_U_2022,
   title={The Lick AGN Monitoring Project 2016: Velocity-resolved H$\beta$ Lags in Luminous Seyfert Galaxies},
   volume={925},
   ISSN={1538-4357},
   url={http://dx.doi.org/10.3847/1538-4357/ac3d26},
   DOI={10.3847/1538-4357/ac3d26},
   number={1},
   journal={The Astrophysical Journal},
   publisher={American Astronomical Society},
   author={U, Vivian and Barth, Aaron J. and Vogler, H. Alexander and Guo, Hengxiao and Treu, Tommaso and Bennert, Vardha N. and Canalizo, Gabriela and Filippenko, Alexei V. and Gates, Elinor and Hamann, Frederick and Joner, Michael D. and Malkan, Matthew A. and Pancoast, Anna and Williams, Peter R. and Woo, Jong-Hak and Abolfathi, Bela and Abramson, L. E. and Armen, Stephen F. and Bae, Hyun-Jin and Bohn, Thomas and Boizelle, Benjamin D. and Bostroem, Azalee and Brandel, Andrew and Brink, Thomas G. and Channa, Sanyum and Cooper, M. C. and Cosens, Maren and Donohue, Edward and Fillingham, Sean P. and González-Buitrago, Diego and Halevi, Goni and Halle, Andrew and Hood, Carol E. and Horne, Keith and Horst, J. Chuck and Kouchkovsky, Maxime de and Kuhn, Benjamin and Kumar, Sahana and Leonard, Douglas C. and Loveland, Donald and Manzano-King, Christina and McHardy, Ian and Michel, Raúl and Olaes, Melanie Kae B. and Park, Daeseong and Park, Songyoun and Pei, Liuyi and Ross, Timothy W. and Runco, Jordan N. and Samuel, Jenna and Sánchez, Javier and Scott, Bryan and Sexton, Remington O. and Shin, Jaejin and Shivvers, Isaac and Spencer, Chance L. and Stahl, Benjamin E. and Stegman, Samantha and Stomberg, Isak and Valenti, Stefano and Villafaña, L. and Walsh, Jonelle L. and Yuk, Heechan and Zheng, WeiKang},
   year={2022},
   month=jan, pages={52} }

@ARTICLE{HBETA_Bentz_2023,
       author = {{Bentz}, Misty C. and {Onken}, Christopher A. and {Street}, Rachel and {Valluri}, Monica},
        title = "{Reverberation Mapping of IC 4329A}",
      journal = {\apj},
         year = 2023,
        month = feb,
       volume = {944},
       number = {1},
          eid = {29},
        pages = {29},
          doi = {10.3847/1538-4357/acab62},
archivePrefix = {arXiv},
       eprint = {2212.05954},
 primaryClass = {astro-ph.GA},
       adsurl = {https://ui.adsabs.harvard.edu/abs/2023ApJ...944...29B}
}

@article{HBETA_Martinez-Aldama_2019,
   title={Can Reverberation-measured Quasars Be Used for Cosmology?},
   volume={883},
   ISSN={1538-4357},
   url={http://dx.doi.org/10.3847/1538-4357/ab3728},
   DOI={10.3847/1538-4357/ab3728},
   number={2},
   journal={The Astrophysical Journal},
   publisher={American Astronomical Society},
   author={Martínez-Aldama, Mary Loli and Czerny, Bożena and Kawka, Damian and Karas, Vladimir and Panda, Swayamtrupta and Zajaček, Michal and Życki, Piotr T.},
   year={2019},
   month=oct, pages={170} }

@article{HBETA_Bentz_2009,
   title={THE RADIUS-LUMINOSITY RELATIONSHIP FOR ACTIVE GALACTIC NUCLEI: THE EFFECT OF HOST-GALAXY STARLIGHT ON LUMINOSITY MEASUREMENTS. II. THE FULL SAMPLE OF REVERBERATION-MAPPED AGNs},
   volume={697},
   ISSN={1538-4357},
   url={http://dx.doi.org/10.1088/0004-637X/697/1/160},
   DOI={10.1088/0004-637x/697/1/160},
   number={1},
   journal={The Astrophysical Journal},
   publisher={American Astronomical Society},
   author={Bentz, Misty C. and Peterson, Bradley M. and Netzer, Hagai and Pogge, Richard W. and Vestergaard, Marianne},
   year={2009},
   month=apr, pages={160–181} }

@article{HBETA_Bentz_2014,
   title={THE MASS OF THE CENTRAL BLACK HOLE IN THE NEARBY SEYFERT GALAXY NGC 5273},
   volume={796},
   ISSN={1538-4357},
   url={http://dx.doi.org/10.1088/0004-637X/796/1/8},
   DOI={10.1088/0004-637x/796/1/8},
   number={1},
   journal={The Astrophysical Journal},
   publisher={American Astronomical Society},
   author={Bentz, Misty C. and Horenstein, Daniel and Bazhaw, Craig and Manne-Nicholas, Emily R. and Ou-Yang, Benjamin J. and Anderson, Matthew and Jones, Jeremy and Norris, Ryan P. and Parks, J. Robert and Saylor, Dicy and Teems, Katherine G. and Turner, Clay},
   year={2014},
   month=oct, pages={8} }

@article{HBETA_Pei_2014,
   title={REVERBERATION MAPPING OF THEKEPLERFIELD AGN KA1858+4850},
   volume={795},
   ISSN={1538-4357},
   url={http://dx.doi.org/10.1088/0004-637X/795/1/38},
   DOI={10.1088/0004-637x/795/1/38},
   number={1},
   journal={The Astrophysical Journal},
   publisher={American Astronomical Society},
   author={Pei, Liuyi and Barth, Aaron J. and Aldering, Greg S. and Briley, Michael M. and Carroll, Carla J. and Carson, Daniel J. and Cenko, S. Bradley and Clubb, Kelsey I. and Cohen, Daniel P. and Cucchiara, Antonino and Desjardins, Tyler D. and Edelson, Rick and Fang, Jerome J. and Fedrow, Joseph M. and Filippenko, Alexei V. and Fox, Ori D. and Furniss, Amy and Gates, Elinor L. and Gregg, Michael and Gustafson, Scott and Horst, J. Chuck and Joner, Michael D. and Kelly, Patrick L. and Lacy, Mark and Laney, C. David and Leonard, Douglas C. and Li, Weidong and Malkan, Matthew A. and Margon, Bruce and Neeleman, Marcel and Nguyen, My L. and Prochaska, J. Xavier and Ross, Nathaniel R. and Sand, David J. and Searcy, Kinchen J. and Shivvers, Isaac S. and Silverman, Jeffrey M. and Smith, Graeme H. and Suzuki, Nao and Smith, Krista Lynne and Tytler, David and Werk, Jessica K. and Worseck, Gábor},
   year={2014},
   month=oct, pages={38} }

@article{HBETA_Lu_2016,
   title={REVERBERATION MAPPING OF THE BROAD-LINE REGION IN NGC 5548: EVIDENCE FOR RADIATION PRESSURE?},
   volume={827},
   ISSN={1538-4357},
   url={http://dx.doi.org/10.3847/0004-637X/827/2/118},
   DOI={10.3847/0004-637x/827/2/118},
   number={2},
   journal={The Astrophysical Journal},
   publisher={American Astronomical Society},
   author={Lu, Kai-Xing and Du, Pu and Hu, Chen and Li, Yan-Rong and Zhang, Zhi-Xiang and Wang, Kai and Huang, Ying-Ke and Bi, Shao-Lan and Bai, Jin-Ming and Ho, Luis C. and Wang, Jian-Min},
   year={2016},
   month=aug, pages={118} }

@article{HBETA_Bentz_2016a,
   title={A REVERBERATION-BASED BLACK HOLE MASS FOR MCG-06-30-15},
   volume={830},
   ISSN={1538-4357},
   url={http://dx.doi.org/10.3847/0004-637X/830/2/136},
   DOI={10.3847/0004-637x/830/2/136},
   number={2},
   journal={The Astrophysical Journal},
   publisher={American Astronomical Society},
   author={Bentz, Misty C. and Cackett, Edward M. and Crenshaw, D. Michael and Horne, Keith and Street, Rachel and Ou-Yang, Benjamin},
   year={2016},
   month=oct, pages={136} }

@article{HBETA_Bentz_2016b,
   title={A LOW-MASS BLACK HOLE IN THE NEARBY SEYFERT GALAXY UGC 06728},
   volume={831},
   ISSN={1538-4357},
   url={http://dx.doi.org/10.3847/0004-637X/831/1/2},
   DOI={10.3847/0004-637x/831/1/2},
   number={1},
   journal={The Astrophysical Journal},
   publisher={American Astronomical Society},
   author={Bentz, Misty C. and Batiste, Merida and Seals, James and Garcia, Karen and Naray, Rachel Kuzio de and Peters, Wesley and Anderson, Matthew D. and Jones, Jeremy and Lester, Kathryn and Machuca, Camilo and Parks, J. Robert and Pope, Crystal L. and Revalski, Mitchell and Roberts, Caroline A. and Saylor, Dicy and Sevrinsky, R. Andrew and Turner, Clay},
   year={2016},
   month=oct, pages={2} }

@article{HBETA_Fausnaugh_2017,
   title={Reverberation Mapping of Optical Emission Lines in Five Active Galaxies},
   volume={840},
   ISSN={1538-4357},
   url={http://dx.doi.org/10.3847/1538-4357/aa6d52},
   DOI={10.3847/1538-4357/aa6d52},
   number={2},
   journal={The Astrophysical Journal},
   publisher={American Astronomical Society},
   author={Fausnaugh, M. M. and Grier, C. J. and Bentz, M. C. and Denney, K. D. and Rosa, G. De and Peterson, B. M. and Kochanek, C. S. and Pogge, R. W. and Adams, S. M. and Barth, A. J. and Beatty, Thomas G. and Bhattacharjee, A. and Borman, G. A. and Boroson, T. A. and Bottorff, M. C. and Brown, Jacob E. and Brown, Jonathan S. and Brotherton, M. S. and Coker, C. T. and Crawford, S. M. and Croxall, K. V. and Eftekharzadeh, Sarah and Eracleous, Michael and Joner, M. D. and Henderson, C. B. and Holoien, T. W.-S. and Horne, Keith and Hutchison, T. and Kaspi, Shai and Kim, S. and King, Anthea L. and Li, Miao and Lochhaas, Cassandra and Ma, Zhiyuan and MacInnis, F. and Manne-Nicholas, E. R. and Mason, M. and Montuori, Carmen and Mosquera, Ana and Mudd, Dale and Musso, R. and Nazarov, S. V. and Nguyen, M. L. and Okhmat, D. N. and Onken, Christopher A. and Ou-Yang, B. and Pancoast, A. and Pei, L. and Penny, Matthew T. and Poleski, Radosław and Rafter, Stephen and Romero-Colmenero, E. and Runnoe, Jessie and Sand, David J. and Schimoia, Jaderson S. and Sergeev, S. G. and Shappee, B. J. and Simonian, Gregory V. and Somers, Garrett and Spencer, M. and Starkey, D. A. and Stevens, Daniel J. and Tayar, Jamie and Treu, T. and Valenti, Stefano and Saders, J. Van and Villanueva Jr., S. and Villforth, C. and Weiss, Yaniv and Winkler, H. and Zhu, W.},
   year={2017},
   month=may, pages={97} }

@article{HBETA_Zhang_2019,
   title={Kinematics of the Broad-line Region of 3C 273 from a 10 yr Reverberation Mapping Campaign},
   volume={876},
   ISSN={1538-4357},
   url={http://dx.doi.org/10.3847/1538-4357/ab1099},
   DOI={10.3847/1538-4357/ab1099},
   number={1},
   journal={The Astrophysical Journal},
   publisher={American Astronomical Society},
   author={Zhang, Zhi-Xiang and Du, Pu and Smith, Paul S. and Zhao, Yulin and Hu, Chen and Xiao, Ming and Li, Yan-Rong and Huang, Ying-Ke and Wang, Kai and Bai, Jin-Ming and Ho, Luis C. and Wang, Jian-Min},
   year={2019},
   month=may, pages={49} }

@article{HBETA_Rakshit_2019,
   title={The Seoul National University AGN Monitoring Project. II. BLR Size and Black Hole Mass of Two AGNs},
   volume={886},
   ISSN={1538-4357},
   url={http://dx.doi.org/10.3847/1538-4357/ab49fd},
   DOI={10.3847/1538-4357/ab49fd},
   number={2},
   journal={The Astrophysical Journal},
   publisher={American Astronomical Society},
   author={Rakshit, Suvendu and Woo, Jong-Hak and Gallo, Elena and Hodges-Kluck, Edmund and Shin, Jaejin and Jeon, Yiseul and Bae, Hyun-Jin and Baldassare, Vivienne and Cho, Hojin and Cho, Wanjin and Foord, Adi and Kang, Daeun and Kang, Wonseok and Karouzos, Marios and Kim, Minjin and Kim, Taewoo and Le, Huynh Anh N. and Park, Daeseong and Park, Songyoun and Son, Donghoon and Sung, Hyun-il and Bennert, Vardha N. and Malkan, Matthew A.},
   year={2019},
   month=nov, pages={93} }

@article{HBETA_Li_2021,
   title={Reverberation Mapping of Two Luminous Quasars: The Broad-line Region Structure and Black Hole Mass},
   volume={920},
   ISSN={1538-4357},
   url={http://dx.doi.org/10.3847/1538-4357/ac116e},
   DOI={10.3847/1538-4357/ac116e},
   number={1},
   journal={The Astrophysical Journal},
   publisher={American Astronomical Society},
   author={Li, Sha-Sha and Yang, Sen and Yang, Zi-Xu and Chen, Yong-Jie and Songsheng, Yu-Yang and Liu, He-Zhen and Du, Pu and Luo, Bin and Yu, Zhe and Hu, Chen and Jiang, Bo-Wei and Bao, Dong-Wei and Guo, Wei-Jian and Zhang, Zhi-Xiang and Li, Yan-Rong and Xiao, Ming and Lu, Kai-Xing and Ho, Luis C. and Bai, Jin-Ming and Bian, Wei-Hao and Aceituno, Jesús and Minezaki, Takeo and Horne, Keith and Kokubo, Mitsuru and Wang, Jian-Min},
   year={2021},
   month=oct, pages={9} }

@ARTICLE{HBETA_Barth_2013,
       author = {{Barth}, Aaron J. and {Pancoast}, Anna and {Bennert}, Vardha N. and {Brewer}, Brendon J. and {Canalizo}, Gabriela and {Filippenko}, Alexei V. and {Gates}, Elinor L. and {Greene}, Jenny E. and {Li}, Weidong and {Malkan}, Matthew A. and {Sand}, David J. and {Stern}, Daniel and {Treu}, Tommaso and {Woo}, Jong-Hak and {Assef}, Roberto J. and {Bae}, Hyun-Jin and {Buehler}, Tabitha and {Cenko}, S. Bradley and {Clubb}, Kelsey I. and {Cooper}, Michael C. and {Diamond-Stanic}, Aleksandar M. and {H{\"o}nig}, Sebastian F. and {Joner}, Michael D. and {Laney}, C. David and {Lazarova}, Mariana S. and {Nierenberg}, A.~M. and {Silverman}, Jeffrey M. and {Tollerud}, Erik J. and {Walsh}, Jonelle L.},
        title = "{The Lick AGN Monitoring Project 2011: Fe II Reverberation from the Outer Broad-line Region}",
      journal = {\apj},
         year = 2013,
        month = jun,
       volume = {769},
       number = {2},
          eid = {128},
        pages = {128},
          doi = {10.1088/0004-637X/769/2/128},
archivePrefix = {arXiv},
       eprint = {1304.4643},
 primaryClass = {astro-ph.CO},
       adsurl = {https://ui.adsabs.harvard.edu/abs/2013ApJ...769..128B}
}

@article{MgII_Metzroth_2006,
   title={The Mass of the Central Black Hole in the Seyfert Galaxy NGC 4151},
   volume={647},
   ISSN={1538-4357},
   url={http://dx.doi.org/10.1086/505525},
   DOI={10.1086/505525},
   number={2},
   journal={The Astrophysical Journal},
   publisher={American Astronomical Society},
   author={Metzroth, Kyle G. and Onken, Christopher A. and Peterson, Bradley M.},
   year={2006},
   month=aug, pages={901–909} }

@article{MgII_Lira_2018,
   title={Reverberation Mapping of Luminous Quasars at High z},
   volume={865},
   ISSN={1538-4357},
   url={http://dx.doi.org/10.3847/1538-4357/aada45},
   DOI={10.3847/1538-4357/aada45},
   number={1},
   journal={The Astrophysical Journal},
   publisher={American Astronomical Society},
   author={Lira, Paulina and Kaspi, Shai and Netzer, Hagai and Botti, Ismael and Morrell, Nidia and Mejía-Restrepo, Julián and Sánchez-Sáez, Paula and Martínez-Palomera, Jorge and López, Paula},
   year={2018},
   month=sep, pages={56} }

@article{MgII_Czerny_2019,
   title={Time Delay Measurement of Mg ii Line in CTS C30.10 with SALT},
   volume={880},
   ISSN={1538-4357},
   url={http://dx.doi.org/10.3847/1538-4357/ab2913},
   DOI={10.3847/1538-4357/ab2913},
   number={1},
   journal={The Astrophysical Journal},
   publisher={American Astronomical Society},
   author={Czerny, Bożena and Olejak, Aleksandra and Rałowski, Mateusz and Kozłowski, Szymon and Aldama, Mary Loli Martinez and Zajacek, Michal and Pych, Wojtek and Hryniewicz, Krzysztof and Pietrzyński, Grzegorz and Figaredo, Catalina Sobrino and Haas, Martin and Średzińska, Justyna and Krupa, Magdalena and Kurcz, Agnieszka and Udalski, Andrzej and Gorski, Marek and Karas, Vladimir and Panda, Swayamtrupta and Sniegowska, Marzena and Naddaf, Mohammad-Hassan and Bilicki, Maciej and Sarna, Marek},
   year={2019},
   month=jul, pages={46} }

@article{MgII_Zajacek_2020,
   title={Time-delay Measurement of Mg ii Broad-line Response for the Highly Accreting Quasar HE 0413-4031: Implications for the Mg ii–based Radius–Luminosity Relation},
   volume={896},
   ISSN={1538-4357},
   url={http://dx.doi.org/10.3847/1538-4357/ab94ae},
   DOI={10.3847/1538-4357/ab94ae},
   number={2},
   journal={The Astrophysical Journal},
   publisher={American Astronomical Society},
   author={Zajaček, Michal and Czerny, Bożena and Martinez–Aldama, Mary Loli and Rałowski, Mateusz and Olejak, Aleksandra and Panda, Swayamtrupta and Hryniewicz, Krzysztof and Śniegowska, Marzena and Naddaf, Mohammad-Hassan and Pych, Wojtek and Pietrzyński, Grzegorz and Figaredo, C. Sobrino and Haas, Martin and Średzińska, Justyna and Krupa, Magdalena and Kurcz, Agnieszka and Udalski, Andrzej and Gorski, Marek and Sarna, Marek},
   year={2020},
   month=jun, pages={146} }

@article{MgII_Zajacek_2021,
   title={Time Delay of Mg ii Emission Response for the Luminous Quasar HE 0435-4312: toward Application of the High-accretor Radius–Luminosity Relation in Cosmology},
   volume={912},
   ISSN={1538-4357},
   url={http://dx.doi.org/10.3847/1538-4357/abe9b2},
   DOI={10.3847/1538-4357/abe9b2},
   number={1},
   journal={The Astrophysical Journal},
   publisher={American Astronomical Society},
   author={Zajaček, Michal and Czerny, Bożena and Martinez–Aldama, Mary Loli and Rałowski, Mateusz and Olejak, Aleksandra and Przyłuski, Robert and Panda, Swayamtrupta and Hryniewicz, Krzysztof and Śniegowska, Marzena and Naddaf, Mohammad-Hassan and Prince, Raj and Pych, Wojtek and Pietrzyński, Grzegorz and Sobrino Figaredo, Catalina and Haas, Martin and Średzińska, Justyna and Krupa, Magdalena and Kurcz, Agnieszka and Udalski, Andrzej and Karas, Vladimír and Sarna, Marek and Worters, Hannah L. and Sefako, Ramotholo R. and Genade, Anja},
   year={2021},
   month=apr, pages={10} }

@article{MgII_Shen_2016,
   title={THE SLOAN DIGITAL SKY SURVEY REVERBERATION MAPPING PROJECT: FIRST BROAD-LINE H$\beta$ AND Mg ii LAGS AT z>=0.3 FROM SIX-MONTH SPECTROSCOPY},
   volume={818},
   ISSN={1538-4357},
   url={http://dx.doi.org/10.3847/0004-637X/818/1/30},
   DOI={10.3847/0004-637x/818/1/30},
   number={1},
   journal={The Astrophysical Journal},
   publisher={American Astronomical Society},
   author={Shen, Yue and Horne, Keith and Grier, C. J. and Peterson, Bradley M. and Denney, Kelly D. and Trump, Jonathan R. and Sun, Mouyuan and Brandt, W. N. and Kochanek, Christopher S. and Dawson, Kyle S. and Green, Paul J. and Greene, Jenny E. and Hall, Patrick B. and Ho, Luis C. and Jiang, Linhua and Kinemuchi, Karen and McGreer, Ian D. and Petitjean, Patrick and Richards, Gordon T. and Schneider, Donald P. and Strauss, Michael A. and Tao, Charling and Wood-Vasey, W. M. and Zu, Ying and Pan, Kaike and Bizyaev, Dmitry and Ge, Jian and Oravetz, Daniel and Simmons, Audrey},
   year={2016},
   month=feb, pages={30} }

@article{MgII_Shen_2019,
   title={The Sloan Digital Sky Survey Reverberation Mapping Project: Sample Characterization},
   volume={241},
   ISSN={1538-4365},
   url={http://dx.doi.org/10.3847/1538-4365/ab074f},
   DOI={10.3847/1538-4365/ab074f},
   number={2},
   journal={The Astrophysical Journal Supplement Series},
   publisher={American Astronomical Society},
   author={Shen, Yue and Hall, Patrick B. and Horne, Keith and Zhu, Guangtun and McGreer, Ian and Simm, Torben and Trump, Jonathan R. and Kinemuchi, Karen and Brandt, W. N. and Green, Paul J. and Grier, C. J. and Guo, Hengxiao and Ho, Luis C. and Homayouni, Yasaman and Jiang, Linhua and Li, Jennifer I-Hsiu and Morganson, Eric and Petitjean, Patrick and Richards, Gordon T. and Schneider, Donald P. and Starkey, D. A. and Wang, Shu and Chambers, Ken and Kaiser, Nick and Kudritzki, Rolf-Peter and Magnier, Eugene and Waters, Christopher},
   year={2019},
   month=apr, pages={34} }

@ARTICLE{OzDES-Yu_2019,
       author = {{Yu}, Z. and {Kochanek}, C.~S. and {Peterson}, B.~M. and {Zu}, Y. and {Brandt}, W.~N. and {Cackett}, E.~M. and {Fausnaugh}, M.~M. and {McHardy}, I.~M.},
        title = "{On reverberation mapping lag uncertainties}",
      journal = {\mnras},
         year = 2020,
        month = feb,
       volume = {491},
       number = {4},
        pages = {6045-6064},
          doi = {10.1093/mnras/stz3464},
archivePrefix = {arXiv},
       eprint = {1909.03072},
 primaryClass = {astro-ph.GA},
       adsurl = {https://ui.adsabs.harvard.edu/abs/2020MNRAS.491.6045Y}
}

@article{OzDES-Hoormann_2019,
   title={CIV black hole mass measurements with the Australian Dark Energy Survey (OzDES)},
   volume={487},
   ISSN={1365-2966},
   url={http://dx.doi.org/10.1093/mnras/stz1539},
   DOI={10.1093/mnras/stz1539},
   number={3},
   journal={Monthly Notices of the Royal Astronomical Society},
   publisher={Oxford University Press (OUP)},
   author={Hoormann, J K and Martini, P and Davis, T M and King, A and Lidman, C and Mudd, D and Sharp, R and Sommer, N E and Tucker, B E and Yu, Z and Allam, S and Asorey, J and Avila, S and Banerji, M and Brooks, D and Buckley-Geer, E and Burke, D L and Calcino, J and Carnero-Rosell, A and Carollo, D and Carrasco-Kind, M and Carretero, J and Castander, F J and Childress, M and De-Vicente, J and Desai, S and Diehl, H T and Doel, P and Flaugher, B and Fosalba, P and Frieman, J and García-Bellido, J and Gerdes, D W and Gruen, D and Gutierrez, G and Hartley, W G and Hinton, S R and Hollowood, D L and Honscheid, K and Hoyle, B and James, D J and Krause, E and Kuehn, K and Kuropatkin, N and Lewis, G F and Lima, M and Macaulay, E and Maia, M A G and Menanteau, F and Miller, C J and Miquel, R and Möller, A and Plazas, A A and Romer, A K and Roodman, A and Sanchez, E and Scarpine, V and Schubnell, M and Serrano, S and Sevilla-Noarbe, I and Smith, M and Smith, R C and Soares-Santos, M and Sobreira, F and Suchyta, E and Swann, E and Swanson, M E C and Tarle, G and Uddin, S A},
   year={2019},
   month=jun, pages={3650–3663} }

@article{OzDES-Penton_2021,
   title={OzDES reverberation mapping program: Lag recovery reliability for 6-yr C<scp>iv</scp> analysis},
   volume={509},
   ISSN={1365-2966},
   url={http://dx.doi.org/10.1093/mnras/stab3027},
   DOI={10.1093/mnras/stab3027},
   number={3},
   journal={Monthly Notices of the Royal Astronomical Society},
   publisher={Oxford University Press (OUP)},
   author={Penton, A and Malik, U and Davis, T M and Martini, P and Yu, Z and Sharp, R and Lidman, C and Tucker, B E and Hoormann, J K and Aguena, M and Allam, S and Annis, J and Asorey, J and Bacon, D and Bertin, E and Bhargava, S and Brooks, D and Calcino, J and Carnero Rosell, A and Carollo, D and Carrasco Kind, M and Carretero, J and Costanzi, M and da Costa, L N and Pereira, M E S and De Vicente, J and Diehl, H T and Eifler, T F and Everett, S and Ferrero, I and Fosalba, P and Frieman, J and García-Bellido, J and Gaztanaga, E and Gerdes, D W and Gruen, D and Gruendl, R A and Gschwend, J and Gutierrez, G and Hinton, S R and Hollowood, D L and Honscheid, K and James, D J and Kim, A G and Kuehn, K and Kuropatkin, N and Maia, M A G and Marshall, J L and Menanteau, F and Miquel, R and Morgan, R and Möller, A and Palmese, A and Paz-Chinchón, F and Plazas, A A and Romer, A K and Sanchez, E and Scarpine, V and Scolnic, D and Serrano, S and Smith, M and Suchyta, E and Swanson, M E C and Tarle, G and To, C and Uddin, S A and Varga, T N and Wester, W and Wilkinson, R D and Lewis, G},
   year={2021},
   month=oct, pages={4008–4023} }

@ARTICLE{OzDES-Penton_2025,
       author = {{Penton}, A. and {McDougall}, H. and {Davis}, T.~M. and {Yu}, Z. and {Malik}, U. and {Martini}, P. and {Tucker}, B.~E. and {Lidman}, C. and {Lewis}, G.~F. and {Sharp}, R. and {Aguena}, M. and {Allam}, S. and {Andrade-Oliveira}, F. and {Asorey}, J. and {Bacon}, D. and {Bocquet}, S. and {Brooks}, D. and {Camilleri}, R. and {Carnero Rosell}, A. and {Carollo}, D. and {Carr}, A. and {Carretero}, J. and {Cheng}, T.~Y. and {da Costa}, L.~N. and {da Silva Pereira}, M.~E. and {De Vicente}, J. and {Desai}, S. and {Everett}, S. and {Garc{\'\i}a-Bellido}, J. and {Glazebrook}, K. and {Gruen}, D. and {Gutierrez}, G. and {Hinton}, S.~R. and {Hollowood}, D.~L. and {Honscheid}, K. and {Kuehn}, K. and {Lahav}, O. and {Lee}, S. and {March}, M. and {Marshall}, J.~L. and {Mena-Fern{\'a}ndez}, J. and {Miquel}, R. and {Myles}, J. and {Ogando}, R.~L.~C. and {Plazas Malag{\'o}n}, A.~A. and {Porredon}, A. and {Rodriguez-Monroy}, M. and {Romer}, A.~K. and {Sanchez}, E. and {Sanchez Cid}, D. and {Smith}, M. and {Suchyta}, E. and {Swanson}, M.~E.~C. and {Vikram}, V. and {Weaverdyck}, N.},
        title = "{OzDES Reverberation Mapping Program: CIV lags from six years of data}",
      journal = {arXiv e-prints},
         year = 2026,
        month = dec,
          eid = {arXiv:2512.01260},
        pages = {arXiv:2512.01260},
          doi = {10.48550/arXiv.2512.01260},
archivePrefix = {arXiv},
       eprint = {2512.01260},
 primaryClass = {astro-ph.GA},
       adsurl = {https://ui.adsabs.harvard.edu/abs/2025arXiv251201260P}
}

@article{OzDES-Yu_2021,
   title={OzDES Reverberation Mapping Programme: the first Mg <scp>ii</scp> lags from 5 yr of monitoring},
   volume={507},
   ISSN={1365-2966},
   url={http://dx.doi.org/10.1093/mnras/stab2244},
   DOI={10.1093/mnras/stab2244},
   number={3},
   journal={Monthly Notices of the Royal Astronomical Society},
   publisher={Oxford University Press (OUP)},
   author={{Yu}, Zhefu and Martini, Paul and Penton, A and Davis, T M and Malik, U and Lidman, C and Tucker, B E and Sharp, R and Kochanek, C S and Peterson, B M and Aguena, M and Allam, S and Andrade-Oliveira, F and Annis, J and Asorey, J and Bertin, E and Brooks, D and Burke, D L and Calcino, J and Carnero Rosell, A and Carollo, D and Carrasco Kind, M and Costanzi, M and da Costa, L N and da Silva Pereira, M E S and Diehl, H T and Everett, S and Ferrero, I and Flaugher, B and Frieman, J and García-Bellido, J and Gaztanaga, E and Gerdes, D W and Gruen, D and Gruendl, R A and Gschwend, J and Gutierrez, G and Hinton, S R and Hollowood, D L and James, D J and Kim, A G and Kron, R and Kuehn, K and Kuropatkin, N and Lewis, G F and Maia, M A G and March, M and Marshall, J L and Menanteau, F and Miquel, R and Morgan, R and Möller, A and Palmese, A and Paz-Chinchón, F and Plazas, A A and Sanchez, E and Scarpine, V and Serrano, S and Sevilla-Noarbe, I and Smith, M and Soares-Santos, M and Suchyta, E and Tarle, G and Thomas, D and To, C and Tucker, D L},
   year={2021},
   month=aug, pages={3771–3788} }

@article{OzDES-Yu_2023,
   title={OzDES Reverberation Mapping Programme: Mg <scp>ii</scp> lags and R-L relation},
   volume={522},
   ISSN={1365-2966},
   url={http://dx.doi.org/10.1093/mnras/stad1224},
   DOI={10.1093/mnras/stad1224},
   number={3},
   journal={Monthly Notices of the Royal Astronomical Society},
   publisher={Oxford University Press (OUP)},
   author={{Yu}, Zhefu and Martini, Paul and Penton, A and Davis, T M and Kochanek, C S and Lewis, G F and Lidman, C and Malik, U and Sharp, R and Tucker, B E and Aguena, M and Annis, J and Bertin, E and Bocquet, S and Brooks, D and Carnero Rosell, A and Carollo, D and Carrasco Kind, M and Carretero, J and Costanzi, M and da Costa, L N and Pereira, M E S and De Vicente, J and Diehl, H T and Doel, P and Everett, S and Ferrero, I and García-Bellido, J and Gatti, M and Gerdes, D W and Gruen, D and Gruendl, R A and Gschwend, J and Gutierrez, G and Hinton, S R and Hollowood, D L and Honscheid, K and James, D J and Kuehn, K and Mena-Fernández, J and Menanteau, F and Miquel, R and Nichol, B and Paz-Chinchón, F and Pieres, A and Plazas Malagón, A A and Raveri, M and Romer, A K and Sanchez, E and Scarpine, V and Sevilla-Noarbe, I and Smith, M and Suchyta, E and Swanson, M E C and Tarle, G and Vincenzi, M and Walker, A R and Weaverdyck, N},
   year={2023},
   month=apr, pages={4132–4147} }

@ARTICLE{OzDES-Malik_2023,
       author = {{Malik}, U. and {Sharp}, R. and {Penton}, A. and {Yu}, Zhefu and {Martini}, P. and {Lidman}, C. and {Tucker}, B.~E. and {Davis}, T.~M. and {Lewis}, G.~F. and {Aguena}, M. and {Allam}, S. and {Alves}, O. and {Andrade-Oliveira}, F. and {Asorey}, J. and {Bacon}, D. and {Bertin}, E. and {Bocquet}, S. and {Brooks}, D. and {Burke}, D.~L. and {Carnero Rosell}, A. and {Carollo}, D. and {Carrasco Kind}, M. and {Carretero}, J. and {Costanzi}, M. and {da Costa}, L.~N. and {Pereira}, M.~E.~S. and {De Vicente}, J. and {Desai}, S. and {Diehl}, H.~T. and {Doel}, P. and {Everett}, S. and {Ferrero}, I. and {Frieman}, J. and {Garc{\'\i}a-Bellido}, J. and {Gerdes}, D.~W. and {Gruen}, D. and {Gruendl}, R.~A. and {Gschwend}, J. and {Hinton}, S.~R. and {Hollowood}, D.~L. and {Honscheid}, K. and {James}, D.~J. and {Kuehn}, K. and {Marshall}, J.~L. and {Mena-Fern{\'a}ndez}, J. and {Menanteau}, F. and {Miquel}, R. and {Ogando}, R.~L.~C. and {Palmese}, A. and {Paz-Chinch{\'o}n}, F. and {Pieres}, A. and {Plazas Malag{\'o}n}, A.~A. and {Raveri}, M. and {Rodriguez-Monroy}, M. and {Romer}, A.~K. and {Sanchez}, E. and {Scarpine}, V. and {Sevilla-Noarbe}, I. and {Smith}, M. and {Soares-Santos}, M. and {Suchyta}, E. and {Swanson}, M.~E.~C. and {Tarle}, G. and {Taylor}, G. and {Tucker}, D.~L. and {Weaverdyck}, N. and {Wilkinson}, R.~D.},
        title = "{OzDES Reverberation Mapping Program: H{\ensuremath{\beta}} lags from the 6-yr survey}",
      journal = {\mnras},
         year = 2023,
        month = apr,
       volume = {520},
       number = {2},
        pages = {2009-2023},
          doi = {10.1093/mnras/stad145},
archivePrefix = {arXiv},
       eprint = {2210.03977},
 primaryClass = {astro-ph.GA},
       adsurl = {https://ui.adsabs.harvard.edu/abs/2023MNRAS.520.2009M}
}

@ARTICLE{OzDES-Malik_2022,
       author = {{Malik}, U. and {Sharp}, Rob and {Martini}, Paul and {Davis}, Tamara M. and {Tucker}, Brad E. and {Yu}, Zhefu and {Penton}, Andrew and {Lewis}, Geraint F. and {Calcino}, Josh},
        title = "{Observational window effects on multi-object reverberation mapping}",
      journal = {\mnras},
         year = 2022,
        month = nov,
       volume = {516},
       number = {3},
        pages = {3238-3253},
          doi = {10.1093/mnras/stac2263},
archivePrefix = {arXiv},
       eprint = {2203.04518},
 primaryClass = {astro-ph.GA},
       adsurl = {https://ui.adsabs.harvard.edu/abs/2022MNRAS.516.3238M}
}

@ARTICLE{OzDES-Malik_2024,
       author = {{Malik}, U. and {Sharp}, R. and {Penton}, A. and {Yu}, Z. and {Martini}, P. and {Tucker}, B.~E. and {Davis}, T.~M. and {Lewis}, G.~F. and {Lidman}, C. and {Aguena}, M. and {Alves}, O. and {Annis}, J. and {Asorey}, J. and {Bacon}, D. and {Brooks}, D. and {Carnero Rosell}, A. and {Carretero}, J. and {Cheng}, T. -Y. and {da Costa}, L.~N. and {Pereira}, M.~E.~S. and {De Vicente}, J. and {Doel}, P. and {Ferrero}, I. and {Frieman}, J. and {Giannini}, G. and {Gruen}, D. and {Gruendl}, R.~A. and {Hinton}, S.~R. and {Hollowood}, D.~L. and {James}, D.~J. and {Kuehn}, K. and {Marshall}, J.~L. and {Mena-Fern{\'a}ndez}, J. and {Menanteau}, F. and {Miquel}, R. and {Ogando}, R.~L.~C. and {Palmese}, A. and {Pieres}, A. and {Plazas Malag{\'o}n}, A.~A. and {Reil}, K. and {Romer}, A.~K. and {Sanchez}, E. and {Schubnell}, M. and {Smith}, M. and {Suchyta}, E. and {Swanson}, M.~E.~C. and {Tarle}, G. and {To}, C. and {Weaverdyck}, N. and {Wiseman}, P.},
        title = "{OzDES Reverberation Mapping Program: Stacking analysis with H{\ensuremath{\beta}}, Mg II, and C IV}",
      journal = {\mnras},
         year = 2024,
        month = jun,
       volume = {531},
       number = {1},
        pages = {163-182},
          doi = {10.1093/mnras/stae1154},
archivePrefix = {arXiv},
       eprint = {2405.06100},
 primaryClass = {astro-ph.GA},
       adsurl = {https://ui.adsabs.harvard.edu/abs/2024MNRAS.531..163M}
}

@article{OZDES-DR2-Lidman_2020,
   title={OzDES multi-object fibre spectroscopy for the Dark Energy Survey: results and second data release},
   volume={496},
   ISSN={1365-2966},
   url={http://dx.doi.org/10.1093/mnras/staa1341},
   DOI={10.1093/mnras/staa1341},
   number={1},
   journal={Monthly Notices of the Royal Astronomical Society},
   publisher={Oxford University Press (OUP)},
   author={Lidman, C and Tucker, B E and Davis, T M and Uddin, S A and Asorey, J and Bolejko, K and Brout, D and Calcino, J and Carollo, D and Carr, A and Childress, M and Hoormann, J K and Foley, R J and Galbany, L and Glazebrook, K and Hinton, S R and Kessler, R and Kim, A G and King, A and Kremin, A and Kuehn, K and Lagattuta, D and Lewis, G F and Macaulay, E and Malik, U and March, M and Martini, P and Möller, A and Mudd, D and Nichol, R C and Panther, F and Parkinson, D and Pursiainen, M and Sako, M and Swann, E and Scalzo, R and Scolnic, D and Sharp, R and Smith, M and Sommer, N E and Sullivan, M and Webb, S and Wiseman, P and Yu, Z and Yuan, F and Zhang, B and Abbott, T M C and Aguena, M and Allam, S and Annis, J and Avila, S and Bertin, E and Bhargava, S and Brooks, D and Carnero Rosell, A and Carrasco Kind, M and Carretero, J and Castander, F J and Costanzi, M and da Costa, L N and De Vicente, J and Doel, P and Eifler, T F and Everett, S and Fosalba, P and Frieman, J and García-Bellido, J and Gaztanaga, E and Gruen, D and Gruendl, R A and Gschwend, J and Gutierrez, G and Hartley, W G and Hollowood, D L and Honscheid, K and James, D J and Kuropatkin, N and Li, T S and Lima, M and Lin, H and Maia, M A G and Marshall, J L and Melchior, P and Menanteau, F and Miquel, R and Palmese, A and Paz-Chinchón, F and Plazas, A A and Roodman, A and Rykoff, E S and Sanchez, E and Santiago, B and Scarpine, V and Schubnell, M and Serrano, S and Sevilla-Noarbe, I and Suchyta, E and Swanson, M E C and Tarle, G and Tucker, D L and Varga, T N and Walker, A R and Wester, W and Wilkinson, R D},
   year={2020},
   month=may, pages={19–35} }

@article{OZDES-DR1-Childress_2017,
   title={OzDES multifibre spectroscopy for the Dark Energy Survey: 3-yr results and first data release},
   volume={472},
   ISSN={1365-2966},
   url={http://dx.doi.org/10.1093/mnras/stx1872},
   DOI={10.1093/mnras/stx1872},
   number={1},
   journal={Monthly Notices of the Royal Astronomical Society},
   publisher={Oxford University Press (OUP)},
   author={Childress, M. J. and Lidman, C. and Davis, T. M. and Tucker, B. E. and Asorey, J. and Yuan, F. and Abbott, T. M. C. and Abdalla, F. B. and Allam, S. and Annis, J. and Banerji, M. and Benoit-Lévy, A. and Bernard, S. R. and Bertin, E. and Brooks, D. and Buckley-Geer, E. and Burke, D. L. and Carnero Rosell, A. and Carollo, D. and Carrasco Kind, M. and Carretero, J. and Castander, F. J. and Cunha, C. E. and da Costa, L. N. and D’Andrea, C. B. and Doel, P. and Eifler, T. F. and Evrard, A. E. and Flaugher, B. and Foley, R. J. and Fosalba, P. and Frieman, J. and García-Bellido, J. and Glazebrook, K. and Goldstein, D. A. and Gruen, D. and Gruendl, R. A. and Gschwend, J. and Gupta, R. R. and Gutierrez, G. and Hinton, S.R. and Hoormann, J. K. and James, D. J. and Kessler, R. and Kim, A. G. and King, A. L. and Kovacs, E. and Kuehn, K. and Kuhlmann, S. and Kuropatkin, N. and Lagattuta, D. J. and Lewis, G. F. and Li, T. S. and Lima, M. and Lin, H. and Macaulay, E. and Maia, M. A. G. and Marriner, J. and March, M. and Marshall, J. L. and Martini, P. and McMahon, R. G. and Menanteau, F. and Miquel, R. and Moller, A. and Morganson, E. and Mould, J. and Mudd, D. and Muthukrishna, D. and Nichol, R. C. and Nord, B. and Ogando, R. L. C. and Ostrovski, F. and Parkinson, D. and Plazas, A. A. and Reed, S. L. and Reil, K. and Romer, A. K. and Rykoff, E. S. and Sako, M. and Sanchez, E. and Scarpine, V. and Schindler, R. and Schubnell, M. and Scolnic, D. and Sevilla-Noarbe, I. and Seymour, N. and Sharp, R. and Smith, M. and Soares-Santos, M. and Sobreira, F. and Sommer, N. E. and Spinka, H. and Suchyta, E. and Sullivan, M. and Swanson, M. E. C. and Tarle, G. and Uddin, S. A. and Walker, A. R. and Wester, W. and Zhang, B. R.},
   year={2017},
   month=jul, pages={273–288} }

@article{OzDES-DR0-Yuan_2015,
   title={OzDES multifibre spectroscopy for the Dark Energy Survey: first-year operation and results},
   volume={452},
   ISSN={1365-2966},
   url={http://dx.doi.org/10.1093/mnras/stv1507},
   DOI={10.1093/mnras/stv1507},
   number={3},
   journal={Monthly Notices of the Royal Astronomical Society},
   publisher={Oxford University Press (OUP)},
   author={Yuan, Fang and Lidman, C. and Davis, T. M. and Childress, M. and Abdalla, F. B. and Banerji, M. and Buckley-Geer, E. and Carnero Rosell, A. and Carollo, D. and Castander, F. J. and D’Andrea, C. B. and Diehl, H. T. and Cunha, C. E and Foley, R. J. and Frieman, J. and Glazebrook, K. and Gschwend, J. and Hinton, S. and Jouvel, S. and Kessler, R. and Kim, A. G. and King, A. L. and Kuehn, K. and Kuhlmann, S. and Lewis, G. F. and Lin, H. and Martini, P. and McMahon, R. G. and Mould, J. and Nichol, R. C. and Norris, R. P. and O’Neill, C. R. and Ostrovski, F. and Papadopoulos, A. and Parkinson, D. and Reed, S. and Romer, A. K. and Rooney, P. J. and Rozo, E. and Rykoff, E. S. and Sako, M. and Scalzo, R. and Schmidt, B. P. and Scolnic, D. and Seymour, N. and Sharp, R. and Sobreira, F. and Sullivan, M. and Thomas, R. C. and Tucker, D. and Uddin, S. A. and Wechsler, R. H. and Wester, W. and Wilcox, H. and Zhang, B. and Abbott, T. and Allam, S. and Bauer, A. H. and Benoit-Lévy, A. and Bertin, E. and Brooks, D. and Burke, D. L. and Carrasco Kind, M. and Covarrubias, R. and Crocce, M. and da Costa, L. N. and DePoy, D. L. and Desai, S. and Doel, P. and Eifler, T. F. and Evrard, A. E. and Fausti Neto, A. and Flaugher, B. and Fosalba, P. and Gaztanaga, E. and Gerdes, D. and Gruen, D. and Gruendl, R. A. and Honscheid, K. and James, D. and Kuropatkin, N. and Lahav, O. and Li, T. S. and Maia, M. A. G. and Makler, M. and Marshall, J. and Miller, C. J. and Miquel, R. and Ogando, R. and Plazas, A. A. and Roodman, A. and Sanchez, E. and Scarpine, V. and Schubnell, M. and Sevilla-Noarbe, I. and Smith, R. C. and Soares-Santos, M. and Suchyta, E. and Swanson, M. E. C. and Tarle, G. and Thaler, J. and Walker, A. R.},
   year={2015},
   month=jul, pages={3047–3063} }

@ARTICLE{OzDES-Spec-Smith2018,
       author = {{Smith}, M. and {D'Andrea}, C.~B. and {Sullivan}, M. and {M{\"o}ller}, A. and {Nichol}, R.~C. and {Thomas}, R.~C. and {Kim}, A.~G. and {Sako}, M. and {Castander}, F.~J. and {Filippenko}, A.~V. and {Foley}, R.~J. and {Galbany}, L. and {Gonz{\'a}lez-Gait{\'a}n}, S. and {Kasai}, E. and {Kirshner}, R.~P. and {Lidman}, C. and {Scolnic}, D. and {Brout}, D. and {Davis}, T.~M. and {Gupta}, R.~R. and {Hinton}, S.~R. and {Kessler}, R. and {Lasker}, J. and {Macaulay}, E. and {Wolf}, R.~C. and {Zhang}, B. and {Asorey}, J. and {Avelino}, A. and {Bassett}, B.~A. and {Calcino}, J. and {Carollo}, D. and {Casas}, R. and {Challis}, P. and {Childress}, M. and {Clocchiatti}, A. and {Crawford}, S. and {Frohmaier}, C. and {Glazebrook}, K. and {Goldstein}, D.~A. and {Graham}, M.~L. and {Hoormann}, J.~K. and {Kuehn}, K. and {Lewis}, G.~F. and {Mandel}, K.~S. and {Morganson}, E. and {Muthukrishna}, D. and {Nugent}, P. and {Pan}, Y. -C. and {Pursiainen}, M. and {Sharp}, R. and {Sommer}, N.~E. and {Swann}, E. and {Thomas}, B.~P. and {Tucker}, B.~E. and {Uddin}, S.~A. and {Wiseman}, P. and {Zheng}, W. and {Abbott}, T.~M.~C. and {Annis}, J. and {Avila}, S. and {Bechtol}, K. and {Bernstein}, G.~M. and {Bertin}, E. and {Brooks}, D. and {Burke}, D.~L. and {Carnero Rosell}, A. and {Carrasco Kind}, M. and {Carretero}, J. and {Cunha}, C.~E. and {da Costa}, L.~N. and {Davis}, C. and {De Vicente}, J. and {Diehl}, H.~T. and {Eifler}, T.~F. and {Estrada}, J. and {Frieman}, J. and {Garc{\'\i}a-Bellido}, J. and {Gaztanaga}, E. and {Gerdes}, D.~W. and {Gruen}, D. and {Gruendl}, R.~A. and {Gschwend}, J. and {Gutierrez}, G. and {Hartley}, W.~G. and {Hollowood}, D.~L. and {Honscheid}, K. and {Hoyle}, B. and {James}, D.~J. and {Johnson}, M.~W.~G. and {Johnson}, M.~D. and {Kuropatkin}, N. and {Li}, T.~S. and {Lima}, M. and {Maia}, M.~A.~G. and {March}, M. and {Marshall}, J.~L. and {Martini}, P. and {Menanteau}, F. and {Miller}, C.~J. and {Miquel}, R. and {Neilsen}, E. and {Ogando}, R.~L.~C. and {Plazas}, A.~A. and {Romer}, A.~K. and {Sanchez}, E. and {Scarpine}, V. and {Schubnell}, M. and {Serrano}, S. and {Sevilla-Noarbe}, I. and {Soares-Santos}, M. and {Sobreira}, F. and {Suchyta}, E. and {Tarle}, G. and {Tucker}, D.~L. and {Wester}, W.},
        title = "{First Cosmology Results using Supernovae Ia from the Dark Energy Survey: Survey Overview, Performance, and Supernova Spectroscopy}",
      journal = {\aj},
         year = 2020,
        month = dec,
       volume = {160},
       number = {6},
          eid = {267},
        pages = {267},
          doi = {10.3847/1538-3881/abc01b},
archivePrefix = {arXiv},
       eprint = {1811.09565},
 primaryClass = {astro-ph.CO},
       adsurl = {https://ui.adsabs.harvard.edu/abs/2020AJ....160..267S}
}

@ARTICLE{Flaugher_2015,
       author = {{Flaugher}, B. and {Diehl}, H.~T. and {Honscheid}, K. and {Abbott}, T.~M.~C. and {Alvarez}, O. and {Angstadt}, R. and {Annis}, J.~T. and {Antonik}, M. and {Ballester}, O. and {Beaufore}, L. and {Bernstein}, G.~M. and {Bernstein}, R.~A. and {Bigelow}, B. and {Bonati}, M. and {Boprie}, D. and {Brooks}, D. and {Buckley-Geer}, E.~J. and {Campa}, J. and {Cardiel-Sas}, L. and {Castander}, F.~J. and {Castilla}, J. and {Cease}, H. and {Cela-Ruiz}, J.~M. and {Chappa}, S. and {Chi}, E. and {Cooper}, C. and {da Costa}, L.~N. and {Dede}, E. and {Derylo}, G. and {DePoy}, D.~L. and {de Vicente}, J. and {Doel}, P. and {Drlica-Wagner}, A. and {Eiting}, J. and {Elliott}, A.~E. and {Emes}, J. and {Estrada}, J. and {Fausti Neto}, A. and {Finley}, D.~A. and {Flores}, R. and {Frieman}, J. and {Gerdes}, D. and {Gladders}, M.~D. and {Gregory}, B. and {Gutierrez}, G.~R. and {Hao}, J. and {Holland}, S.~E. and {Holm}, S. and {Huffman}, D. and {Jackson}, C. and {James}, D.~J. and {Jonas}, M. and {Karcher}, A. and {Karliner}, I. and {Kent}, S. and {Kessler}, R. and {Kozlovsky}, M. and {Kron}, R.~G. and {Kubik}, D. and {Kuehn}, K. and {Kuhlmann}, S. and {Kuk}, K. and {Lahav}, O. and {Lathrop}, A. and {Lee}, J. and {Levi}, M.~E. and {Lewis}, P. and {Li}, T.~S. and {Mandrichenko}, I. and {Marshall}, J.~L. and {Martinez}, G. and {Merritt}, K.~W. and {Miquel}, R. and {Mu{\~n}oz}, F. and {Neilsen}, E.~H. and {Nichol}, R.~C. and {Nord}, B. and {Ogando}, R. and {Olsen}, J. and {Palaio}, N. and {Patton}, K. and {Peoples}, J. and {Plazas}, A.~A. and {Rauch}, J. and {Reil}, K. and {Rheault}, J. -P. and {Roe}, N.~A. and {Rogers}, H. and {Roodman}, A. and {Sanchez}, E. and {Scarpine}, V. and {Schindler}, R.~H. and {Schmidt}, R. and {Schmitt}, R. and {Schubnell}, M. and {Schultz}, K. and {Schurter}, P. and {Scott}, L. and {Serrano}, S. and {Shaw}, T.~M. and {Smith}, R.~C. and {Soares-Santos}, M. and {Stefanik}, A. and {Stuermer}, W. and {Suchyta}, E. and {Sypniewski}, A. and {Tarle}, G. and {Thaler}, J. and {Tighe}, R. and {Tran}, C. and {Tucker}, D. and {Walker}, A.~R. and {Wang}, G. and {Watson}, M. and {Weaverdyck}, C. and {Wester}, W. and {Woods}, R. and {Yanny}, B. and {DES Collaboration}},
        title = "{The Dark Energy Camera}",
      journal = {\aj},
         year = 2015,
        month = nov,
       volume = {150},
       number = {5},
          eid = {150},
        pages = {150},
          doi = {10.1088/0004-6256/150/5/150},
archivePrefix = {arXiv},
       eprint = {1504.02900},
 primaryClass = {astro-ph.IM},
       adsurl = {https://ui.adsabs.harvard.edu/abs/2015AJ....150..150F}
}

@ARTICLE{Lewis_2002_2df,
       author = {{Lewis}, I.~J. and {Cannon}, R.~D. and {Taylor}, K. and {Glazebrook}, K. and {Bailey}, J.~A. and {Baldry}, I.~K. and {Barton}, J.~R. and {Bridges}, T.~J. and {Dalton}, G.~B. and {Farrell}, T.~J. and {Gray}, P.~M. and {Lankshear}, A. and {McCowage}, C. and {Parry}, I.~R. and {Sharples}, R.~M. and {Shortridge}, K. and {Smith}, G.~A. and {Stevenson}, J. and {Straede}, J.~O. and {Waller}, L.~G. and {Whittard}, J.~D. and {Wilcox}, J.~K. and {Willis}, K.~C.},
        title = "{The Anglo-Australian Observatory 2dF facility}",
      journal = {\mnras},
         year = 2002,
        month = jun,
       volume = {333},
       number = {2},
        pages = {279-299},
          doi = {10.1046/j.1365-8711.2002.05333.x},
archivePrefix = {arXiv},
       eprint = {astro-ph/0202175},
 primaryClass = {astro-ph},
       adsurl = {https://ui.adsabs.harvard.edu/abs/2002MNRAS.333..279L}
}

@article{McDougall_2025_LITMUS, 
    title={LITMUS: Bayesian lag recovery in reverberation mapping with fast differentiable models}, 
    volume={43}, 
    DOI={10.1017/pasa.2026.10149}, 
    journal={Publications of the Astronomical Society of Australia}, 
    author={McDougall, Hugh Gareth and Pope, Benjamin and Davis, Tamara M.}, year={2026}, pages={e018}
}

@ARTICLE{SDSS-Shen_2023,
       author = {{Shen}, Yue and {Grier}, Catherine J. and {Horne}, Keith and {Stone}, Zachary and {Li}, Jennifer I. and {Yang}, Qian and {Homayouni}, Yasaman and {Trump}, Jonathan R. and {Anderson}, Scott F. and {Brandt}, W.~N. and {Hall}, Patrick B. and {Ho}, Luis C. and {Jiang}, Linhua and {Petitjean}, Patrick and {Schneider}, Donald P. and {Tao}, Charling and {Donnan}, Fergus. R. and {AlSayyad}, Yusra and {Bershady}, Matthew A. and {Blanton}, Michael R. and {Bizyaev}, Dmitry and {Bundy}, Kevin and {Chen}, Yuguang and {Davis}, Megan C. and {Dawson}, Kyle and {Fan}, Xiaohui and {Greene}, Jenny E. and {Gr{\"o}ller}, Hannes and {Guo}, Yucheng and {Ibarra-Medel}, H{\'e}ctor and {Jiang}, Yuanzhe and {Keenan}, Ryan P. and {Kollmeier}, Juna A. and {Lejoly}, Cassandra and {Li}, Zefeng and {de la Macorra}, Axel and {Moe}, Maxwell and {Nie}, Jundan and {Rossi}, Graziano and {Smith}, Paul S. and {Tee}, Wei Leong and {Weijmans}, Anne-Marie and {Xu}, Jiachuan and {Yue}, Minghao and {Zhou}, Xu and {Zhou}, Zhimin and {Zou}, Hu},
        title = "{The Sloan Digital Sky Survey Reverberation Mapping Project: Key Results}",
      journal = {\apjs},
         year = 2024,
        month = jun,
       volume = {272},
       number = {2},
          eid = {26},
        pages = {26},
          doi = {10.3847/1538-4365/ad3936},
archivePrefix = {arXiv},
       eprint = {2305.01014},
 primaryClass = {astro-ph.GA},
       adsurl = {https://ui.adsabs.harvard.edu/abs/2024ApJS..272...26S}
}

@ARTICLE{SDSS-Grier_2017,
       author = {{Grier}, C.~J. and {Trump}, J.~R. and {Shen}, Yue and {Horne}, Keith and {Kinemuchi}, Karen and {McGreer}, Ian D. and {Starkey}, D.~A. and {Brandt}, W.~N. and {Hall}, P.~B. and {Kochanek}, C.~S. and {Chen}, Yuguang and {Denney}, K.~D. and {Greene}, Jenny E. and {Ho}, L.~C. and {Homayouni}, Y. and {I-Hsiu Li}, Jennifer and {Pei}, Liuyi and {Peterson}, B.~M. and {Petitjean}, P. and {Schneider}, D.~P. and {Sun}, Mouyuan and {AlSayyad}, Yusura and {Bizyaev}, Dmitry and {Brinkmann}, Jonathan and {Brownstein}, Joel R. and {Bundy}, Kevin and {Dawson}, K.~S. and {Eftekharzadeh}, Sarah and {Fernandez-Trincado}, J.~G. and {Gao}, Yang and {Hutchinson}, Timothy A. and {Jia}, Siyao and {Jiang}, Linhua and {Oravetz}, Daniel and {Pan}, Kaike and {Paris}, Isabelle and {Ponder}, Kara A. and {Peters}, Christina and {Rogerson}, Jesse and {Simmons}, Audrey and {Smith}, Robyn and {Wang}, Ran},
        title = "{The Sloan Digital Sky Survey Reverberation Mapping Project: H{\ensuremath{\alpha}} and H{\ensuremath{\beta}} Reverberation Measurements from First-year Spectroscopy and Photometry}",
      journal = {\apj},
         year = 2017,
        month = dec,
       volume = {851},
       number = {1},
          eid = {21},
        pages = {21},
          doi = {10.3847/1538-4357/aa98dc},
archivePrefix = {arXiv},
       eprint = {1711.03114},
 primaryClass = {astro-ph.GA},
       adsurl = {https://ui.adsabs.harvard.edu/abs/2017ApJ...851...21G}
}

@article{Burke_2017,
   title={Forward Global Photometric Calibration of the Dark Energy Survey},
   volume={155},
   ISSN={1538-3881},
   url={http://dx.doi.org/10.3847/1538-3881/aa9f22},
   DOI={10.3847/1538-3881/aa9f22},
   number={1},
   journal={The Astronomical Journal},
   publisher={American Astronomical Society},
   author={Burke, D. L. and Rykoff, E. S. and Allam, S. and Annis, J. and Bechtol, K. and Bernstein, G. M. and Drlica-Wagner, A. and Finley, D. A. and Gruendl, R. A. and James, D. J. and Kent, S. and Kessler, R. and Kuhlmann, S. and Lasker, J. and Li, T. S. and Scolnic, D. and Smith, J. and Tucker, D. L. and Wester, W. and Yanny, B. and Abbott, T. M. C. and Abdalla, F. B. and Benoit-Lévy, A. and Bertin, E. and Rosell, A. Carnero and Kind, M. Carrasco and Carretero, J. and Cunha, C. E. and D’Andrea, C. B. and da Costa, L. N. and Desai, S. and Diehl, H. T. and Doel, P. and Estrada, J. and García-Bellido, J. and Gruen, D. and Gutierrez, G. and Honscheid, K. and Kuehn, K. and Kuropatkin, N. and Maia, M. A. G. and March, M. and Marshall, J. L. and Melchior, P. and Menanteau, F. and Miquel, R. and Plazas, A. A. and Sako, M. and Sanchez, E. and Scarpine, V. and Schindler, R. and Sevilla-Noarbe, I. and Smith, M. and Smith, R. C. and Soares-Santos, M. and Sobreira, F. and Suchyta, E. and Tarle, G. and Walker, A. R.},
   year={2017},
   month=dec, pages={41} }

@ARTICLE{Shakura_1973,
       author = {{Shakura}, N.~I. and {Sunyaev}, R.~A.},
        title = "{Black holes in binary systems. Observational appearance.}",
      journal = {\aap},
         year = 1973,
        month = jan,
       volume = {24},
        pages = {337-355},
       adsurl = {https://ui.adsabs.harvard.edu/abs/1973A\&A....24..337S}
}

@article{King_2015,
   title={How big can a black hole grow?},
   volume={456},
   ISSN={1745-3933},
   url={http://dx.doi.org/10.1093/mnrasl/slv186},
   DOI={10.1093/mnrasl/slv186},
   number={1},
   journal={Monthly Notices of the Royal Astronomical Society: Letters},
   publisher={Oxford University Press (OUP)},
   author={King, Andrew},
   year={2015},
   month=dec, pages={L109–L112} 
}

@ARTICLE{Peterson_1993,
       author = {{Peterson}, Bradley M.},
        title = "{Reverberation Mapping of Active Galactic Nuclei}",
      journal = {\pasp},
         year = 1993,
        month = mar,
       volume = {105},
        pages = {247},
          doi = {10.1086/133140},
       adsurl = {https://ui.adsabs.harvard.edu/abs/1993PASP..105..247P}
}

@ARTICLE{Press_Rybicki_1989,
       author = {{Press}, William H. and {Rybicki}, George B.},
        title = "{Fast Algorithm for Spectral Analysis of Unevenly Sampled Data}",
      journal = {\apj},
         year = 1989,
        month = mar,
       volume = {338},
        pages = {277},
          doi = {10.1086/167197},
       adsurl = {https://ui.adsabs.harvard.edu/abs/1989ApJ...338..277P}
}

@ARTICLE{Blandford_McKee_1982,
       author = {{Blandford}, R.~D. and {McKee}, C.~F.},
        title = "{Reverberation mapping of the emission line regions of Seyfert galaxies and quasars.}",
      journal = {\apj},
         year = 1982,
        month = apr,
       volume = {255},
        pages = {419-439},
          doi = {10.1086/159843},
       adsurl = {https://ui.adsabs.harvard.edu/abs/1982ApJ...255..419B}
}

@ARTICLE{Denney_2012,
       author = {{Denney}, K.~D.},
        title = "{Are Outflows Biasing Single-epoch C IV Black Hole Mass Estimates?}",
      journal = {\apj},
         year = 2012,
        month = nov,
       volume = {759},
       number = {1},
          eid = {44},
        pages = {44},
          doi = {10.1088/0004-637X/759/1/44},
archivePrefix = {arXiv},
       eprint = {1208.3465},
 primaryClass = {astro-ph.CO},
       adsurl = {https://ui.adsabs.harvard.edu/abs/2012ApJ...759...44D}
}

@article{Runnoe_2012,
   title={Updating quasar bolometric luminosity corrections: Updating quasar bolometric corrections},
   volume={422},
   ISSN={0035-8711},
   url={http://dx.doi.org/10.1111/j.1365-2966.2012.20620.x},
   DOI={10.1111/j.1365-2966.2012.20620.x},
   number={1},
   journal={Monthly Notices of the Royal Astronomical Society},
   publisher={Oxford University Press (OUP)},
   author={Runnoe, Jessie C. and Brotherton, Michael S. and Shang, Zhaohui},
   year={2012},
   month=feb, pages={478–493} }

@article{Netzer_2019,
   title={Bolometric correction factors for active galactic nuclei},
   volume={488},
   ISSN={1365-2966},
   url={http://dx.doi.org/10.1093/mnras/stz2016},
   DOI={10.1093/mnras/stz2016},
   number={4},
   journal={Monthly Notices of the Royal Astronomical Society},
   publisher={Oxford University Press (OUP)},
   author={Netzer, Hagai},
   year={2019},
   month=jul, pages={5185–5191} }

@ARTICLE{Richards_2006,
       author = {{Richards}, Gordon T. and {Lacy}, Mark and {Storrie-Lombardi}, Lisa J. and {Hall}, Patrick B. and {Gallagher}, S.~C. and {Hines}, Dean C. and {Fan}, Xiaohui and {Papovich}, Casey and {Vanden Berk}, Daniel E. and {Trammell}, George B. and {Schneider}, Donald P. and {Vestergaard}, Marianne and {York}, Donald G. and {Jester}, Sebastian and {Anderson}, Scott F. and {Budav{\'a}ri}, Tam{\'a}s and {Szalay}, Alexander S.},
        title = "{Spectral Energy Distributions and Multiwavelength Selection of Type 1 Quasars}",
      journal = {\apjs},
         year = 2006,
        month = oct,
       volume = {166},
       number = {2},
        pages = {470-497},
          doi = {10.1086/506525},
archivePrefix = {arXiv},
       eprint = {astro-ph/0601558},
 primaryClass = {astro-ph},
       adsurl = {https://ui.adsabs.harvard.edu/abs/2006ApJS..166..470R}
}

@article{Kaspi_2007,
   title={Reverberation Mapping of High‐Luminosity Quasars: First Results},
   volume={659},
   ISSN={1538-4357},
   url={http://dx.doi.org/10.1086/512094},
   DOI={10.1086/512094},
   number={2},
   journal={The Astrophysical Journal},
   publisher={American Astronomical Society},
   author={Kaspi, Shai and Brandt, W. N. and Maoz, Dan and Netzer, Hagai and Schneider, Donald P. and Shemmer, Ohad},
   year={2007},
   month=apr, pages={997–1007} }

@article{Mejia_Restrepo_2018,
   title={Can we improve Civ-based single-epoch black hole mass estimations?},
   volume={478},
   ISSN={1365-2966},
   url={http://dx.doi.org/10.1093/mnras/sty1086},
   DOI={10.1093/mnras/sty1086},
   number={2},
   journal={Monthly Notices of the Royal Astronomical Society},
   publisher={Oxford University Press (OUP)},
   author={Mejía-Restrepo, J E and Trakhtenbrot, B and Lira, P and Netzer, H},
   year={2018},
   month=may, pages={1929–1941} }

@article{Banados_2017,
   title={An 800-million-solar-mass black hole in a significantly neutral Universe at a redshift of 7.5},
   volume={553},
   ISSN={1476-4687},
   url={http://dx.doi.org/10.1038/nature25180},
   DOI={10.1038/nature25180},
   number={7689},
   journal={Nature},
   publisher={Springer Science and Business Media LLC},
   author={Bañados, Eduardo and Venemans, Bram P. and Mazzucchelli, Chiara and Farina, Emanuele P. and Walter, Fabian and Wang, Feige and Decarli, Roberto and Stern, Daniel and Fan, Xiaohui and Davies, Frederick B. and Hennawi, Joseph F. and Simcoe, Robert A. and Turner, Monica L. and Rix, Hans-Walter and Yang, Jinyi and Kelson, Daniel D. and Rudie, Gwen C. and Winters, Jan Martin},
   year={2017},
   month=dec, pages={473–476} }

@article{Mejia_Restrepo_2016,
   title={Active galactic nuclei atz∼ 1.5 – II. Black hole mass estimation by means of broad emission lines},
   volume={460},
   ISSN={1365-2966},
   url={http://dx.doi.org/10.1093/mnras/stw568},
   DOI={10.1093/mnras/stw568},
   number={1},
   journal={Monthly Notices of the Royal Astronomical Society},
   publisher={Oxford University Press (OUP)},
   author={Mejía-Restrepo, J. E. and Trakhtenbrot, B. and Lira, P. and Netzer, H. and Capellupo, D. M.},
   year={2016},
   month=mar, pages={187–211} }

@article{Woo_2015,
   title={THE BLACK HOLE MASS–STELLAR VELOCITY DISPERSION RELATION OF NARROW-LINE SEYFERT 1 GALAXIES},
   volume={801},
   ISSN={1538-4357},
   url={http://dx.doi.org/10.1088/0004-637X/801/1/38},
   DOI={10.1088/0004-637x/801/1/38},
   number={1},
   journal={The Astrophysical Journal},
   publisher={American Astronomical Society},
   author={Woo, Jong-Hak and Yoon, Yosep and Park, Songyoun and Park, Daeseong and Kim, Sang Chul},
   year={2015},
   month=feb, pages={38} }

@article{Lu_2019,
   title={Supermassive Black Holes with High Accretion Rates in Active Galactic Nuclei. X. Optical Variability Characteristics},
   volume={877},
   ISSN={1538-4357},
   url={http://dx.doi.org/10.3847/1538-4357/ab16e8},
   DOI={10.3847/1538-4357/ab16e8},
   number={1},
   journal={The Astrophysical Journal},
   publisher={American Astronomical Society},
   author={Lu, Kai-Xing and Huang, Ying-Ke and Zhang, Zhi-Xiang and Wang, Kai and Du, Pu and Hu, Chen and Xiao, Ming and Li, Yan-Rong and Bai, Jin-Ming and Bian, Wei-Hao and Yuan, Ye-Fei and Ho, Luis C. and Wang, Jian-Min},
   year={2019},
   month=may, pages={23} }

@ARTICLE{Kelly_2009,
       author = {{Kelly}, Brandon C. and {Bechtold}, Jill and {Siemiginowska}, Aneta},
        title = "{Are the Variations in Quasar Optical Flux Driven by Thermal Fluctuations?}",
      journal = {\apj},
         year = 2009,
        month = jun,
       volume = {698},
       number = {1},
        pages = {895-910},
          doi = {10.1088/0004-637X/698/1/895},
archivePrefix = {arXiv},
       eprint = {0903.5315},
 primaryClass = {astro-ph.CO},
       adsurl = {https://ui.adsabs.harvard.edu/abs/2009ApJ...698..895K}
}

@ARTICLE{Kozlowski_2010a,
       author = {{Koz{\l}owski}, Szymon and {Kochanek}, Christopher S. and {Udalski}, A. and {Wyrzykowski}, {\L}. and {Soszy{\'n}ski}, I. and {Szyma{\'n}ski}, M.~K. and {Kubiak}, M. and {Pietrzy{\'n}ski}, G. and {Szewczyk}, O. and {Ulaczyk}, K. and {Poleski}, R. and {OGLE Collaboration}},
        title = "{Quantifying Quasar Variability as Part of a General Approach to Classifying Continuously Varying Sources}",
      journal = {\apj},
         year = 2010,
        month = jan,
       volume = {708},
       number = {2},
        pages = {927-945},
          doi = {10.1088/0004-637X/708/2/927},
archivePrefix = {arXiv},
       eprint = {0909.1326},
 primaryClass = {astro-ph.CO},
       adsurl = {https://ui.adsabs.harvard.edu/abs/2010ApJ...708..927K}
}

@ARTICLE{Denney_2009,
       author = {{Denney}, K.~D. and {Peterson}, B.~M. and {Pogge}, R.~W. and {Adair}, A. and {Atlee}, D.~W. and {Au-Yong}, K. and {Bentz}, M.~C. and {Bird}, J.~C. and {Brokofsky}, D.~J. and {Chisholm}, E. and {Comins}, M.~L. and {Dietrich}, M. and {Doroshenko}, V.~T. and {Eastman}, J.~D. and {Efimov}, Y.~S. and {Ewald}, S. and {Ferbey}, S. and {Gaskell}, C.~M. and {Hedrick}, C.~H. and {Jackson}, K. and {Klimanov}, S.~A. and {Klimek}, E.~S. and {Kruse}, A.~K. and {Lad{\'e}route}, A. and {Lamb}, J.~B. and {Leighly}, K. and {Minezaki}, T. and {Nazarov}, S.~V. and {Onken}, C.~A. and {Petersen}, E.~A. and {Peterson}, P. and {Poindexter}, S. and {Sakata}, Y. and {Schlesinger}, K.~J. and {Sergeev}, S.~G. and {Skolski}, N. and {Stieglitz}, L. and {Tobin}, J.~J. and {Unterborn}, C. and {Vestergaard}, M. and {Watkins}, A.~E. and {Watson}, L.~C. and {Yoshii}, Y.},
        title = "{Diverse Kinematic Signatures from Reverberation Mapping of the Broad-Line Region in AGNs}",
      journal = {\apjl},
         year = 2009,
        month = oct,
       volume = {704},
       number = {2},
        pages = {L80-L84},
          doi = {10.1088/0004-637X/704/2/L80},
archivePrefix = {arXiv},
       eprint = {0908.0327},
 primaryClass = {astro-ph.CO},
       adsurl = {https://ui.adsabs.harvard.edu/abs/2009ApJ...704L..80D}
}

@ARTICLE{Grier_2013a,
       author = {{Grier}, C.~J. and {Peterson}, B.~M. and {Horne}, Keith and {Bentz}, M.~C. and {Pogge}, R.~W. and {Denney}, K.~D. and {De Rosa}, G. and {Martini}, Paul and {Kochanek}, C.~S. and {Zu}, Y. and {Shappee}, B. and {Siverd}, R. and {Beatty}, T.~G. and {Sergeev}, S.~G. and {Kaspi}, S. and {Araya Salvo}, C. and {Bird}, J.~C. and {Bord}, D.~J. and {Borman}, G.~A. and {Che}, X. and {Chen}, C. and {Cohen}, S.~A. and {Dietrich}, M. and {Doroshenko}, V.~T. and {Efimov}, Yu. S. and {Free}, N. and {Ginsburg}, I. and {Henderson}, C.~B. and {King}, A.~L. and {Mogren}, K. and {Molina}, M. and {Mosquera}, A.~M. and {Nazarov}, S.~V. and {Okhmat}, D.~N. and {Pejcha}, O. and {Rafter}, S. and {Shields}, J.~C. and {Skowron}, J. and {Szczygiel}, D.~M. and {Valluri}, M. and {van Saders}, J.~L.},
        title = "{The Structure of the Broad-line Region in Active Galactic Nuclei. I. Reconstructed Velocity-delay Maps}",
      journal = {\apj},
         year = 2013,
        month = feb,
       volume = {764},
       number = {1},
          eid = {47},
        pages = {47},
          doi = {10.1088/0004-637X/764/1/47},
archivePrefix = {arXiv},
       eprint = {1210.2397},
 primaryClass = {astro-ph.CO},
       adsurl = {https://ui.adsabs.harvard.edu/abs/2013ApJ...764...47G}
}

@ARTICLE{ICCF-Gaskell_1987,
       author = {{Gaskell}, C. Martin and {Peterson}, Bradley M.},
        title = "{The Accuracy of Cross-Correlation Estimates of Quasar Emission-Line Region Sizes}",
      journal = {\apjs},
         year = 1987,
        month = sep,
       volume = {65},
        pages = {1},
          doi = {10.1086/191216},
       adsurl = {https://ui.adsabs.harvard.edu/abs/1987ApJS...65....1G}
}

@misc{PyCCF-Mouyan_2018,
       author = {{Sun}, Mouyuan and {Grier}, C.~J. and {Peterson}, B.~M.},
        title = "{PyCCF: Python Cross Correlation Function for reverberation mapping studies}",
 howpublished = {Astrophysics Source Code Library, record ascl:1805.032},
         year = 2018,
        month = may,
          eid = {ascl:1805.032},
       adsurl = {https://ui.adsabs.harvard.edu/abs/2018ascl.soft05032S}
}

@MISC{JAVELIN-Zu_2010,
       author = {{Zu}, Ying and {Kochanek}, C.~S. and {Peterson}, Bradley M.},
        title = "{JAVELIN: Just Another Vehicle for Estimating Lags In Nuclei}",
 howpublished = {Astrophysics Source Code Library, record ascl:1010.007},
         year = 2010,
        month = oct,
          eid = {ascl:1010.007},
        pages = {ascl:1010.007},
archivePrefix = {ascl},
       eprint = {1010.007},
       adsurl = {https://ui.adsabs.harvard.edu/abs/2010ascl.soft10007Z}
}

@article{emcee-Foreman_Mackey_2013,
	doi = {10.1086/670067},
	url = {https://doi.org/10.1086%2F670067},
	year = 2013,
	month = {mar},
	publisher = {{IOP} Publishing},
	volume = {125},
	number = {925},
	pages = {306--312},
	author = {Daniel Foreman-Mackey and David W. Hogg and Dustin Lang and Jonathan Goodman},
	title = {\texttt{emcee}: The {MCMC} Hammer},
	journal = {Publications of the Astronomical Society of the Pacific}
}

@misc{numpyro,
  doi = {10.48550/ARXIV.1912.11554},
  url = {https://arxiv.org/abs/1912.11554},
  author = {Phan, Du and Pradhan, Neeraj and Jankowiak, Martin},
  title = {Composable Effects for Flexible and Accelerated Probabilistic Programming in NumPyro},
  publisher = {arXiv},
  year = {2019},
  copyright = {arXiv.org perpetual, non-exclusive license}
}

@misc{pyroa-ferus_2021,
       author = {{Donnan}, Fergus},
        title = "{PyROA: Modeling quasar light curves}",
 howpublished = {Astrophysics Source Code Library, record ascl:2107.012},
         year = 2021,
        month = jul,
          eid = {ascl:2107.012},
       adsurl = {https://ui.adsabs.harvard.edu/abs/2021ascl.soft07012D}
}

@article{Starkey_2015_CREAM,
   title={Accretion disc time lag distributions: applying CREAM to simulated AGN light curves},
   volume={456},
   ISSN={1365-2966},
   url={http://dx.doi.org/10.1093/mnras/stv2744},
   DOI={10.1093/mnras/stv2744},
   number={2},
   journal={Monthly Notices of the Royal Astronomical Society},
   publisher={Oxford University Press (OUP)},
   author={Starkey, D. A. and Horne, Keith and Villforth, C.},
   year={2015},
   month=dec, pages={1960–1973} }

@ARTICLE{AffineInvariant-Goodman_2010,
       author = {{Goodman}, Jonathan and {Weare}, Jonathan},
        title = "{Ensemble samplers with affine invariance}",
      journal = {Communications in Applied Mathematics and Computational Science},
         year = 2010,
        month = jan,
       volume = {5},
       number = {1},
        pages = {65-80},
          doi = {10.2140/camcos.2010.5.65},
       adsurl = {https://ui.adsabs.harvard.edu/abs/2010CAMCS...5...65G}
}

@ARTICLE{Template_Vestergaard_2001,
       author = {{Vestergaard}, M. and {Wilkes}, B.~J.},
        title = "{An Empirical Ultraviolet Template for Iron Emission in Quasars as Derived from I Zwicky 1}",
      journal = {\apjs},
         year = 2001,
        month = may,
       volume = {134},
       number = {1},
        pages = {1-33},
          doi = {10.1086/320357},
archivePrefix = {arXiv},
       eprint = {astro-ph/0104320},
 primaryClass = {astro-ph},
       adsurl = {https://ui.adsabs.harvard.edu/abs/2001ApJS..134....1V}
}

@ARTICLE{Template_Tzuzuki_2006,
       author = {{Tsuzuki}, Yumihiko and {Kawara}, Kimiaki and {Yoshii}, Yuzuru and {Oyabu}, Shinki and {Tanab{\'e}}, Toshihiko and {Matsuoka}, Yoshiki},
        title = "{Fe II Emission in 14 Low-Redshift Quasars. I. Observations}",
      journal = {\apj},
         year = 2006,
        month = oct,
       volume = {650},
       number = {1},
        pages = {57-79},
          doi = {10.1086/506376},
archivePrefix = {arXiv},
       eprint = {astro-ph/0606040},
 primaryClass = {astro-ph},
       adsurl = {https://ui.adsabs.harvard.edu/abs/2006ApJ...650...57T}
}

@ARTICLE{Template_Salviander_2007,
       author = {{Salviander}, S. and {Shields}, G.~A. and {Gebhardt}, K. and {Bonning}, E.~W.},
        title = "{The Black Hole Mass-Galaxy Bulge Relationship for QSOs in the Sloan Digital Sky Survey Data Release 3}",
      journal = {\apj},
         year = 2007,
        month = jun,
       volume = {662},
       number = {1},
        pages = {131-144},
          doi = {10.1086/513086},
archivePrefix = {arXiv},
       eprint = {astro-ph/0612568},
 primaryClass = {astro-ph},
       adsurl = {https://ui.adsabs.harvard.edu/abs/2007ApJ...662..131S}
}

@article{EHT_2019,
   title={First M87 Event Horizon Telescope Results. I. The Shadow of the Supermassive Black Hole},
   volume={875},
   ISSN={2041-8213},
   url={http://dx.doi.org/10.3847/2041-8213/ab0ec7},
   DOI={10.3847/2041-8213/ab0ec7},
   number={1},
   journal={The Astrophysical Journal Letters},
   publisher={American Astronomical Society},
   author={Akiyama, Kazunori and Alberdi, Antxon and Alef, Walter and Asada, Keiichi and Azulay, Rebecca and Baczko, Anne-Kathrin and Ball, David and Baloković, Mislav and Barrett, John and Bintley, Dan and Blackburn, Lindy and Boland, Wilfred and Bouman, Katherine L. and Bower, Geoffrey C. and Bremer, Michael and Brinkerink, Christiaan D. and Brissenden, Roger and Britzen, Silke and Broderick, Avery E. and Broguiere, Dominique and Bronzwaer, Thomas and Byun, Do-Young and Carlstrom, John E. and Chael, Andrew and Chan, Chi-kwan and Chatterjee, Shami and Chatterjee, Koushik and Chen, Ming-Tang and Chen 陈, Yongjun 永军 and Cho, Ilje and Christian, Pierre and Conway, John E. and Cordes, James M. and Crew, Geoffrey B. and Cui, Yuzhu and Davelaar, Jordy and Laurentis, Mariafelicia De and Deane, Roger and Dempsey, Jessica and Desvignes, Gregory and Dexter, Jason and Doeleman, Sheperd S. and Eatough, Ralph P. and Falcke, Heino and Fish, Vincent L. and Fomalont, Ed and Fraga-Encinas, Raquel and Freeman, William T. and Friberg, Per and Fromm, Christian M. and Gómez, José L. and Galison, Peter and Gammie, Charles F. and García, Roberto and Gentaz, Olivier and Georgiev, Boris and Goddi, Ciriaco and Gold, Roman and Gu 顾, Minfeng 敏峰 and Gurwell, Mark and Hada, Kazuhiro and Hecht, Michael H. and Hesper, Ronald and Ho 何, Luis C. 子山 and Ho, Paul and Honma, Mareki and Huang, Chih-Wei L. and Huang 黄, Lei 磊 and Hughes, David H. and Ikeda, Shiro and Inoue, Makoto and Issaoun, Sara and James, David J. and Jannuzi, Buell T. and Janssen, Michael and Jeter, Britton and Jiang 江, Wu 悟 and Johnson, Michael D. and Jorstad, Svetlana and Jung, Taehyun and Karami, Mansour and Karuppusamy, Ramesh and Kawashima, Tomohisa and Keating, Garrett K. and Kettenis, Mark and Kim, Jae-Young and Kim, Junhan and Kim, Jongsoo and Kino, Motoki and Koay, Jun Yi and Koch, Patrick M. and Koyama, Shoko and Kramer, Michael and Kramer, Carsten and Krichbaum, Thomas P. and Kuo, Cheng-Yu and Lauer, Tod R. and Lee, Sang-Sung and Li 李, Yan-Rong 彦荣 and Li 李, Zhiyuan 志远 and Lindqvist, Michael and Liu, Kuo and Liuzzo, Elisabetta and Lo, Wen-Ping and Lobanov, Andrei P. and Loinard, Laurent and Lonsdale, Colin and Lu 路, Ru-Sen 如森 and MacDonald, Nicholas R. and Mao 毛, Jirong 基荣 and Markoff, Sera and Marrone, Daniel P. and Marscher, Alan P. and Martí-Vidal, Iván and Matsushita, Satoki and Matthews, Lynn D. and Medeiros, Lia and Menten, Karl M. and Mizuno, Yosuke and Mizuno, Izumi and Moran, James M. and Moriyama, Kotaro and Moscibrodzka, Monika and Müller, Cornelia and Nagai, Hiroshi and Nagar, Neil M. and Nakamura, Masanori and Narayan, Ramesh and Narayanan, Gopal and Natarajan, Iniyan and Neri, Roberto and Ni, Chunchong and Noutsos, Aristeidis and Okino, Hiroki and Olivares, Héctor and Ortiz-León, Gisela N. and Oyama, Tomoaki and Özel, Feryal and Palumbo, Daniel C. M. and Patel, Nimesh and Pen, Ue-Li and Pesce, Dominic W. and Piétu, Vincent and Plambeck, Richard and PopStefanija, Aleksandar and Porth, Oliver and Prather, Ben and Preciado-López, Jorge A. and Psaltis, Dimitrios and Pu, Hung-Yi and Ramakrishnan, Venkatessh and Rao, Ramprasad and Rawlings, Mark G. and Raymond, Alexander W. and Rezzolla, Luciano and Ripperda, Bart and Roelofs, Freek and Rogers, Alan and Ros, Eduardo and Rose, Mel and Roshanineshat, Arash and Rottmann, Helge and Roy, Alan L. and Ruszczyk, Chet and Ryan, Benjamin R. and Rygl, Kazi L. J. and Sánchez, Salvador and Sánchez-Arguelles, David and Sasada, Mahito and Savolainen, Tuomas and Schloerb, F. Peter and Schuster, Karl-Friedrich and Shao, Lijing and Shen 沈, Zhiqiang 志强 and Small, Des and Sohn, Bong Won and SooHoo, Jason and Tazaki, Fumie and Tiede, Paul and Tilanus, Remo P. J. and Titus, Michael and Toma, Kenji and Torne, Pablo and Trent, Tyler and Trippe, Sascha and Tsuda, Shuichiro and Bemmel, Ilse van and van Langevelde, Huib Jan and van Rossum, Daniel R. and Wagner, Jan and Wardle, John and Weintroub, Jonathan and Wex, Norbert and Wharton, Robert and Wielgus, Maciek and Wong, George N. and Wu 吴, Qingwen 庆文 and Young, Ken and Young, André and Younsi, Ziri and Yuan 袁, Feng 峰 and Yuan 袁, Ye-Fei 业飞 and Zensus, J. Anton and Zhao, Guangyao and Zhao, Shan-Shan and Zhu, Ziyan and Algaba, Juan-Carlos and Allardi, Alexander and Amestica, Rodrigo and Anczarski, Jadyn and Bach, Uwe and Baganoff, Frederick K. and Beaudoin, Christopher and Benson, Bradford A. and Berthold, Ryan and Blanchard, Jay M. and Blundell, Ray and Bustamente, Sandra and Cappallo, Roger and Castillo-Domínguez, Edgar and Chang, Chih-Cheng and Chang, Shu-Hao and Chang, Song-Chu and Chen, Chung-Chen and Chilson, Ryan and Chuter, Tim C. and Rosado, Rodrigo Córdova and Coulson, Iain M. and Crawford, Thomas M. and Crowley, Joseph and David, John and Derome, Mark and Dexter, Matthew and Dornbusch, Sven and Dudevoir, Kevin A. and Dzib, Sergio A. and Eckart, Andreas and Eckert, Chris and Erickson, Neal R. and Everett, Wendeline B. and Faber, Aaron and Farah, Joseph R. and Fath, Vernon and Folkers, Thomas W. and Forbes, David C. and Freund, Robert and Gómez-Ruiz, Arturo I. and Gale, David M. and Gao, Feng and Geertsema, Gertie and Graham, David A. and Greer, Christopher H. and Grosslein, Ronald and Gueth, Frédéric and Haggard, Daryl and Halverson, Nils W. and Han, Chih-Chiang and Han, Kuo-Chang and Hao, Jinchi and Hasegawa, Yutaka and Henning, Jason W. and Hernández-Gómez, Antonio and Herrero-Illana, Rubén and Heyminck, Stefan and Hirota, Akihiko and Hoge, James and Huang, Yau-De and Impellizzeri, C. M. Violette and Jiang, Homin and Kamble, Atish and Keisler, Ryan and Kimura, Kimihiro and Kono, Yusuke and Kubo, Derek and Kuroda, John and Lacasse, Richard and Laing, Robert A. and Leitch, Erik M. and Li, Chao-Te and Lin, Lupin C.-C. and Liu, Ching-Tang and Liu, Kuan-Yu and Lu, Li-Ming and Marson, Ralph G. and Martin-Cocher, Pierre L. and Massingill, Kyle D. and Matulonis, Callie and McColl, Martin P. and McWhirter, Stephen R. and Messias, Hugo and Meyer-Zhao, Zheng and Michalik, Daniel and Montaña, Alfredo and Montgomerie, William and Mora-Klein, Matias and Muders, Dirk and Nadolski, Andrew and Navarro, Santiago and Neilsen, Joseph and Nguyen, Chi H. and Nishioka, Hiroaki and Norton, Timothy and Nowak, Michael A. and Nystrom, George and Ogawa, Hideo and Oshiro, Peter and Oyama, Tomoaki and Parsons, Harriet and Paine, Scott N. and Peñalver, Juan and Phillips, Neil M. and Poirier, Michael and Pradel, Nicolas and Primiani, Rurik A. and Raffin, Philippe A. and Rahlin, Alexandra S. and Reiland, George and Risacher, Christopher and Ruiz, Ignacio and Sáez-Madaín, Alejandro F. and Sassella, Remi and Schellart, Pim and Shaw, Paul and Silva, Kevin M. and Shiokawa, Hotaka and Smith, David R. and Snow, William and Souccar, Kamal and Sousa, Don and Sridharan, T. K. and Srinivasan, Ranjani and Stahm, William and Stark, Anthony A. and Story, Kyle and Timmer, Sjoerd T. and Vertatschitsch, Laura and Walther, Craig and Wei, Ta-Shun and Whitehorn, Nathan and Whitney, Alan R. and Woody, David P. and Wouterloot, Jan G. A. and Wright, Melvin and Yamaguchi, Paul and Yu, Chen-Yu and Zeballos, Milagros and Zhang, Shuo and Ziurys, Lucy},
   year={2019},
   month=apr, pages={L1} }

@ARTICLE{Ferrarese_2000,
       author = {{Ferrarese}, Laura and {Merritt}, David},
        title = "{A Fundamental Relation between Supermassive Black Holes and Their Host Galaxies}",
      journal = {\apjl},
         year = 2000,
        month = aug,
       volume = {539},
       number = {1},
        pages = {L9-L12},
          doi = {10.1086/312838},
archivePrefix = {arXiv},
       eprint = {astro-ph/0006053},
 primaryClass = {astro-ph},
       adsurl = {https://ui.adsabs.harvard.edu/abs/2000ApJ...539L...9F}
}

@article{Kaspi_2000,
   title={Reverberation Measurements for 17 Quasars and the Size‐Mass‐Luminosity Relations in Active Galactic Nuclei},
   volume={533},
   ISSN={1538-4357},
   url={http://dx.doi.org/10.1086/308704},
   DOI={10.1086/308704},
   number={2},
   journal={The Astrophysical Journal},
   publisher={American Astronomical Society},
   author={Kaspi, Shai and Smith, Paul S. and Netzer, Hagai and Maoz, Dan and Jannuzi, Buell T. and Giveon, Uriel},
   year={2000},
   month=apr, pages={631–649} }

@ARTICLE{Gebhardt_2000,
       author = {{Gebhardt}, Karl and {Kormendy}, John and {Ho}, Luis C. and {Bender}, Ralf and {Bower}, Gary and {Dressler}, Alan and {Faber}, S.~M. and {Filippenko}, Alexei V. and {Green}, Richard and {Grillmair}, Carl and {Lauer}, Tod R. and {Magorrian}, John and {Pinkney}, Jason and {Richstone}, Douglas and {Tremaine}, Scott},
        title = "{Black Hole Mass Estimates from Reverberation Mapping and from Spatially Resolved Kinematics}",
      journal = {\apjl},
         year = 2000,
        month = nov,
       volume = {543},
       number = {1},
        pages = {L5-L8},
          doi = {10.1086/318174},
archivePrefix = {arXiv},
       eprint = {astro-ph/0007123},
 primaryClass = {astro-ph},
       adsurl = {https://ui.adsabs.harvard.edu/abs/2000ApJ...543L...5G}
}

@article{Cackett_2021,
   title={Reverberation mapping of active galactic nuclei: From X-ray corona to dusty torus},
   volume={24},
   ISSN={2589-0042},
   url={http://dx.doi.org/10.1016/j.isci.2021.102557},
   DOI={10.1016/j.isci.2021.102557},
   number={6},
   journal={iScience},
   publisher={Elsevier BV},
   author={Cackett, Edward M. and Bentz, Misty C. and Kara, Erin},
   year={2021},
   month=jun, pages={102557} }

@article{Peterson_1999,
   title={Steps toward Determination of the Size and Structure of the Broad‐Line Region in Active Galactic Nuclei. XV. Long‐Term Optical Monitoring of NGC 5548},
   volume={510},
   ISSN={1538-4357},
   url={http://dx.doi.org/10.1086/306604},
   DOI={10.1086/306604},
   number={2},
   journal={The Astrophysical Journal},
   publisher={American Astronomical Society},
   author={Peterson, Bradley M. and Barth, A. J. and Berlind, P. and Bertram, R. and Bischoff, K. and Bochkarev, N. G. and Burenkov, A. N. and Cheng, F.‐Z. and Dietrich, M. and Filippenko, A. V. and Giannuzzo, E. and Ho, L. C. and Huchra, J. P. and Hunley, J. and Kaspi, S. and Kollatschny, W. and Leonard, D. C. and Malkov, Yu. F. and Matheson, T. and Mignoli, M. and Nelson, B. and Papaderos, P. and Peters, J. and Pogge, R. W. and Pronik, V. I. and Sergeev, S. G. and Sergeeva, E. A. and Shapovalova, A. I. and Stirpe, G. M. and Tokarz, S. and Wagner, R. M. and Wanders, I. and Wei, J.‐Y. and Wilkes, B. J. and Wu, H. and Xue, S.‐J. and Zou, Z.‐L.},
   year={1999},
   month=jan, pages={659–668} }

@article{Denney_2006,
   title={The Mass of the Black Hole in the Seyfert 1 Galaxy NGC 4593 from Reverberation Mapping},
   volume={653},
   ISSN={1538-4357},
   url={http://dx.doi.org/10.1086/508533},
   DOI={10.1086/508533},
   number={1},
   journal={The Astrophysical Journal},
   publisher={American Astronomical Society},
   author = {{Denney}, Kelly D. and {Bentz}, Misty C. and {Peterson}, Bradley M. and {Pogge}, Richard W. and {Cackett}, Edward M. and {Dietrich}, Matthias and {Fogel}, Jeffrey K.~J. and {Ghosh}, Himel and {Horne}, Keith D. and {Kuehn}, Charles and {Minezaki}, Takeo and {Onken}, Christopher A. and {Pronik}, Vladimir I. and {Richstone}, Douglas O. and {Sergeev}, Sergey G. and {Vestergaard}, Marianne and {Walker}, Matthew G. and {Yoshii}, Yuzuru},
   year={2006},
   month=dec, pages={152–158} }

@article{Kozlowski_2016,
	doi = {10.3847/0004-637x/826/2/118},
  
	url = {https://doi.org/10.3847%2F0004-637x%2F826%2F2%2F118},
  
	year = 2016,
	month = {jul},
  
	publisher = {American Astronomical Society},
  
	volume = {826},
  
	number = {2},
  
	pages = {118},
  
	author = {Szymon Koz{\l}owski},
  
	title = {{REVISITING} {STOCHASTIC} {VARIABILITY} {OF} {AGNs} {WITH} {STRUCTURE} {FUNCTIONS}
},
  
	journal = {The Astrophysical Journal}
}

@article{MacLeod_2010,
	doi = {10.1088/0004-637x/721/2/1014},
  
	url = {https://doi.org/10.1088%2F0004-637x%2F721%2F2%2F1014},
  
	year = 2010,
	month = {sep},
  
	publisher = {American Astronomical Society},
  
	volume = {721},
  
	number = {2},
  
	pages = {1014--1033},
  
	author = {C. L. MacLeod and {\v{Z}
}. Ivezi{\'{c}} and C. S. Kochanek and S. Koz{\l}owski and B. Kelly and E. Bullock and A. Kimball and B. Sesar and D. Westman and K. Brooks and R. Gibson and A. C. Becker and W. H. de Vries},
  
	title = {{MODELING} {THE} {TIME} {VARIABILITY} {OF} {SDSS} {STRIPE} 82 {QUASARS} {AS} A {DAMPED} {RANDOM} {WALK}},
  
	journal = {The Astrophysical Journal}
}

@article{Zu_2013,
	doi = {10.1088/0004-637x/765/2/106},
  
	url = {https://doi.org/10.1088%2F0004-637x%2F765%2F2%2F106},
  
	year = 2013,
	month = {feb},
  
	publisher = {American Astronomical Society},
  
	volume = {765},
  
	number = {2},
  
	pages = {106},
  
	author = {Ying Zu and C. S. Kochanek and Szymon Koz{\l}owski and Andrzej Udalski},
  
	title = {{IS} {QUASAR} {OPTICAL} {VARIABILITY} A {DAMPED} {RANDOM} {WALK}?},
  
	journal = {The Astrophysical Journal}
}

@article{Read_2019,
   title={The performance of photometric reverberation mapping at high redshift and the reliability of damped random walk models},
   volume={492},
   ISSN={1365-2966},
   url={http://dx.doi.org/10.1093/mnras/stz3574},
   DOI={10.1093/mnras/stz3574},
   number={3},
   journal={Monthly Notices of the Royal Astronomical Society},
   publisher={Oxford University Press (OUP)},
   author={Read, S C and Smith, D J B and Jarvis, M J and Gürkan, G},
   year={2019},
   month=dec, pages={3940–3959} }

@article{Volonteri_2021,
   title={The origins of massive black holes},
   volume={3},
   ISSN={2522-5820},
   url={http://dx.doi.org/10.1038/s42254-021-00364-9},
   DOI={10.1038/s42254-021-00364-9},
   number={11},
   journal={Nature Reviews Physics},
   publisher={Springer Science and Business Media LLC},
   author={Volonteri, Marta and Habouzit, Mélanie and Colpi, Monica},
   year={2021},
   month=sep, pages={732–743} }

@ARTICLE{Urry_1995,
       author = {{Urry}, C. Megan and {Padovani}, Paolo},
        title = "{Unified Schemes for Radio-Loud Active Galactic Nuclei}",
      journal = {\pasp},
         year = 1995,
        month = sep,
       volume = {107},
        pages = {803},
          doi = {10.1086/133630},
archivePrefix = {arXiv},
       eprint = {astro-ph/9506063},
 primaryClass = {astro-ph},
       adsurl = {https://ui.adsabs.harvard.edu/abs/1995PASP..107..803U}
}

@article{Guy_2005,
   title={SALT: a spectral adaptive light curve template  for type Ia supernovae},
   volume={443},
   ISSN={1432-0746},
   url={http://dx.doi.org/10.1051/0004-6361:20053025},
   DOI={10.1051/0004-6361:20053025},
   number={3},
   journal={Astronomy \& amp; Astrophysics},
   publisher={EDP Sciences},
   author={Guy, J. and Astier, P. and Nobili, S. and Regnault, N. and Pain, R.},
   year={2005},
   month=nov, pages={781–791} }

@article{Madau_2014,
   title={Cosmic Star-Formation History},
   volume={52},
   ISSN={1545-4282},
   url={http://dx.doi.org/10.1146/annurev-astro-081811-125615},
   DOI={10.1146/annurev-astro-081811-125615},
   number={1},
   journal={Annual Review of Astronomy and Astrophysics},
   publisher={Annual Reviews},
   author={Madau, Piero and Dickinson, Mark},
   year={2014},
   month=aug, pages={415–486} }

@ARTICLE{Grier_2013b,
       author = {{Grier}, C.~J. and {Martini}, P. and {Watson}, L.~C. and {Peterson}, B.~M. and {Bentz}, M.~C. and {Dasyra}, K.~M. and {Dietrich}, M. and {Ferrarese}, L. and {Pogge}, R.~W. and {Zu}, Y.},
        title = "{Stellar Velocity Dispersion Measurements in High-luminosity Quasar Hosts and Implications for the AGN Black Hole Mass Scale}",
      journal = {\apj},
         year = 2013,
        month = aug,
       volume = {773},
       number = {2},
          eid = {90},
        pages = {90},
          doi = {10.1088/0004-637X/773/2/90},
archivePrefix = {arXiv},
       eprint = {1305.2447},
 primaryClass = {astro-ph.CO},
       adsurl = {https://ui.adsabs.harvard.edu/abs/2013ApJ...773...90G}
}

@article{Kormendy_2013,
   title={Coevolution (Or Not) of Supermassive Black Holes and Host Galaxies},
   volume={51},
   ISSN={1545-4282},
   url={http://dx.doi.org/10.1146/annurev-astro-082708-101811},
   DOI={10.1146/annurev-astro-082708-101811},
   number={1},
   journal={Annual Review of Astronomy and Astrophysics},
   publisher={Annual Reviews},
   author={Kormendy, John and Ho, Luis C.},
   year={2013},
   month=aug, pages={511–653} }

@ARTICLE{SDSS-Shen_2015,
       author = {{Shen}, Yue and {Brandt}, W.~N. and {Dawson}, Kyle S. and {Hall}, Patrick B. and {McGreer}, Ian D. and {Anderson}, Scott F. and {Chen}, Yuguang and {Denney}, Kelly D. and {Eftekharzadeh}, Sarah and {Fan}, Xiaohui and {Gao}, Yang and {Green}, Paul J. and {Greene}, Jenny E. and {Ho}, Luis C. and {Horne}, Keith and {Jiang}, Linhua and {Kelly}, Brandon C. and {Kinemuchi}, Karen and {Kochanek}, Christopher S. and {P{\^a}ris}, Isabelle and {Peters}, Christina M. and {Peterson}, Bradley M. and {Petitjean}, Patrick and {Ponder}, Kara and {Richards}, Gordon T. and {Schneider}, Donald P. and {Seth}, Anil and {Smith}, Robyn N. and {Strauss}, Michael A. and {Tao}, Charling and {Trump}, Jonathan R. and {Wood-Vasey}, W.~M. and {Zu}, Ying and {Eisenstein}, Daniel J. and {Pan}, Kaike and {Bizyaev}, Dmitry and {Malanushenko}, Viktor and {Malanushenko}, Elena and {Oravetz}, Daniel},
        title = "{The Sloan Digital Sky Survey Reverberation Mapping Project: Technical Overview}",
      journal = {\apjs},
         year = 2015,
        month = jan,
       volume = {216},
       number = {1},
          eid = {4},
        pages = {4},
          doi = {10.1088/0067-0049/216/1/4},
archivePrefix = {arXiv},
       eprint = {1408.5970},
 primaryClass = {astro-ph.IM},
       adsurl = {https://ui.adsabs.harvard.edu/abs/2015ApJS..216....4S}
}

@ARTICLE{Schodel_2002,
       author = {{Sch{\"o}del}, R. and {Ott}, T. and {Genzel}, R. and {Hofmann}, R. and {Lehnert}, M. and {Eckart}, A. and {Mouawad}, N. and {Alexander}, T. and {Reid}, M.~J. and {Lenzen}, R. and {Hartung}, M. and {Lacombe}, F. and {Rouan}, D. and {Gendron}, E. and {Rousset}, G. and {Lagrange}, A. -M. and {Brandner}, W. and {Ageorges}, N. and {Lidman}, C. and {Moorwood}, A.~F.~M. and {Spyromilio}, J. and {Hubin}, N. and {Menten}, K.~M.},
        title = "{A star in a 15.2-year orbit around the supermassive black hole at the centre of the Milky Way}",
      journal = {\nat},
         year = 2002,
        month = oct,
       volume = {419},
       number = {6908},
        pages = {694-696},
          doi = {10.1038/nature01121},
archivePrefix = {arXiv},
       eprint = {astro-ph/0210426},
 primaryClass = {astro-ph},
       adsurl = {https://ui.adsabs.harvard.edu/abs/2002Natur.419..694S}
}

@ARTICLE{Pancoast_2014a,
       author = {{Pancoast}, Anna and {Brewer}, Brendon J. and {Treu}, Tommaso},
        title = "{Modelling reverberation mapping data - I. Improved geometric and dynamical models and comparison with cross-correlation results}",
      journal = {\mnras},
         year = 2014,
        month = dec,
       volume = {445},
       number = {3},
        pages = {3055-3072},
          doi = {10.1093/mnras/stu1809},
archivePrefix = {arXiv},
       eprint = {1407.2941},
 primaryClass = {astro-ph.GA},
       adsurl = {https://ui.adsabs.harvard.edu/abs/2014MNRAS.445.3055P}
}

@ARTICLE{Pancoast_2014b,
       author = {{Pancoast}, Anna and {Brewer}, Brendon J. and {Treu}, Tommaso and {Park}, Daeseong and {Barth}, Aaron J. and {Bentz}, Misty C. and {Woo}, Jong-Hak},
        title = "{Modelling reverberation mapping data - II. Dynamical modelling of the Lick AGN Monitoring Project 2008 data set}",
      journal = {\mnras},
         year = 2014,
        month = dec,
       volume = {445},
       number = {3},
        pages = {3073-3091},
          doi = {10.1093/mnras/stu1419},
archivePrefix = {arXiv},
       eprint = {1311.6475},
 primaryClass = {astro-ph.CO},
       adsurl = {https://ui.adsabs.harvard.edu/abs/2014MNRAS.445.3073P}
}

@ARTICLE{GravityPlus_2022,
       author = {{Gravity+ Collaboration} and {Abuter}, R. and {Alarcon}, P. and {Allouche}, F. and {Amorim}, A. and {Bailet}, C. and {Bedigan}, H. and {Berdeu}, A. and {Berger}, J. -P. and {Berio}, P. and {Bigioli}, A. and {Blaho}, R. and {Boebion}, O. and {Bolzer}, M. -L. and {Bonnet}, H. and {Bourdarot}, G. and {Bourget}, P. and {Brandner}, W. and {Cardenas}, C. and {Conzelmann}, R. and {Comin}, M. and {Cl{\'e}net}, Y. and {Courtney-Barrer}, B. and {Dallilar}, Y. and {Davies}, R. and {Defr{\`e}re}, D. and {Delboulb{\'e}}, A. and {Delplancke-Str{\"o}bele}, F. and {Dembet}, R. and {de Zeeuw}, T. and {Drescher}, A. and {Eckart}, A. and {{\'E}douard}, C. and {Eisenhauer}, F. and {Fabricius}, M. and {Feuchtgruber}, H. and {Finger}, G. and {F{\"o}rster Schreiber}, N.~M. and {Fuenteseca}, E. and {Garcia}, E. and {Garcia}, P. and {Gao}, F. and {Gendron}, E. and {Genzel}, R. and {Gil}, J.~P. and {Gillessen}, S. and {Gomes}, T. and {Gont{\'e}}, F. and {Gouvret}, C. and {Guajardo}, P. and {Guidolin}, I. and {Guieu}, S. and {Guzmann}, R. and {Hackenberg}, W. and {Haddad}, N. and {Hartl}, M. and {Haubois}, X. and {Hau{\ss}mann}, F. and {Hei{\ss}el}, G. and {Henning}, T. and {Hippler}, S. and {H{\"o}nig}, S. and {Horrobin}, M. and {Hubin}, N. and {Jacqmart}, E. and {Jocou}, L. and {Kaufer}, A. and {Kervella}, P. and {Kirchbauer}, J. -P. and {Kolb}, J. and {Korhonen}, H. and {Kreidberg}, L. and {Krempl}, P. and {Lacour}, S. and {Lagarde}, S. and {Lai}, O. and {Lapeyr{\`e}re}, V. and {Laugier}, R. and {Le Bouquin}, J. -B. and {Leftley}, J. and {L{\'e}na}, P. and {Lewis}, S. and {Lutz}, D. and {Magnard}, Y. and {Mang}, F. and {Marcotto}, A. and {Maurel}, D. and {M{\'e}rand}, A. and {Millour}, F. and {More}, N. and {Nowacki}, H. and {Nowak}, M. and {Oberti}, S. and {Olivares}, F. and {Ott}, T. and {Pallanca}, L. and {Paumard}, T. and {Perraut}, K. and {Perrin}, G. and {Petrov}, R. and {Pfuhl}, O. and {Pourr{\'e}}, N. and {Rabien}, S. and {Rau}, C. and {Riquelme}, M. and {Robbe-Dubois}, S. and {Rochat}, S. and {Salman}, M. and {Scherbarth}, M. and {Sch{\"o}ller}, M. and {Schubert}, J. and {Schuhler}, N. and {Shangguan}, J. and {Shchekaturov}, P. and {Shimizu}, T. and {Scheithauer}, S. and {Sevin}, A. and {Soenke}, C. and {Soulez}, F. and {Spang}, A. and {Stadler}, E. and {Straubmeier}, C. and {Sturm}, E. and {Sykes}, C. and {Tacconi}, L. and {Tischer}, H. and {Tristram}, K. and {Vincent}, F. and {von Fellenberg}, S. and {Uysal}, S. and {Widmann}, F. and {Wieprecht}, E. and {Wiezorrek}, E. and {Woillez}, J. and {Yaz{\i}c{\i}}, {\c{S}}. and {Zins}, G.},
        title = "{The GRAVITY+ Project: Towards All-sky, Faint-Science, High-Contrast Near-Infrared Interferometry at the VLTI}",
      journal = {The Messenger},
         year = 2022,
        month = dec,
       volume = {189},
        pages = {17-22},
          doi = {10.18727/0722-6691/5285},
archivePrefix = {arXiv},
       eprint = {2301.08071},
 primaryClass = {astro-ph.IM},
       adsurl = {https://ui.adsabs.harvard.edu/abs/2022Msngr.189...17A}
}

@Article{Lewis_2023,
    author={Lewis, Geraint F.
    and Brewer, Brendon J.},
    title={Detection of the cosmological time dilation of high-redshift quasars},
    journal={Nature Astronomy},
    year={2023},
    month={Oct},
    day={01},
    volume={7},
    number={10},
    pages={1265-1269},
    issn={2397-3366},
    doi={10.1038/s41550-023-02029-2},
    url={https://doi.org/10.1038/s41550-023-02029-2}
}

@book{VanRossum_2009_python,
 author = {Van Rossum, Guido and Drake, Fred L.},
 title = {Python 3 Reference Manual},
 year = {2009},
 isbn = {1441412697},
 publisher = {CreateSpace},
 address = {Scotts Valley, CA}
}

@Article{harris_2020_numpy,
 title         = {Array programming with {NumPy}},
 author        = {Charles R. Harris and K. Jarrod Millman and St{\'{e}}fan J.
                 van der Walt and Ralf Gommers and Pauli Virtanen and David
                 Cournapeau and Eric Wieser and Julian Taylor and Sebastian
                 Berg and Nathaniel J. Smith and Robert Kern and Matti Picus
                 and Stephan Hoyer and Marten H. van Kerkwijk and Matthew
                 Brett and Allan Haldane and Jaime Fern{\'{a}}ndez del
                 R{\'{i}}o and Mark Wiebe and Pearu Peterson and Pierre
                 G{\'{e}}rard-Marchant and Kevin Sheppard and Tyler Reddy and
                 Warren Weckesser and Hameer Abbasi and Christoph Gohlke and
                 Travis E. Oliphant},
 year          = {2020},
 month         = sep,
 journal       = {Nature},
 volume        = {585},
 number        = {7825},
 pages         = {357--362},
 doi           = {10.1038/s41586-020-2649-2},
 publisher     = {Springer Science and Business Media {LLC}},
 url           = {https://doi.org/10.1038/s41586-020-2649-2}
}

@Article{Hunter_2007_matplotlib,
  Author    = {Hunter, J. D.},
  Title     = {Matplotlib: A 2D graphics environment},
  Journal   = {Computing in Science \& Engineering},
  Volume    = {9},
  Number    = {3},
  Pages     = {90--95},
  publisher = {IEEE COMPUTER SOC},
  doi       = {10.1109/MCSE.2007.55},
  year      = 2007
}

@ARTICLE{Hinton_2016_chainconsumer,
      author = {{Hinton}, S.~R.},
       title = "{ChainConsumer}",
     journal = {The Journal of Open Source Software},
        year = 2016,
       month = aug,
      volume = 1,
         eid = {00045},
       pages = {00045},
         doi = {10.21105/joss.00045},
      adsurl = {http://adsabs.harvard.edu/abs/2016JOSS....1...45H},
   }

@article{Fine_2012_stacking,
    author = {Fine, S. and Shanks, T. and Croom, S. M. and Green, P. and Kelly, B. C. and Berger, E. and Chornock, R. and Burgett, W. S. and Magnier, E. A. and Price, P. A.},
    title = {Composite reverberation mapping},
    journal = {Monthly Notices of the Royal Astronomical Society},
    volume = {427},
    number = {4},
    pages = {2701-2710},
    year = {2012},
    month = {12},
    issn = {0035-8711},
    doi = {10.1111/j.1365-2966.2012.21248.x},
    url = {https://doi.org/10.1111/j.1365-2966.2012.21248.x},
    eprint = {https://academic.oup.com/mnras/article-pdf/427/4/2701/2919484/427-4-2701.pdf},
}

@article{Fine_2013_stacking,
    author = {Fine, S. and Shanks, T. and Green, P. and Kelly, B. C. and Croom, S. M. and Webster, R. L. and Berger, E. and Chornock, R. and Burgett, W. S. and Chambers, K. C. and Kaiser, N. and Price, P. A.},
    title = {Stacked reverberation mapping},
    journal = {Monthly Notices of the Royal Astronomical Society: Letters},
    volume = {434},
    number = {1},
    pages = {L16-L20},
    year = {2013},
    month = {06},
    issn = {1745-3925},
    doi = {10.1093/mnrasl/slt069},
    url = {https://doi.org/10.1093/mnrasl/slt069},
    eprint = {https://academic.oup.com/mnrasl/article-pdf/434/1/L16/54658147/mnrasl_434_1_l16.pdf},
}

@ARTICLE{Li_2017_stacking,
       author = {{Li}, Jennifer and {Shen}, Yue and {Horne}, Keith and {Brandt}, W.~N. and {Greene}, Jenny E. and {Grier}, C.~J. and {Ho}, Luis C. and {Kochanek}, Chris and {Schneider}, Donald P. and {Trump}, Jonathan R. and {Dawson}, Kyle S. and {Pan}, Kaike and {Bizyaev}, Dmitry and {Oravetz}, Daniel and {Simmons}, Audrey and {Malanushenko}, Elena},
        title = "{The Sloan Digital Sky Survey Reverberation Mapping Project: Composite Lags at z {\ensuremath{\leq}} 1}",
      journal = {\apj},
         year = 2017,
        month = sep,
       volume = {846},
       number = {1},
          eid = {79},
        pages = {79},
          doi = {10.3847/1538-4357/aa845d},
archivePrefix = {arXiv},
       eprint = {1712.02366},
 primaryClass = {astro-ph.GA},
       adsurl = {https://ui.adsabs.harvard.edu/abs/2017ApJ...846...79L}
}

@ARTICLE{Villafana_2023_dynamical,
       author = {{Villafa{\~n}a}, Lizvette and {Williams}, Peter R. and {Treu}, Tommaso and {Brewer}, Brendon J. and {Barth}, Aaron J. and {U}, Vivian and {Bennert}, Vardha N. and {Guo}, Hengxiao and {Bentz}, Misty C. and {Canalizo}, Gabriela and {Filippenko}, Alexei V. and {Gates}, Elinor and {Joner}, Michael D. and {Malkan}, Matthew A. and {Woo}, Jong-Hak and {Abolfathi}, Bela and {Bohn}, Thomas and {Bostroem}, K. Azalee and {Brandel}, Andrew and {Brink}, Thomas G. and {Channa}, Sanyum and {Cosens}, Maren and {Donohue}, Edward and {Halevi}, Goni and {Hood}, Carol E. and {Horst}, J. Chuck and {de Kouchkovsky}, Maxime and {Kuhn}, Benjamin and {Leonard}, Douglas C. and {Michel}, Ra{\'u}l and {B. Olaes}, Melanie Kae and {Park}, Daeseong and {Runco}, Jordan N. and {Sexton}, Remington O. and {Shivvers}, Isaac and {Spencer}, Chance L. and {Stahl}, Benjamin E. and {Stegman}, Samantha and {Walsh}, Jonelle L. and {Zheng}, WeiKang},
        title = "{What Does the Geometry of the H{\ensuremath{\beta}} BLR Depend On?}",
      journal = {\apj},
         year = 2023,
        month = may,
       volume = {948},
       number = {2},
          eid = {95},
        pages = {95},
          doi = {10.3847/1538-4357/accc84},
archivePrefix = {arXiv},
       eprint = {2304.06764},
 primaryClass = {astro-ph.GA},
       adsurl = {https://ui.adsabs.harvard.edu/abs/2023ApJ...948...95V}
}

@ARTICLE{Shen_2024_dynamical,
       author = {{Shen}, Yue and {Grier}, Catherine J. and {Horne}, Keith and {Stone}, Zachary and {Li}, Jennifer I. and {Yang}, Qian and {Homayouni}, Yasaman and {Trump}, Jonathan R. and {Anderson}, Scott F. and {Brandt}, W.~N. and {Hall}, Patrick B. and {Ho}, Luis C. and {Jiang}, Linhua and {Petitjean}, Patrick and {Schneider}, Donald P. and {Tao}, Charling and {Donnan}, Fergus. R. and {AlSayyad}, Yusra and {Bershady}, Matthew A. and {Blanton}, Michael R. and {Bizyaev}, Dmitry and {Bundy}, Kevin and {Chen}, Yuguang and {Davis}, Megan C. and {Dawson}, Kyle and {Fan}, Xiaohui and {Greene}, Jenny E. and {Gr{\"o}ller}, Hannes and {Guo}, Yucheng and {Ibarra-Medel}, H{\'e}ctor and {Jiang}, Yuanzhe and {Keenan}, Ryan P. and {Kollmeier}, Juna A. and {Lejoly}, Cassandra and {Li}, Zefeng and {de la Macorra}, Axel and {Moe}, Maxwell and {Nie}, Jundan and {Rossi}, Graziano and {Smith}, Paul S. and {Tee}, Wei Leong and {Weijmans}, Anne-Marie and {Xu}, Jiachuan and {Yue}, Minghao and {Zhou}, Xu and {Zhou}, Zhimin and {Zou}, Hu},
        title = "{The Sloan Digital Sky Survey Reverberation Mapping Project: Key Results}",
      journal = {\apjs},
         year = 2024,
        month = jun,
       volume = {272},
       number = {2},
          eid = {26},
        pages = {26},
          doi = {10.3847/1538-4365/ad3936},
archivePrefix = {arXiv},
       eprint = {2305.01014},
 primaryClass = {astro-ph.GA},
       adsurl = {https://ui.adsabs.harvard.edu/abs/2024ApJS..272...26S}
}

@ARTICLE{shen_2011_stellarcontamination,
       author = {{Shen}, Yue and {Richards}, Gordon T. and {Strauss}, Michael A. and {Hall}, Patrick B. and {Schneider}, Donald P. and {Snedden}, Stephanie and {Bizyaev}, Dmitry and {Brewington}, Howard and {Malanushenko}, Viktor and {Malanushenko}, Elena and {Oravetz}, Dan and {Pan}, Kaike and {Simmons}, Audrey},
        title = "{A Catalog of Quasar Properties from Sloan Digital Sky Survey Data Release 7}",
      journal = {\apjs},
         year = 2011,
        month = jun,
       volume = {194},
       number = {2},
          eid = {45},
        pages = {45},
          doi = {10.1088/0067-0049/194/2/45},
archivePrefix = {arXiv},
       eprint = {1006.5178},
 primaryClass = {astro-ph.CO},
       adsurl = {https://ui.adsabs.harvard.edu/abs/2011ApJS..194...45S}
}

@ARTICLE{Guo_2020_Photoionization,
       author = {{Guo}, Hengxiao and {Shen}, Yue and {He}, Zhicheng and {Wang}, Tinggui and {Liu}, Xin and {Wang}, Shu and {Sun}, Mouyuan and {Yang}, Qian and {Kong}, Minzhi and {Sheng}, Zhenfeng},
        title = "{Understanding Broad Mg II Variability in Quasars with Photoionization: Implications for Reverberation Mapping and Changing-look Quasars}",
      journal = {\apj},
         year = 2020,
        month = jan,
       volume = {888},
       number = {2},
          eid = {58},
        pages = {58},
          doi = {10.3847/1538-4357/ab5db0},
archivePrefix = {arXiv},
       eprint = {1907.06669},
 primaryClass = {astro-ph.GA},
       adsurl = {https://ui.adsabs.harvard.edu/abs/2020ApJ...888...58G}
}

@article{Fanidakis_2011_downsizing,
   title={The evolution of active galactic nuclei across cosmic time: what is downsizing?: The evolution of AGNs across cosmic time},
   volume={419},
   ISSN={0035-8711},
   url={http://dx.doi.org/10.1111/j.1365-2966.2011.19931.x},
   DOI={10.1111/j.1365-2966.2011.19931.x},
   number={4},
   journal={Monthly Notices of the Royal Astronomical Society},
   publisher={Oxford University Press (OUP)},
   author={Fanidakis, N. and Baugh, C. M. and Benson, A. J. and Bower, R. G. and Cole, S. and Done, C. and Frenk, C. S. and Hickox, R. C. and Lacey, C. and del P. Lagos, C.},
   year={2011},
   month=dec, pages={2797–2820} }

@ARTICLE{Barger_2005_downsizing,
       author = {{Barger}, A.~J. and {Cowie}, L.~L. and {Mushotzky}, R.~F. and {Yang}, Y. and {Wang}, W.-H. and {Steffen}, A.~T. and {Capak}, P.},
        title = "{The Cosmic Evolution of Hard X-Ray-selected Active Galactic Nuclei}",
      journal = {\aj},
         year = 2005,
        month = feb,
       volume = {129},
       number = {2},
        pages = {578-609},
          doi = {10.1086/426915},
archivePrefix = {arXiv},
       eprint = {astro-ph/0410527},
 primaryClass = {astro-ph},
       adsurl = {https://ui.adsabs.harvard.edu/abs/2005AJ....129..578B}
}

@ARTICLE{Vestergaard_2009_downsizing,
       author = {{Vestergaard}, M. and {Osmer}, Patrick S.},
        title = "{Mass Functions of the Active Black Holes in Distant Quasars from the Large Bright Quasar Survey, the Bright Quasar Survey, and the Color-selected Sample of the SDSS Fall Equatorial Stripe}",
      journal = {\apj},
         year = 2009,
        month = jul,
       volume = {699},
       number = {1},
        pages = {800-816},
          doi = {10.1088/0004-637X/699/1/800},
archivePrefix = {arXiv},
       eprint = {0904.3348},
 primaryClass = {astro-ph.CO},
       adsurl = {https://ui.adsabs.harvard.edu/abs/2009ApJ...699..800V}
}

@ARTICLE{Kelly_2010_downsizing,
       author = {{Kelly}, Brandon C. and {Vestergaard}, Marianne and {Fan}, Xiaohui and {Hopkins}, Philip and {Hernquist}, Lars and {Siemiginowska}, Aneta},
        title = "{Constraints on Black Hole Growth, Quasar Lifetimes, and Eddington Ratio Distributions from the SDSS Broad-line Quasar Black Hole Mass Function}",
      journal = {\apj},
         year = 2010,
        month = aug,
       volume = {719},
       number = {2},
        pages = {1315-1334},
          doi = {10.1088/0004-637X/719/2/1315},
archivePrefix = {arXiv},
       eprint = {1006.3561},
 primaryClass = {astro-ph.CO},
       adsurl = {https://ui.adsabs.harvard.edu/abs/2010ApJ...719.1315K}
}

%%%%%%%%%%%%%%%%%%%%%%%%%%%%%%%%%%%%%%%%%%%%%%%%%%

%%%%%%%%%%%%%%%%% APPENDICES %%%%%%%%%%%%%%%%%%%%%

\end{document}